\documentclass[a4paper,10pt]{2020SCGE}

\usepackage{bookmark}
\usepackage{color}
\usepackage{enumerate}
\usepackage{multirow}
\usepackage{amsmath}
\usepackage{lscape}
\usepackage{graphicx,times}
\usepackage{longtable}

\newcommand\eff{\rm eff}

\newcommand\R{\rm R}
\newcommand\K{\rm K}
\newcommand\orb{\rm orb}

\begin{document}

\ensubject{subject}

\ArticleType{Article}
\Year{2022}
\Month{February}
\Vol{65}
\No{2}
\DOI{10.1007/s11433-021-1809-8 }
\ArtNo{229711}
\ReceiveDate{September 6, 2021}
\AcceptDate{November 5, 2021}
\OnlineDate{December 29, 2021}

\title{Compact Object Candidates with K/M-dwarf Companions from LAMOST Low-resolution Survey}
{Compact Object Candidates with K/M-dwarf Companions from LAMOST Low-resolution Survey}

\author[1]{Hui-Jun Mu}{}
\author[2]{Wei-Min Gu}{{guwm@xmu.edu.cn}}%
\author[2]{Tuan Yi}{{astroyi@stu.xmu.edu.cn}}
\author[2]{Ling-Lin Zheng}{}
\author[3]{Hao Sou}{}
\author[4]{Zhong-Rui Bai}{{zrbai@bao.ac.cn}}
\author[4]{Hao-Tong Zhang}{}
\author[4]{\\Ya-Juan Lei}{}
\author[1]{Cheng-Ming Li}{}

\AuthorMark{H-J Mu}

\AuthorCitation{H.-J. Mu, W.-M. Gu, T. Yi, L.-L. Zheng, H. Sou, Z.-R. Bai, H.-T. Zhang, Y.-J. Lei, and C.-M. Li,}

\address[1]{International Laboratory for Quantum Functional Materials of Henan, and School of Physics and Microelectronics,
\\Zhengzhou University, Zhengzhou, Henan 450001, China}
\address[2]{Department of Astronomy, Xiamen University, Xiamen, Fujian 361005, China}
\address[3]{CAS Key Laboratory for Research in Galaxies and Cosmology, Department of Astronomy, University of Science
\\and Technology of China, Hefei, Anhui 230026, China}
\address[4]{National Astronomical Observatories, Chinese Academy of Sciences, Beijing 100012, China}

\abstract{
Searching for compact objects (black holes, neutron stars, or white dwarfs)
in the Milky Way is essential for understanding the stellar evolution history,
the physics of compact objects, and the structure of our Galaxy.
Compact objects in binaries with a luminous stellar companion are perfect targets for optical observations.
Candidate compact objects can be achieved by monitoring the radial velocities of the companion star.
However, most of the spectroscopic telescopes usually obtain stellar spectra at a relatively low efficiency,
which makes a sky survey for millions of stars practically impossible.
The efficiency of a large-scale spectroscopic survey, the Large Sky Area Multi-Object Fiber Spectroscopy Telescope (LAMOST),
presents a specific opportunity to search for compact object candidates,
i.e., simply from the spectroscopic observations.
Late-type K/M stars are the most abundant populations in our Galaxy.
Owing to the relatively large Keplerian velocities in the close
binaries with a K/M-dwarf companion,
a hidden compact object could be discovered and followed-up more easily.
In this study,
compact object candidates with K/M-dwarf companions are investigated with
the LAMOST low-resolution stellar spectra.
Based on the LAMOST Data Release 5,
we obtained a sample of $56$ binaries, each containing a K/M-dwarf with
a large radial velocity variation $\Delta V_{\R} > 150~{\rm km~s}^{-1}$.
Complemented with the photometric information from the Transiting Exoplanet Survey Satellite,
we derived a sample of $35$ compact object candidates,
among which, the orbital periods of $16$ sources were revealed by the light curves.
Considering two sources as examples,
we confirmed that a compact object existed in the two systems
by fitting the radial velocity curve.
This study demonstrates the principle
and the power of searching for compact objects through LAMOST.
}

\bigskip
\keywords{Spectroscopic binaries, Binary stars, Close binaries, Radial velocities, Stellar dynamics and kinematics}
\medskip
\PACS{97.80.Fk, 97.80.-d, 97.80.Fk, 98.62.Py, 98.10.+z}

\maketitle


\noindent

\clearpage
\section{Introduction}\label{sec1}

Searching for compact objects (black holes (BH), neutron
stars (NS), or white dwarfs (WD)) in the Milky Way is essential
for understanding the stellar evolution history, the physics
of compact objects, and the structure of our Galaxy.
Compact objects are remnant products at the endpoint of stellar evolution. For progenitors
at different stellar mass ranges,
the type of compact object differs (WD: less
than $8$~solar mass $(M_{\odot})$; NS: $\operatorname{8-25 M_{\odot}}$;
BH:  greater than $25 M_{\odot}$).
Most compact binaries have been discovered via signatures of accretion
\cite{Patterson1984,Joss1984,Ritter2003,Warner2003,Remillard2006,McClintock2006,Casares2014,Corral2016,Tetarenko2016}.
X-rays in X-ray binaries (XRBs) are produced by materials accreted from a
secondary companion star onto a primary star.
Recently, searching for compact objects in binary systems by utilizing
large spectroscopic/photometric surveys has become a well-known strategy.
The advantage of using optical database instead of conventional X-ray surveys is that
it enables one to discover quiescent (noninteracting) systems.
For example, by monitoring the radial velocities from large spectroscopic surveys,
Thompson et al. \cite{Thompson2019} discovered 2MASS J05215658+4359220,
which may contain a noninteracting low-mass BH ($3.3^{+2.8}_{-0.7} M_{\odot}$)
or an unexpectedly high-mass NS;
Liu et al. \cite{Liu2019} reported LB-1,
a wide binary which may contain
an unusually massive BH ($68^{+11}_{-13} M_{\odot}$).
Using a custom Monte Carlo sampler to analyze sparse, noisy, and poorly sampled radial velocities,
Price-Whelan et al. \cite{Price-Whelan2020} found faint companions
in binary systems at different masses:
high masses (BH candidates), low masses (substellar candidates),
and very close separations (mass-transfer candidates).
They identified $40$ candidate noninteracting compact-object companions.
By exploiting multiyear all-sky photometry from All-Sky Automated Survey for Supernovae,
Rowan et al. \cite{Rowan2021} identified $369$ candidates for ellipsoidal variability.
By analyzing photometric light curves,
WD+main sequence (MS) binaries showing eclipsing signals,
such as those presented by Pyrzas et al. \cite{Pyrzas2009} can be detected.
Although there are catalogs of compact objects and candidates
(i.e., refs. \cite{Bradt1983,van1995,Liu2006,Liu2007,Rebassa2016,Ren2018}),
more candidates are demanded by the astronomers to reveal the population.
Owing to the large-scale spectroscopic survey of the
Large Sky Area Multi-Object Fiber Spectroscopic Telescope (LAMOST \cite{Wang1996,Su2004,Cui2012}),
an amazing opportunity for the search for compact objects has been presented.

In this study, we are going to focus on exploiting the
spectroscopic surveys of LAMOST,
which is located at the Xinglong Observatory, northeast of Beijing, China.
It is characterized by both a large field of view (5 degrees field)
and large effective aperture ($\operatorname{3.6-4.9}$ meters).
A total of $4000$ fibers mounted on the focal plane make LAMOST facility a powerful
tool with a high spectral acquisition rate \cite{Deng2012}.
The wavelength coverage is $\operatorname{3650-9000}$~{\AA}.
Around nine million low-resolution
(with a spectral resolution $R\sim1800$) spectra
have been released through the Data Release (DR) 5 of LAMOST
\footnote{See \url{http://dr5.lamost.org/}}.
The database is also publicly available on CasJobs \footnote{See \url{http://sdss.china-vo.org/casjobs/}}.
Luo et al. \cite{Luo2015}\footnote{see \url{http://www.lamost.org/public/dr/algorithms/spectra-anlyse?locale=en}}
introduced the survey design, the observational and instrumental limitations, data reduction and analysis of LAMOST.
The spectral type and basic parameters of the co-added spectra
were derived by LAMOST stellar parameter pipeline \cite{Wu2011}.
Among all the spectra published in DR5,
about six million spectra were assigned with effective temperatures, metallicities,
surface gravities, heliocentric radial velocities, and errors.
Multi-epoch radial velocity measurements are helpful in studying
the physics of binary systems.
For instance,
Qian et al. \cite{Qian2019} presented $\sim 256,000$ spectroscopic binary
or variable candidates with radial velocity variation
$\Delta V_{\R}>10~{\rm km~s}^{-1}$.
Gao et al. \cite{Gao2017} estimated the fraction of binary stars
by repeating spectral observations from LAMOST.
Yi, Sun \& Gu \cite{Yi2019} predicted around $400$ BH binary candidates
could be found by the LAMOST surveys.
Recently, for binaries with unknown orbital periods,
Gu et al. \cite{Gu2019} proposed a method to search for
stellar-mass BH candidates with giant companions from
spectroscopic observations.
For binaries with known orbital periods,
Zheng et al. \cite{Zheng2019} showed that the mass of an optically invisible object
in the binary can be well-constrained.
Since only co-added spectrum on the same observation night was released in LAMOST DR5,
the identified binary systems usually have a longer orbital period ($>1$~\rm day).
Recently, Bai et al. \cite{Bai2021} released corresponding $\operatorname{single-epoch}$ data of all sources
in LAMOST DR5 general catalog (hereafter DR5GC).
They presented the first data release of LAMOST low-resolution
single-epoch spectra
\footnote{see \url{http://dr5.lamost.org/sedr5/}},
typically not less than three exposures at each observation night.
Their catalog was perfectly suitable to study close binaries, in particular,
to search for spectroscopic binaries in short orbital periods ($<1$~\rm day)
using radial velocity methods.
Close binary candidates based on the database
from Bai et al. \cite{Bai2021} would be available at Yuan et al. (in preparation).
We can search for compact object candidates with short periods ($<1$~\rm day)
based on $V_{\R}$ from Bai et al. \cite{Bai2021}.
Moreover, these binaries are more accessible to follow-up dynamical
measurement with spectroscopic observations.

This study focuses on the close binary comprising a compact object
and a K/M-dwarf companion
from the single-lined spectroscopic binaries of DR5GC.
The rest of this study is organized as follows.
The sample and data analyses are presented in sect.~~\ref{sec2}.
The results are presented in sect.~~\ref{sec3}.
Summary and discussion are presented in sect.~~\ref{sec4}.

\section{Sample and Analyses}\label{sec2}

The catalog of single-lined spectroscopic binaries with
multi-epoch observations was presented by Bai et al. \cite{Bai2021}.
A series of frequently-used lines (refer to Table~1 of \cite{Bai2021})
were selected to calculate $V_{\R}$.
The wavelength bands used to calculate K/M-dwarfs are $\operatorname{4550 - 5300}$~{\AA} and $\operatorname{6350-9000}$~{\AA}.
$V_{\R}$ was measured by minimizing $\chi^{2}$ between
the spectrum and its best template.
$V_{\R}$ can be well-estimated from spectroscopic
observations with the signal-to-noise ratio $\rm S/N > 10$ \cite{Bai2021}.
Based on the catalog of Bai et al. \cite{Bai2021},
binaries with K/M-dwarf companions are
selected by containing repeated radial
velocity measurements (at least three times) within $3^{''}$.
In this study, we select candidates according to the following criteria:

\begin{enumerate}[(i)]
\item signal-to-noise ratio $\rm S/N > 10$ is required in the $g$-band;
\item single-lined spectroscopic binary systems only;
\item large radial velocity variation $\Delta V_{\R} > 150~{\rm km~s}^{-1}$;
\item spectral type classified as K/M-dwarfs;
\item not eclipsing binaries.
\end{enumerate}

Although the initial cut of criteria $\operatorname{(i-iii)}$ is effective at excluding
binaries with small $V_{\R}$ variations,
some bad $V_{\R}$ measurements still remained.
Thus, we visually inspect each spectrum by reexamining
the profile (single peak) and center position ($V_{\R}$)
of the major lines in K/M-dwarfs,
e.g., $\rm H\alpha$ and $\rm H\beta$.
Spectra which characterized double-lined features
or poorly measured radial velocities are ruled out.
In a single-lined spectroscopic binary,
an optically visible star and unseen object are denoted
as $M_{1}$ and $M_{2}$, respectively.
There are two possibilities for the object $M_{2}$:
$(\rm I)$, an MS star with relatively low mass
and luminosity (empirically,  for  $L_{2}/L_{1} \lesssim 1/10$ or equivalently, $M_{2}/M_{1} \lesssim 1/2$);
$(\rm II)$, a compact object (WD, NS, or BH).

\begin{figure}[H]
\centering
\includegraphics[width=0.46\textwidth]{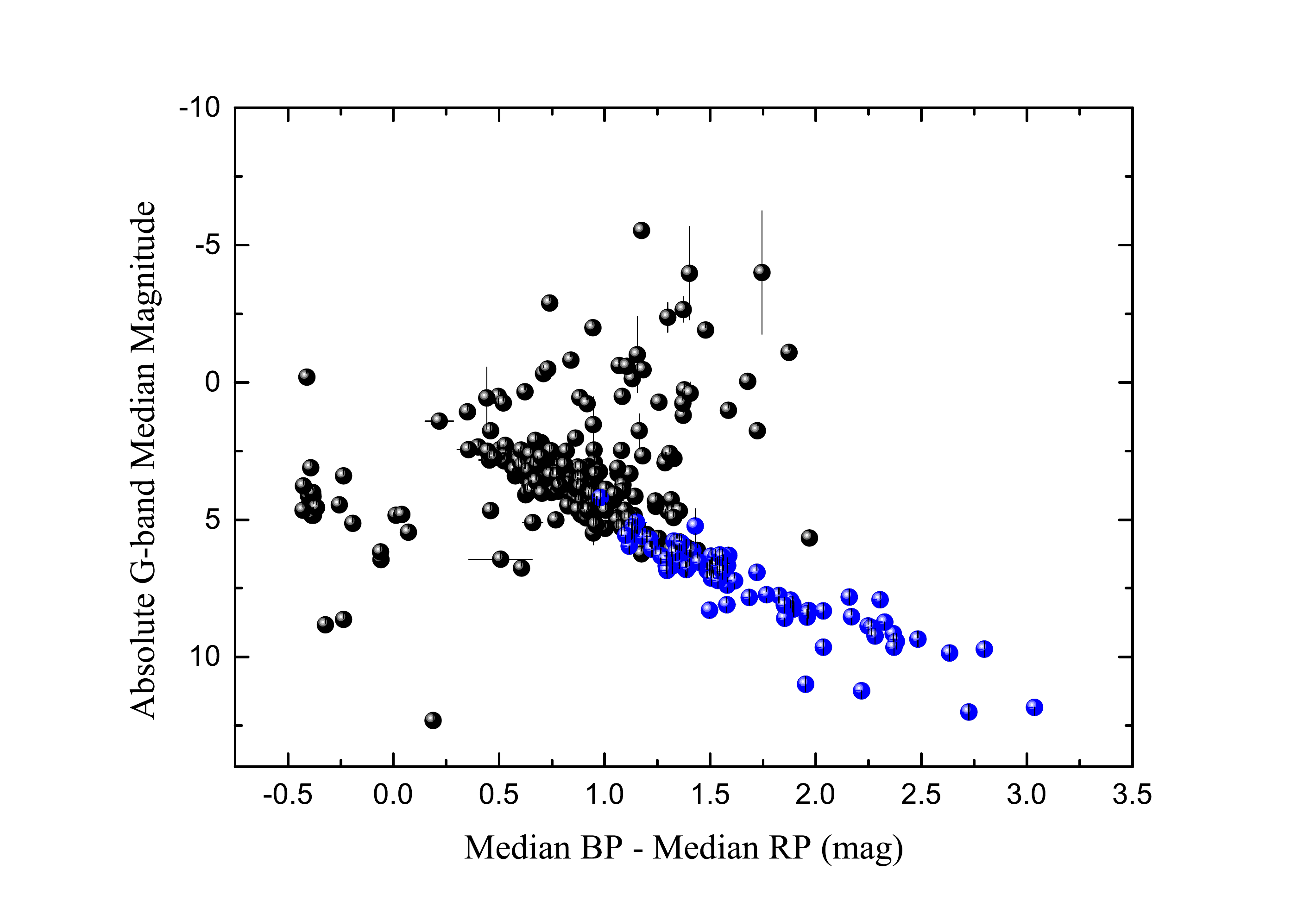}
\caption{
Comparison of $81$ K/M-dwarf binaries (blue circles: the criteria $\operatorname{i-iv}$)
and the total $308$ binaries (all circles: the criteria $\operatorname{i-iii}$)
in a color-magnitude diagram.}
\label{CMD}
\end{figure}

There are two reasons for choosing K/M-dwarfs.
First, K/M-dwarfs have a relatively low mass and thus
generally have large Keplerian velocity in close binaries.
Since the average uncertainty of the radial velocity measurements is approximately
$5~{\rm km~s}^{-1}$ \cite{Deng2012} in the low-resolution spectra,
larger radial velocity variations will reduce
the relative uncertainty for good constraints in our analyses.
Second, whether the unseen object is a compact candidate or
a dimmer dwarf is much more easily distinguished due
to the larger discrepancy between
the masses for these two object types. Namely,
an MS star with relatively low mass and luminosity
is far lighter than a BH or an NS.
Hence, it is much easier to identify the unseen object
from the amplitude of radial velocity variation.
There are $308$ binaries meet the criteria $\operatorname{(i-iii)}$ in total.
The sample has been narrowed down to $81$ single-lined binaries
with K/M-dwarf companions after imposing constraints $\operatorname{(i-iv)}$.
We cross-match the $308$ sources with Gaia early data release 3 (Gaia EDR3) \cite{Gaia2020}
using a matching radius of $3^{''}$.
G-band magnitude and median $BP-RP$ color were adopted from Gaia EDR3.
The G-band extinction is taken into account.
Figure~\ref{CMD} shows the locations of $308$ sources in
the color-magnitude diagram.
It is seen that the sources in our sample
are confined to the MS.

Since the Transiting Exoplanet Survey Satellite (TESS~\cite{Ricker2015})
\footnote{See \url{https://mast.stsci.edu/portal/Mashup/Clients/Mast/Portal.html}}
is optimized to observe
low-mass stars that potentially harbor exoplanets,
it is a perfect cross-match with the sample of K/M-dwarfs
in this study. We implement high cadence
photometry from the TESS to exclude contaminations
such as eclipsing binaries and search for periods
of the candidate sources. We use the open-source
tool, \texttt{eleanor}, to extract TESS light curves
\cite{Ricker2015,Feinstein2019}.
The target pixel files ($15 \times 15$ pixels) were
cut out from the TESS full-frame images.
We remove flux data that were distributed outside of $3~\sigma$
and flux with large error bars (top $5\%$).
The period was searched by using the
generalized Lomb-Scargle periodogram \cite{Scargle1982,Zechmeister2009}.
We visually inspect each light curve to exclude $25$ eclipsing binaries.
There are $19$ sources that show ellipsoidal modulation \cite{Morris1985,Morris1993} effects,
i.e., the light curve features a quasi-sinusoidal shape due to the
rotation of the tidally distorted star. Light curves are presented
in the Figures~S1 and S2 of \textcolor{blue}{Supplementary Material}, respectively.
Interestingly, some of the ellipsoidal light curves show a certain
degree of asymmetry between the two maxima, which could originate from some unknown orbital modulations rather than the pure ellipsoidal.
This phenomenon is called the O'Connell effect \cite{OConnell1951}.
Traditionally, these asymmetries are often thought to be explained by
models of dark spots \cite{Bopp1980,Broens2013,Li2014,Li2015},
hot spots \cite{Soonthornthum2003,Virnina2011},
or magnetohydrodynamics \cite{McCartney1999}, which are beyond the scope of this paper.
In addition, there are $37$ sources with no significant photometric periodicity,
which require further analyses from the theoretical perspective (see below).

For simplicity, we assume a circular orbit (eccentricity $e = 0$).
The mass function for the invisible object
\cite{Remillard2006} is
\begin{equation}\label{func}
f(m_2)=\frac{m_{2}^{3} \sin^3 i}{(m_{1}+m_{2})^{2}} =
1.0361\times10^{-7}~\left( \frac{K_{1}}{{\rm km~s}^{-1}} \right)^{3}~
\frac{P_{\orb}}{\rm days},
\end{equation}
where $m_{1}=M_{1}/M_{\odot}$ and $m_{2}=M_{2}/M_{\odot}$
are dimensionless masses ($M_{\odot}$ being the solar mass),
$i$ is the inclination angle of the orbit,
$K_{1}$ is the semi-amplitude of radial velocity curve (in ${\rm km~s}^{-1}$),
and $P_{\orb}$ is the orbital period (in $\rm days$).
If the object $M_2$ is a low-mass MS star, i.e., $M_{2}\lesssim M_{1}/2$.
Then, we can obtain the following constraint for $K_{1}$ based on
Eq.~(\ref{func}):
\begin{equation}\label{KP1}
K_{1} \lesssim 81.24 \left(\frac{m_{1}}{P_{\orb}/\rm days} \right)^{1/3} ~{\rm km~s}^{-1}.
\end{equation}
Obviously, $\Delta V_{\R}/2$ can be regarded as the lower limit of the
semi-amplitude $K_{1}$, i.e., $K_1 \geqslant \Delta V_{\R}/2$,
where $\Delta V_{\R}$ is the largest variation between all
radial velocity measurements for a specific source.
Then, we can derive the following constraint,
\begin{equation}\label{KP}
\frac{\Delta V_{\R}}{2} \lesssim 81.24 \left( \frac{m_{1}}{P_{\orb}/\rm days}
\right)^{1/3}~{\rm km~s}^{-1}.
\end{equation}
Eq.~(\ref{KP}) shows the upper limit for the radial velocity variation
for Case~(I), i.e., a much fainter MS star.
Thus, if $M_{1}$ and $P_{\orb}$ are derived from the optical observations,
and the radial velocity variation is beyond the upper limit, then the
unseen object can be regarded as a compact object candidate.

For $37$ sources with no significant photometric periodicity,
a strict lower limit of orbital period $P_{\orb}^{\min}$ is evaluated as follows.
Based on the assumption that the Roche-lobe radius is not less than
the radius of the K/M-dwarf, i.e., $R_{L1}\geqslant R_{1}$ \cite{Gu2019},
there exists a lower limit for the orbital period \cite{Zheng2019}:
\begin{equation}\label{Pmin}
P_{\orb}^{\min} = 0.369(\rho_{1}/\rho_{\odot})^{-1/2}~\rm{days},
\end{equation}
where $\rho_{\odot}$ is the solar density.
Thus, $P_{\orb}^{\min}$ simply depends on the mass density $\rho_{1}$.
The $\operatorname{mass-radius}$ relation (MRR) for the
low mass ($M_{1} < 1.66 M_{\odot}$) MS stars takes the form
\cite{Demircan1991}:
\begin{equation}\label{MRR}
r_{1} = 1.06 m_{1}^{0.945},
\end{equation}
where $r_{1}=R_{1}/R_{\odot}$.
By combining eqs. (\ref{Pmin}) and (\ref{MRR}), we obtain
\begin{equation}\label{rho}
P_{\orb}^{\min} = 0.403 m_{1}^{0.9175}~\rm{days}.
\end{equation}

To evaluate $P_{\orb}^{\min}$, the mass $M_{1}$
should be estimated in advance.
We estimated $M_{1}$ by using an empirical in $\operatorname{mass-luminosity}$
relation \cite{Henry1993},
where $M_{K}$ is the
absolute magnitude in the K-band
(infrared between 2 and 3 micron) from 2MASS \cite{Cutri2003}.

\section{Results}\label{sec3}

\begin{figure}[H]
\centering
\includegraphics[width=0.46\textwidth]{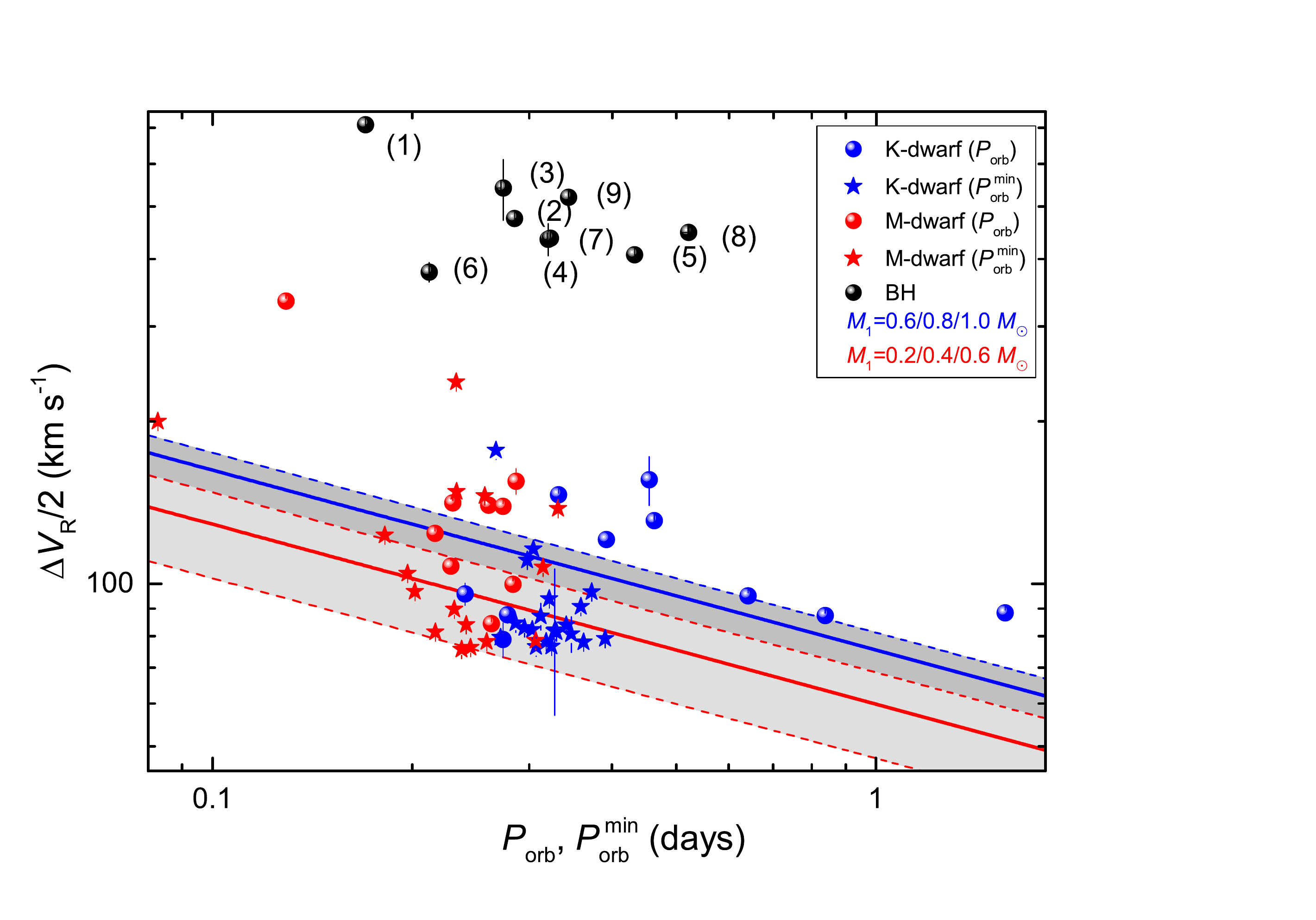}
\caption{Comparison of the observations (circles) with the theoretical
results (lines) in the $\Delta V_{\R}/2 - P_{\orb}$ diagram.
The blue and red circles represent K- and M-dwarfs, respectively.
The blue and red solid lines represent the theoretical upper limits
(eq.~(\ref{KP})) with a median mass $M_{1}=0.8 M_{\odot}$ for K-dwarfs
and $M_{1}=0.4 M_{\odot}$ for M-dwarfs, respectively.
The gray region ($\operatorname{0.6 - 1.0 M_{\odot}}$) and the light gray region
($\operatorname{0.2- 0.6 M_{\odot}}$) are typical mass ranges for K- and M-dwarfs,
respectively.
The black circles represent known BHXBs with MS companions:
(1) XTE J1118+480;
(2) GRS 1009-45;
(3) XTE J1859+226;
(4) XTE J1650-500;
(5) GS 1124-684;
(6) GRO J0422+32;
(7) 3A 0620-003;
(8) H 1705-250;
(9) GS 2000+251.}
\label{F2}
\end{figure}

A comparison of the observations with the theoretical line
(eq.~(\ref{KP})) in
$\Delta V_{\R}/2 - P_{\rm orb}$ diagram is shown in Figure~\ref{F2}.
Here, we adopt a median value $M_1 = 0.8 M_{\odot}$,
with a range of $0.6-1.0 M_{\odot}$ for K-dwarfs;
and a median value $M_1=0.4 M_{\odot}$, with
a range of $0.2-0.6 M_{\odot}$ for M-dwarfs.
We also plot a few dynamically confirmed BHs with short orbital
periods simply for comparison (black circles in Figure~\ref{F2});
data were adopted from the catalog of
Corral-Santana et al. \cite{Corral2016} and Tetarenko et al. \cite{Tetarenko2016}.,

Owing to the insufficient cadence of observations \cite{Gu2019,Zheng2019},
$\Delta V_{\R}/2$ is a lower limit of
the semi-amplitude of the radial velocity curve.
For the sources located below the theoretical lines,
they could not be selected as candidates of the compact objects for now.
$11$ K-dwarfs locate in (or above) the $0.6-1.0$$M_{\odot}$ region,
and $24$ M-dwarfs locate in (or above) the $0.2-0.6$$M_{\odot}$ region.
The parameters for the $35$ sources are shown in Table~\ref{T1}.
The combined spectra of the $35$ sources are presented in the Figure~\textcolor{blue}{S3}.
We fit the spectra of the $35$ sources using a mixture of stellar models provided
by the PyHammer package \cite{Kesseli2017}.
PyHammer has introduced in its version 2 \cite{Roulston2020},
a fully automatic technique to spectral type spectroscopic binaries.
We find that 20 sources are well-fitted by a single star template, whereas $15$
sources could be fitted with combined templates of WD+K/M-dwarf
(the classification results are presented in the last column of Table~\ref{T1}).
Our test shows that the sample is not contaminated by binary stars.
According to the theoretical analyses in sect.~\ref{sec2},
the unseen object $M_2$ of the $35$ sources is unlikely to be
a much fainter MS star with low mass and low luminosity.
Instead, $M_{2}$ could be regarded as a compact object candidate.
The astrometric parallax and the total proper motion from Gaia EDR3
\cite{Gaia2020} are also presented in Table~\ref{T1}.
We collect Gaia astrometric quality flags.
\texttt{ruwe} (the renormalised unit weight error),
\texttt{astrometric\_excess\_noise}
(excess noise of the source) and \texttt{phot\_bp\_rp\_excess\_factor}
(BP/RP flux excess factor) are added in the Table S1 of the \textcolor{blue}{Supplementary Material}.

\begin{table}[t]
\footnotesize
\centering
\caption{Parameters for the compact object candidates with K/M-dwarf companions.}
\label{T1}
\tabcolsep 6.5pt
\begin{threeparttable}
\begin{tabular*}{\textwidth}{lccccccccc}
\toprule
No. & Designation & Type1 & $T_{\eff}$ & $\varpi$ & pm & $\Delta V_{\R}$ & $P_{\orb}$ or $P_{\orb}^{\min}$ &Type2 &Type3 \\
& & &($\K$)&($\rm mas$) &($\rm mas/yr$)&(${\rm km~s}^{-1}$)& $($\rm days$)$& &   \\
a) & b) & c) & d) & e) & f) & g) & h) & i) & j) \\
\hline
(1)&J013008.78+360226.7&M3&3133 $\pm$69 &23.426 $\pm$0.039 &42.712 $\pm$0.042 &169 $\pm$1 &0.2631 &--&M6\\
(2)&J013622.94+210017.9&M0&3740 $\pm$68 &5.868 $\pm$0.021 &6.589 $\pm$0.031 &282 $\pm$2 &0.2300 &EB*&M1\\
(3)&J045242.82+122006.9&K7&3994 $\pm$98 &5.935 $\pm$0.020 &32.738 $\pm$0.030 &292 $\pm$2 &0.3322 &--&K4+DA1.5\\
(4)&J063656.98+412931.3&M0&3861 $\pm$84 &2.695 $\pm$0.055 &40.761 $\pm$0.065 &279 $\pm$6 &0.2605 &EB*&K7+DA2.0\\
(5)&J070307.60+344203.3&M0&3707 $\pm$74 &4.735 $\pm$0.030 &13.140 $\pm$0.041 &248 $\pm$3 &0.2163 &ELL&M2\\
(6)&J072710.26+341730.1&M0&3800 $\pm$83 &2.051 $\pm$0.070 &38.905 $\pm$0.097 &215 $\pm$5 &0.2286 &EB*&M2+DA5.0\\
(7)&J074432.30+395421.5&K3&4647 $\pm$56 &2.033 $\pm$0.030 &20.152 $\pm$0.044 &312 $\pm$32 &0.4552 &EB*&K3+dCK\\
(8)&J091631.61+271614.7&M2&3632 $\pm$70 &3.793 $\pm$0.071 &22.269 $\pm$0.087 &310 $\pm$17 &0.2865 &--&M4\\
(9)&J094653.67+535754.0&K7&5316 $\pm$151 &4.050 $\pm$0.022 &22.338 $\pm$0.028 &174 $\pm$2 &0.8382 &--&K2\\
(10)&J101356.31+272410.8&M3&3504 $\pm$89 &7.469 $\pm$0.064 &57.903 $\pm$0.082 &668 $\pm$8 &0.1290 &WD+M4&M4+DA5.5\\
(11)&J112306.94+400736.7&M0&3682 $\pm$74 &3.147 $\pm$0.038 &31.841 $\pm$0.050 &278 $\pm$3 &0.2738 &EB*&M2\\
(12)&J113713.64+124559.9&K4&4485 $\pm$82 &3.179 $\pm$0.020 &6.578 $\pm$0.025 &190 $\pm$3 &0.6410 &--&K4\\
(13)&J115059.18+272806.0&M2&3700 $\pm$79 &4.896 $\pm$0.044 &17.589 $\pm$0.062 &199 $\pm$3 &0.2838 &--&M3\\
(14)&J120802.64+311103.9&K3&4670 $\pm$29 &11.251 $\pm$0.019 &80.843 $\pm$0.028 &262 $\pm$0 &0.4630 &RotV*&K3+dCK\\
(15)&J121046.90+303403.2&K5&4478 $\pm$53 &1.962 $\pm$0.034 &15.931 $\pm$0.045 &242 $\pm$7 &0.3922 &EB*&K4+DA1.5\\
(16)&J150335.90+224322.7&K3&4916 $\pm$105 &1.437 $\pm$0.029 &11.965 $\pm$0.039 &177 $\pm$2 &1.5643 &--&K2+dCK\\
\hline
(17)&J002909.24+361323.8&M0&3873 $\pm$77 &2.938 $\pm$0.053 &30.568 $\pm$0.062 &168 $\pm$3 & 0.2411&--&M0\\
(18)&J015256.57+384413.4&M3&3673 $\pm$77 &4.354 $\pm$0.055 &23.633 $\pm$0.094 &209 $\pm$3 & 0.1968&WD+M3&M4\\
(19)&J035540.77+381549.9&M3&3323 $\pm$79 &4.644 $\pm$0.063 &50.630 $\pm$0.082 &163 $\pm$2 &  0.2167&--&M5\\
(20)&J035916.33+400732.3&M0&3691 $\pm$80 &2.900 $\pm$0.063 &5.661 $\pm$0.099 &152 $\pm$4 & 0.2447&--&M2+DA3.5\\
(21)&J041004.94+293102.0&M0&3833 $\pm$108 &6.485 $\pm$0.050 &36.240 $\pm$0.072 &215 $\pm$5 & 0.3148&EB*&M0\\
(22)&J041116.70+221522.4&M0&3814 $\pm$70 &5.641 $\pm$0.023 &71.823 $\pm$0.033 &275 $\pm$2 & 0.3315&--&K7+DA2.0\\
(23)&J050854.93+303039.0&K7&3796 $\pm$90 &2.249 $\pm$0.062 &7.299 $\pm$0.094 &353 $\pm$13 & 0.2675&--&K7+DA0.5\\
(24)&J060253.72+003558.4&M2&3612 $\pm$78 &8.501 $\pm$0.026 &21.402 $\pm$0.036 &473 $\pm$2 & 0.2329&--&M3\\
(25)&J060418.87+250218.8&K7&4228 $\pm$139 &1.222 $\pm$0.076 &4.955 $\pm$0.110 &232 $\pm$7 & 0.3044&--&K3+DA1.5\\
(26)&J063023.56+210952.8&M0&3758 $\pm$73 &4.621 $\pm$0.033 &25.715 $\pm$0.041 &291 $\pm$2 & 0.2572&--&M0+DA2.5\\
(27)&J072225.45+220525.8&M3&3386 $\pm$83 &6.671 $\pm$0.042 &71.308 $\pm$0.057 &246 $\pm$3 & 0.1819&--&M6\\
(28)&J081035.34+230444.9&M2&3664 $\pm$91 &3.537 $\pm$0.051 &14.391 $\pm$0.067 &179 $\pm$4 & 0.2314&--&M3\\
(29)&J090826.14+123648.2&M6&3100 $\pm$76 &20.474 $\pm$0.037 &204.311 $\pm$0.049 &400 $\pm$2 & 0.0827&EB*&M6\\
(30)&J093507.99+270049.2&M4&3443 $\pm$78 &5.252 $\pm$0.048 &26.204 $\pm$0.064 &193 $\pm$2 & 0.2019&ELL&M6+DA6.5\\
(31)&J093524.14+110836.2&K0&5149 $\pm$217 &1.1207$\pm$0.0478&15.213$\pm$0.065 &193 $\pm$3 & 0.3726&EB*&K2\\
(32)&J094811.23+552728.2&M0&3858 $\pm$90 &1.326 $\pm$0.058 &8.851 $\pm$0.075 &157 $\pm$5 & 0.3069&--&M1\\
(33)&J104444.64+190229.6&K7&4143 $\pm$145 &2.344 $\pm$0.032 &22.607 $\pm$0.052 &221 $\pm$4 & 0.2978&--&K4+DA1.0\\
(34)&J104531.95+582901.5&M2&3657 $\pm$85 &2.862 $\pm$0.049 &14.964 $\pm$0.056 &296 $\pm$3 & 0.2331&--&M4\\
(35)&J161922.13+081914.3&M0&3867 $\pm$67 &4.158 $\pm$0.027 &41.474 $\pm$0.034 &156 $\pm$2 & 0.2589&--&M0\\
\bottomrule
\end{tabular*}
\begin{tablenotes}
  \item[*]
${\rm a)}$ The serial number of the sources.
${\rm b)}$ Target designation.
${\rm c)}$ Spectral types reported by LAMOST.
We adopt the spectral type in this column throughout the paper to report the number of K/M dwarfs.
${\rm d)}$ The effective temperature.
${\rm e)}$ The parallax from Gaia EDR3.
${\rm f)}$ The total proper motion from Gaia EDR3.
${\rm g)}$ The largest variation of radial velocity measurement
from the single epoch spectra of LAMOST.
${\rm h)}$ The orbital period $P_{\rm orb}$ of the binaries No. (1-16).
$P_{\rm orb}^{\min}$ is adopted for the sources No. (17-35).
${\rm i)}$ The main type of the source from the SIMBAD Database.
EB, ELL, and RotV represent eclipsing binary, ellipsoidal variable star,
and rotational variable star, respectively.
${\rm j)}$ The spectral type from PyHammer classification.
\end{tablenotes}
\end{threeparttable}
\end{table}

The inclination angle is also expected to be smaller than $90^{\circ}$
by assuming a certain distribution
(not random though, due to the selection effects) of the orientations of orbital planes.
To fully solve the orbital parameters of the systems,
more follow-up observations should be conducted,
and the candidates of compact objects should be solidly confirmed.
The reasoning also holds for the sources above the theoretical lines.
Minimum mass functions
(assume $K = \Delta V_{\R}/2$ and $P_{\orb} = P$ or $P_{\orb}^{\min}$)
are shown in Table~\ref{T2}.
Moreover, the estimated mass of the K/M-dwarf companions $M_1$
and the unseen objects $M_2$ are presented in Table~\ref{T2},
where six uniform inclinations in sine
($\sin i=1$, $0.9$, $0.8$, $0.7$, $0.6$ and $0.5$)
are considered.

\begin{table*}[t]
\footnotesize
\caption{Evaluation of $M_2$ for the compact object candidates in our sample.}
\label{T2}
\tabcolsep 6pt
\centering
\begin{tabular*}{0.85\textwidth}{lccccccccc}
 \hline
No.&Designation&$f_{\rm min}$&$M_1$& $M_{2,\sin i=1}$&$M_{2,\sin i=0.9}$ &$M_{2,\sin i=0.8}$&
$M_{2,\sin i=0.7}$&$M_{2,\sin i=0.6}$& $M_{2,\sin i=0.5}$ \\
a) & b) & c) & d) & e) & f) & g) & h) & i) & j)\\\hline
(1)&J013008.78+360226.7&0.02&0.18&0.11&0.13&0.15&0.18&0.23&0.32\\
(2)&J013622.94+210017.9&0.07&0.63&0.42&0.48&0.57&0.70&0.90&1.23\\
(3)&J045242.82+122006.9&0.08&0.68&0.47&0.54&0.64&0.79&1.01&1.39\\
(4)&J063656.98+412931.3&0.07&0.66&0.45&0.52&0.62&0.75&0.96&1.32\\
(5)&J070307.60+344203.3&0.04&0.60&0.34&0.39&0.45&0.55&0.69&0.93\\
(6)&J072710.26+341730.1&0.03&0.58&0.28&0.32&0.37&0.45&0.56&0.74\\
(7)&J074432.30+395421.5&0.18&0.74&0.73&0.85&1.03&1.29&1.70&2.43\\
(8)&J091631.61+271614.7&0.11&0.50&0.47&0.55&0.66&0.83&1.09&1.54\\
(9)&J094653.67+535754.0&0.06&0.42&0.32&0.37&0.44&0.54&0.69&0.96\\
(10)&J101356.31+272410.8&0.50&0.23&0.81&1.02&1.33&1.83&2.71&4.40\\
(11)&J112306.94+400736.7&0.08&0.60&0.43&0.50&0.60&0.74&0.95&1.31\\
(12)&J113713.64+124559.9&0.06&0.78&0.44&0.50&0.59&0.72&0.91&1.22\\
(13)&J115059.18+272806.0&0.03&0.51&0.26&0.30&0.35&0.42&0.52&0.70\\
(14)&J120802.64+311103.9&0.11&0.74&0.57&0.66&0.79&0.97&1.26&1.75\\
(15)&J121046.90+303403.2&0.07&0.80&0.49&0.57&0.67&0.82&1.04&1.41\\
(16)&J150335.90+224322.7&0.11&0.86&0.63&0.73&0.87&1.06&1.37&1.89\\
(17)&J002909.24+361323.8&0.01&0.57&0.21&0.24&0.27&0.33&0.40&0.52\\
(18)&J015256.57+384413.4&0.02&0.46&0.22&0.25&0.30&0.35&0.44&0.59\\
(19)&J035540.77+381549.9&0.01&0.51&0.18&0.20&0.24&0.28&0.34&0.44\\
(20)&J035916.33+400732.3&0.01&0.58&0.19&0.21&0.25&0.29&0.36&0.46\\
(21)&J041004.94+293102.0&0.04&0.76&0.37&0.43&0.50&0.60&0.75&1.00\\
(22)&J041116.70+221522.4&0.09&0.81&0.55&0.64&0.75&0.92&1.18&1.62\\
(23)&J050854.93+303039.0&0.15&0.64&0.62&0.73&0.88&1.11&1.46&2.08\\
(24)&J060253.72+003558.4&0.32&0.55&0.86&1.03&1.28&1.65&2.28&3.44\\
(25)&J060418.87+250218.8&0.05&0.74&0.40&0.46&0.54&0.65&0.82&1.10\\
(26)&J063023.56+210952.8&0.08&0.61&0.45&0.53&0.63&0.77&0.99&1.37\\
(27)&J072225.45+220525.8&0.03&0.42&0.25&0.29&0.34&0.41&0.52&0.71\\
(28)&J081035.34+230444.9&0.02&0.55&0.22&0.25&0.29&0.34&0.42&0.55\\
(29)&J090826.14+123648.2&0.07&0.18&0.22&0.26&0.32&0.41&0.55&0.82\\
(30)&J093507.99+270049.2&0.02&0.47&0.21&0.23&0.27&0.33&0.41&0.54\\
(31)&J093524.14+110836.2&0.03&0.92&0.39&0.45&0.52&0.62&0.77&1.01\\
(32)&J094811.23+552728.2&0.02&0.74&0.25&0.28&0.32&0.38&0.47&0.61\\
(33)&J104444.64+190229.6&0.04&0.72&0.37&0.42&0.49&0.59&0.74&0.99\\
(34)&J104531.95+582901.5&0.08&0.55&0.42&0.49&0.58&0.72&0.93&1.28\\
(35)&J161922.13+081914.3&0.01&0.62&0.21&0.23&0.27&0.32&0.39&0.50\\
\hline
\end{tabular*}
\begin{tablenotes}
\item[*]
${\rm a)}$ and ${\rm b)}$ The serial number and designation of the targets,
which are the same as Table~\ref{T1}.
${\rm c)}$ The minimum mass function.
${\rm d)}$ The estimated mass of the K/M-dwarf companions.
${\rm e)}$-${\rm j)}$ The estimated mass of the unseen objects
with six uniform inclinations in sine,
i.e., $\sin i=1$, $0.9$,
$0.8$, $0.7$, $0.6$ and $0.5$.
\end{tablenotes}
\end{table*}

\begin{figure}[H]
\centering
\includegraphics[angle=0,width=0.5\textwidth,height=0.25\textheight]{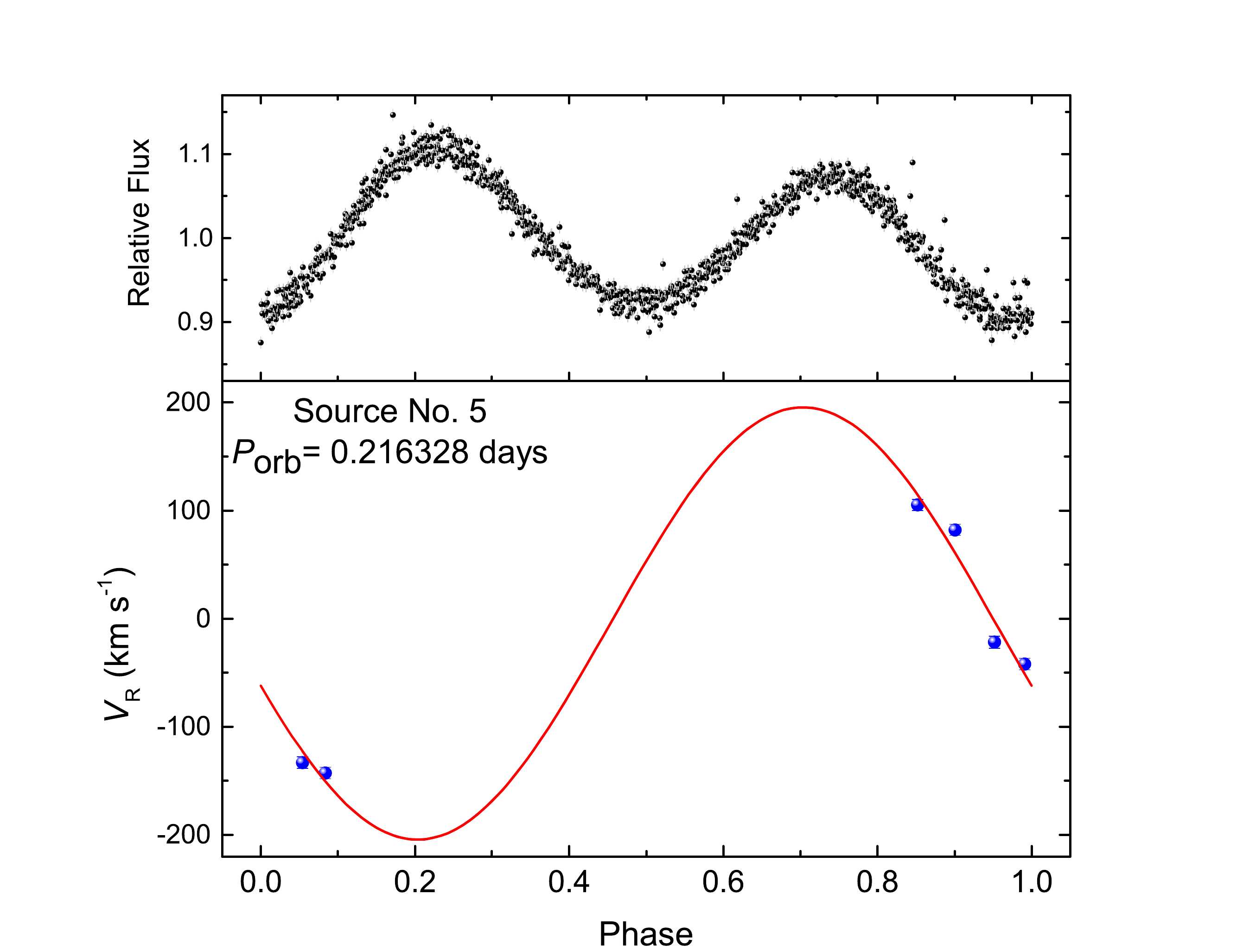}%
\includegraphics[angle=0,width=0.5\textwidth,height=0.25\textheight]{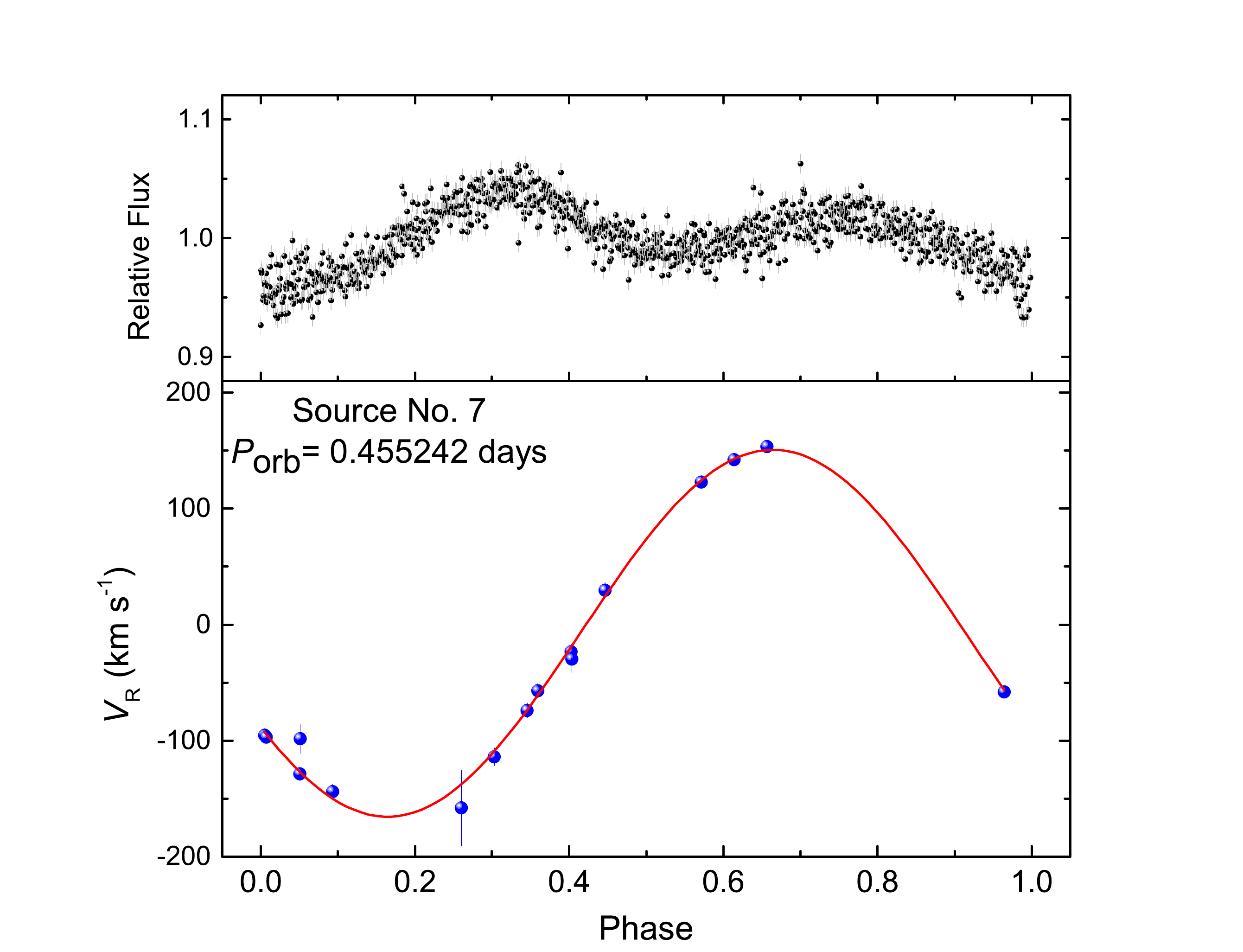}
\caption{V-mag light curve and radial velocity curve of the
example sources (No.5 and No.7).
Upper panel: the folded TESS light curve with the orbital period
derived from the $\operatorname{Lomb-Scargle}$ periodogram.
Lower panel: radial velocities (black dots; some error
bars are too small to be seen) and the $\operatorname{best-fitted}$ radial
velocity curve (red curve).}
\label{F3}
\end{figure}

\begin{figure}[H]
\centering
\includegraphics[angle=0,width=0.5\textwidth,height=0.25\textheight]{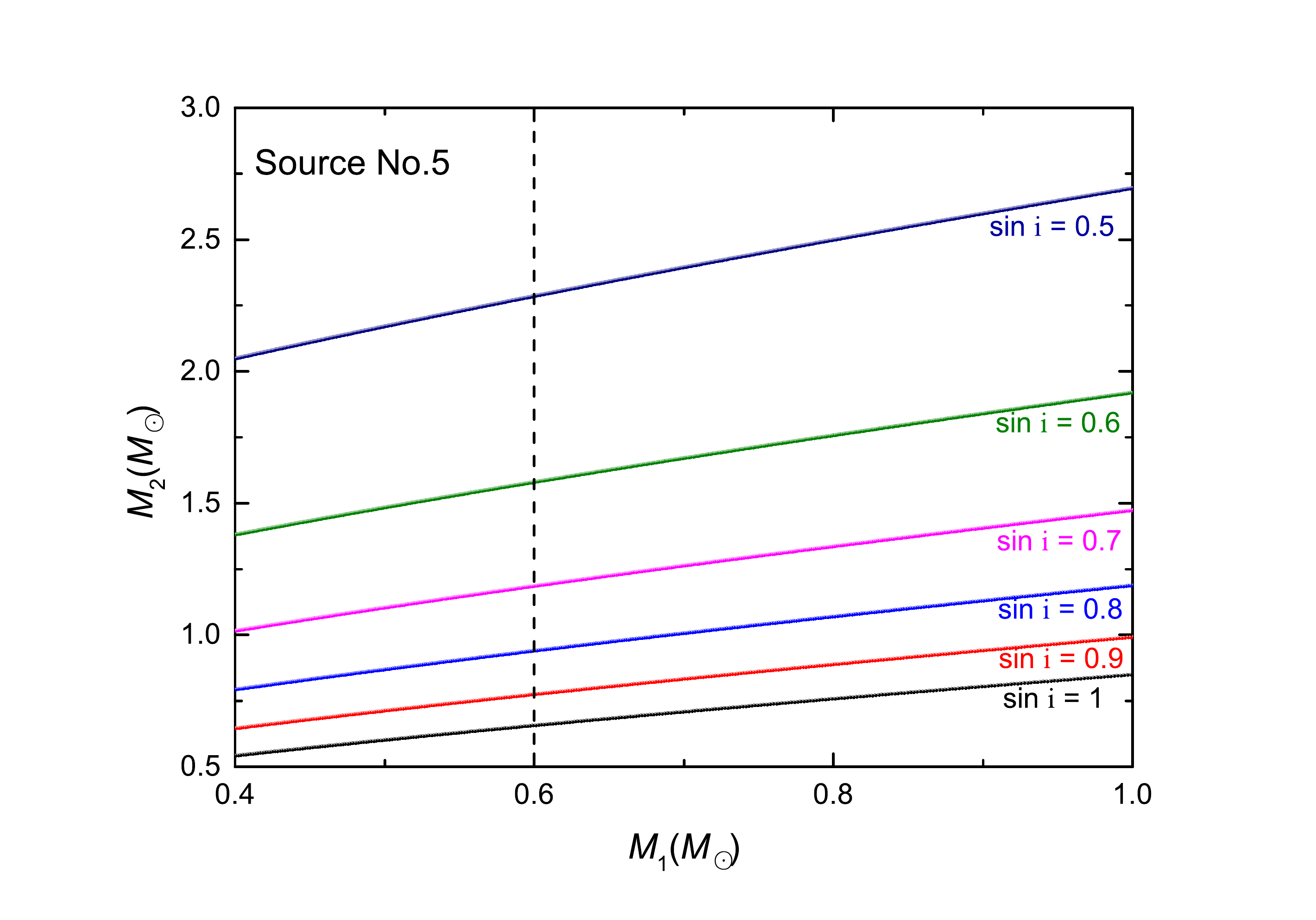}%
\includegraphics[angle=0,width=0.5\textwidth,height=0.25\textheight]{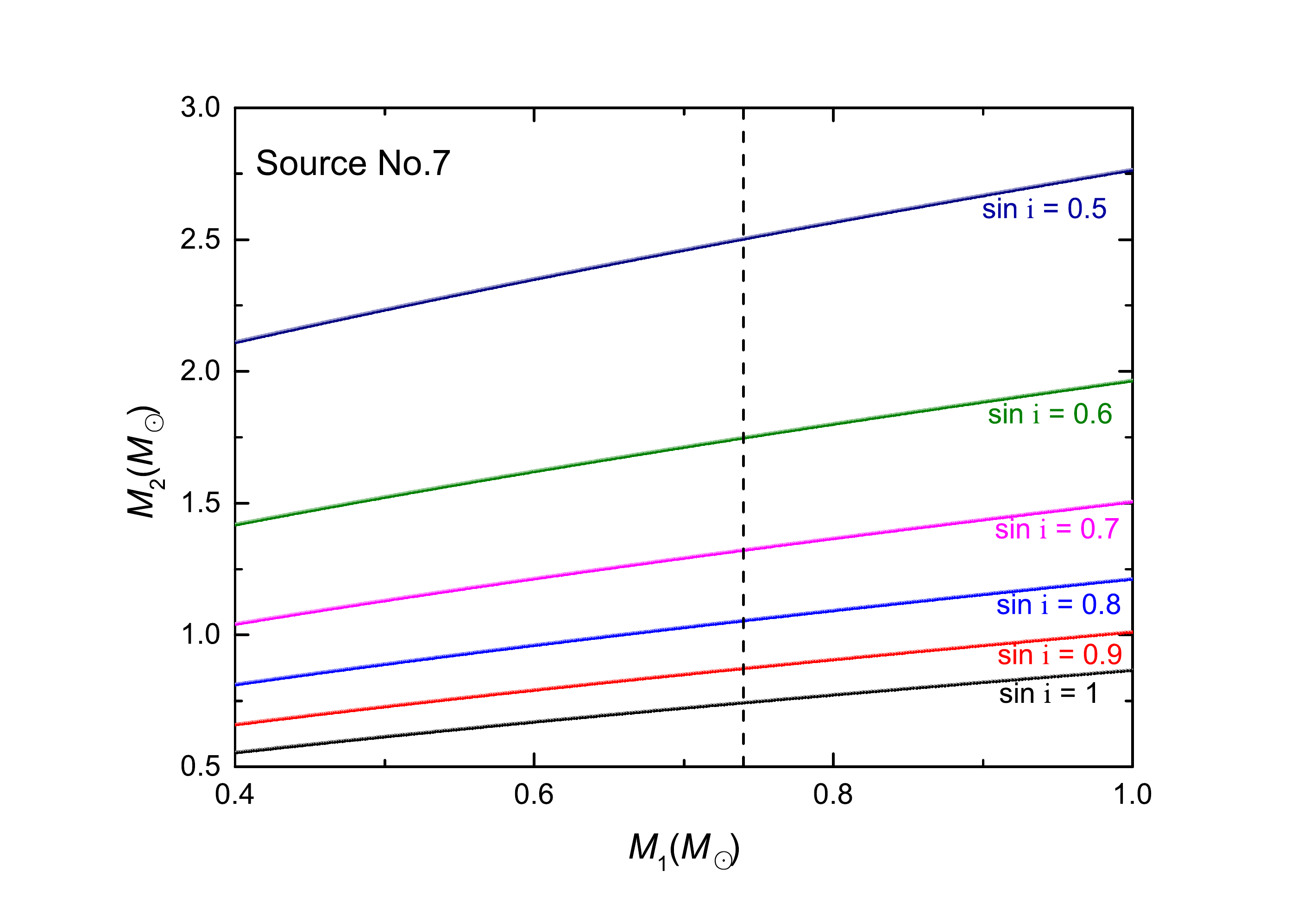}
\caption{By the results of radial velocity curve fitting,
the comparison of $M_{2}$ with $M_{1}$ about the two examples (No.5 and No.7).
Six values of the orbital inclination from $\sin i = 1.0$ to $0.5$ were considered.
The vertical dashed line represents the $M_{1}$ estimated by the empirical $\operatorname{mass-luminosity}$ relation.}
\label{F4}
\end{figure}

As mentioned in the selection process, we have excluded
apparent eclipsing binaries.
We cross-matched our sources with SIMBAD Database
\footnote{See \url{http://simbad.u-strasbg.fr/simbad/}}
using a matching radius of $3^{''}$.
Several sources in our sample were previously known as the EB*WUMa type,
as denoted in the last column in Table~\ref{T1}.
However, the following argument shows that at least some sources may not be eclipsing binaries.
No.5 and No.7 serve as two examples for more detailed analyses.
The targets' observation log is shown in Table~\ref{T3}.
Using the orbital period from the light curve,
we folded and fitted the radial velocity data with a sinusoidal function:
$V_{\rm R}=K_{1} \sin(2 \pi/P_{\orb}~(t-t_{0}))+V_{0}$,
where $K_{1}$ is the radial velocity curve semi-amplitude,
$t_{0}$ is an epoch at orbital phase $0$,
and $V_{0}$ is the radial velocity of the system's center of mass
(with respect to the Earth).
Figure~\ref{F3} shows the phase-folded light curve (upper panels)
and the best-fitted radial velocity curve (lower panels).
The semi-amplitude $K_1 = 195\pm 28~{\rm km~s}^{-1}$ (No.5)
and $158\pm 2~{\rm km~s}^{-1}$ (No.7),
which is larger than the radial velocity variation
$\Delta V_{\R}/2 = 124\pm 4~{\rm km~s}^{-1}$ (No.5)
and $156\pm 17~{\rm km~s}^{-1}$ (No.7) measured by the single-epoch spectra.
The corresponding mass functions of the two sources are
$f(m_2)\sim0.18 M_{\odot}$ (No.5) and $0.19 M_{\odot}$ (No.7), respectively.
For $M_1 \sim0.60-0.65 M_{\odot}$, $M_2$ is derived
as $M_2 \sim 0.70-0.75 M_{\odot}$, which is always higher than $M_1$.
$M_2$ under typical inclination angles are also shown in Figure~\ref{F4}.
Thus, $M_2$ may be regarded as a compact object candidate,
and the system is unlikely an eclipsing binary of the EB*WUMa type.

\begin{table*}[t]
\footnotesize
\caption{Radial velocity measurements of of sources No. 5 and No. 7.}
\label{T3}
\tabcolsep 12pt
\centering
\begin{tabular*}{0.55\textwidth}{ccccc}
 \hline
\multirow{2}{*}{No.}&\multirow{2}{*}{obsdate} &\multirow{2}{*}{$\rm S/N_{r}$}& \multirow{2}{*}{$\rm JD_{0}$}&$V_{\R}$\\
 & & & & $({\rm km~s}^{-1})$\\
a) & b) & c) & d) & e)\\
\hline
\multirow{6}{*}{No.5}&\multirow{3}{*}{2014 Jan. 22}	&	\multirow{3}{*}{59.44}&6680.129167&105.1 	$\pm$	5.3 	\\
  & & &6680.150694&	-21.8 	$\pm$	5.4 	\\
  & & &6680.172917&	-133.1 	$\pm$	5.3 	\\
\cline{2-5}
&\multirow{3}{*}{2015 Mar. 22}	&	\multirow{3}{*}{71.48}&7103.971528&82.1 	$\pm$	5.2 	\\
  & &  &7103.990972&	-42.0 	$\pm$	5.3 	\\
  & &  &7104.011111&	-142.9 	$\pm$	5.3 	\\
\hline
\multirow{16}{*}{No.7}&\multirow{3}{*}{2014 Dec. 12}	&\multirow{3}{*}{32.79}&7004.281250&	-57.1	$\pm$	5.9 	\\
  &  &  &7004.300694&	-23.6	$\pm$	6.5 	\\
  &  &  &7004.320833&	29.4	$\pm$	6.3 	\\
\cline{2-5}
&\multirow{3}{*}{2015 Jan. 02}	&\multirow{3}{*}{27.40}&7025.177083&	-158.2	$\pm$	32.6 	\\
  &  &  &7025.196528&	-114.2	$\pm$	7.7 	\\
  &  &  &7025.215972&	-74.2	$\pm$	6.3 	\\
\cline{2-5}
&2015 Mar. 07	&16.98&7088.976389&	-29.68	$\pm$	11.8 	\\
\cline{2-5}
&\multirow{3}{*}{2015 Mar. 18}	&\multirow{3}{*}{65.93}&7099.978472&	122.8	$\pm$	5.2 	\\
  &  &  &7099.997917&	142.0	$\pm$	5.3 	\\
  &  &  &7100.017361&	153.3	$\pm$	5.4 	\\
\cline{2-5}
&\multirow{3}{*}{2016 Jan. 12}	&\multirow{3}{*}{42.92}&7400.180556&	-95.4	$\pm$	5.6 	\\
  &  &  &7400.201389&	-128.8	$\pm$	5.6 	\\
  &  &  &7400.220833&	-144.2	$\pm$	6.5 	\\
\cline{2-5}
&\multirow{3}{*}{2016 Nov. 10}	&\multirow{3}{*}{97.44}&7703.353472&	-58.0	$\pm$	5.1 	\\
  &  &  &7703.372917&	-97.0	$\pm$	5.1 	\\
  &  &  &7703.393056&	-98.6	$\pm$	12.6 	\\
\hline
\end{tabular*}
\begin{tablenotes}
\item[*]
${\rm a)}$ the number of the source.\\
${\rm b)}$ target observation date (Beijing time).\\
${\rm c)}$ the signal-to-noise ratio in the r-band.\\
${\rm d)}$ Julian Date ($\rm JD -2,450,000$).\\
${\rm e)}$ radial velocity measured by the single epoch spectra.\\
\end{tablenotes}
\end{table*}

In addition, a short comment on three other systems in our sample is presented.
Notably, No.10 \cite{Parsons2015}, No.18 \cite{Law2012} and No.30 \cite{Ren2018}
were previously identified as WD+M-dwarf eclipsing binaries systems, which were re-discovered by our selection procedures.
It is seen from Figure~\textcolor{blue}{S3} that,
No.10 and No.30 are obviously WD+M-dwarfs.
No.2 was identified as (1) an X-ray source according to Chandra's observation;
(2) a UV-excess star cataloged in Bai et al. \cite{Bai2018},
which indicate that the source could be an accreting system
with a visible star filling the Roche lobe.
In addition, a recent study on source No.11, proposed that the system contains an NS or an unusually high-mass WD, by utilizing the spectroscopy of LAMOST  and Palomar 200-inch telescope, and the high-cadence photometry of TESS (Yi, Gu, Zhang et al., under review).

\section{Summary and Discussion}\label{sec4}

In this study, we have focused on the compact object candidates
in close binaries with K/M-dwarf companions.
For the case that the unseen object in the binary is an MS star with lower
luminosity, we have obtained an upper limit for the semi-amplitude of
radial velocity variation (eq.~(\ref{KP})).
In other words, if the radial velocity variation derived from
the LAMOST spectra is beyond the upper limit, the unseen
object can be regarded as a compact object candidate.
We have derived $35$ compact object candidates,
by using only the limited exposures from LAMOST low-resolution survey,
aided with photometric light curves.
This study demonstrates the principle
and power of searching for compact objects through LAMOST
low-resolution survey.

Notably, majority of
K/M binaries, including those with compact companions,
likely have long periods. The systems are
harder to detect because the induced $V_{\R}$ variations are smaller.
Thus the systems are not investigated in the current work.
Candidates in our sample need more careful analysis,
e.g., contaminations from subtle double-line spectroscopic
binaries (e.g., equal mass pairs),
the inaccurate estimation of mass or stellar parameters,
the case of triple systems, or higher-order multiples.
For instance, trying to exclude spectroscopic binaries through visual inspection
and PyHammer \cite{Kesseli2017,Roulston2020}
could not provide a quantitative estimation of the false classification rate.
A more rigorous approach to exclude contaminations should be similar to
the spectral fitting of El-Badry et al. \cite{El-Badry2018}.
However, since the resolution and signal-to-noise ratio of the spectra used
in this study is relatively low, it is difficult to give a very precise classification.
Therefore, some candidates in our sample might be spectroscopic binaries.
These problems can be improved in future works, for instance,
using the higher-resolution spectroscopy from LAMOST
medium-resolution survey (R $\sim$ 7500)\cite{Liu2020}
and the SDSS APOGEE data (R $\sim$ 22000).

Recently, Shao et al. \cite{Shao2019} simulated the Galactic
population of detached BH binaries with normal-star companions.
They showed that it is difficult for conventional models
to produce BH low-mass XRBs. However,
some investigations of massive star evolution suggested
that the BH progenitors have masses as low as $\sim 15 M_{\odot}$
\cite{Sukhbold2016}.
Based on such a result, Shao et al. \cite{Shao2019} showed that
the overall population of detached BH binaries is dominated by
those with relatively low-mass companions.
Shao et al. \cite{Shao2019} also predicted that the total number of
detached BH binaries with MS companions is more than $4,000$,
among which $700$ systems
have companions brighter than $20$ mag.
In this spirit, the compact object candidates in our sample are worth
follow-up observations for precise dynamical measurement.

Despite using the spectroscopic and photometric data from surveys,
astrometric surveys can also be used to search for compact objects in binary systems.
For instance, Gandhi et al. \cite{Gandhi2020} proposed a method of using
the astrometric excess noise,
to discover candidate X-ray emitting sources (accreting binaries).
The newly released data (EDR3) of Gaia
\footnote{\url{https://www.cosmos.esa.int/web/gaia/early-data-release-3}}
will present an unprecedented opportunity for
compact object searching tasks.
A joined study of data from Gaia, LAMOST,
and various photometric surveys can extend the capability of discoveries.

\Acknowledgements{
This work was supported by the National Natural Science Foundation
of China (NSFC) under grants 12103047, 11925301, 12033006, and 12005192,
and the National Key Research
and Development Program of China (2019YFA0405000).
We acknowledge the science research grants from the
China Manned Space Project with NO. CMS-CSST-2021-B07,
acknowledge support from
the Project funded by China Postdoctoral Science Foundation
under grants 2019TQ0288, 2020TQ0287, and 2020M672255, and 2021M702742.
and acknowledge the Natural Science Foundation of Henan Province of China 212300410290.
This work has made use of data products from the Guoshoujing Telescope
(the Large Sky Area Multi-Object Fiber Spectroscopic Telescope, LAMOST).
LAMOST is a National Major Scientific Project built by the Chinese
Academy of Sciences. Funding for the project has been provided by
the National Development and Reform Commission.
LAMOST is operated and managed by the National Astronomical
Observatories, Chinese Academy of Sciences.
We acknowledge the use of public TESS data
from pipelines at the TESS Science Office and at the
TESS Science Processing Operations Center.
We thank Yi-Ze Dong, Yu Bai, Fan Yang, Mou-Yuan Sun,
and Jin-Bo Fu for beneficial discussions during the derivation of our sample,
and thank the referee for helpful suggestions that improved the manuscript.
}

\InterestConflict{The authors declare that they have no conflict of interest.}

\def\apj{{Astrophys.\ J.}}
\def\apjs{{Astrophys.\ J.\ Suppl.\ Ser.}}
\def\apjl{{Astrophys.\ J.\ Lett.}}
\def\mnras{{Mon.\ Notic.\ Roy.\ Astron.\ Soc.}}
\def\raa{{Res.\ Astron.\ Astrophys.}}
\def\araa {{Annu.\ Rev.\ Astron.\ Astrophys.}}
\def\apss {{Astrophys.\ Space.\ Sci.}}
\def\ssp {{Space.\ Sci.\ Rev.}}
\def\aap{{Astron.\& Astrophys.}}

 \clearpage

 \section{Supplementary Material}

 \newcommand{\beginsupplement}{%
         \setcounter{table}{0}
         \renewcommand{\thetable}{S\arabic{table}}%
         \setcounter{figure}{0}
         \renewcommand{\thefigure}{S\arabic{figure}}%
      }

 \beginsupplement

 \begin{figure}[H]
 \centering
 \includegraphics[angle=0,width=0.49\textwidth,height=0.2\textheight]{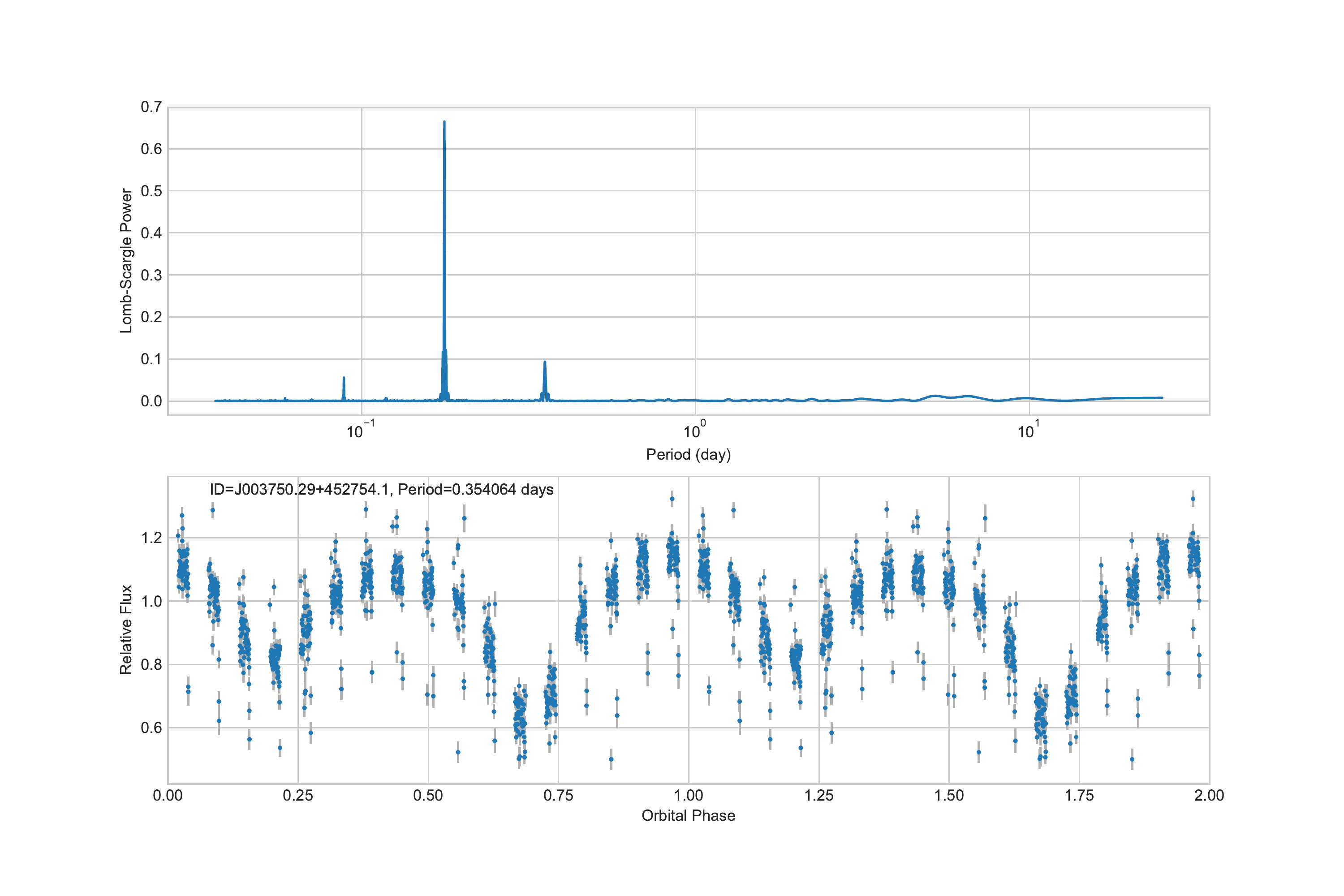}%
 \includegraphics[angle=0,width=0.49\textwidth,height=0.2\textheight]{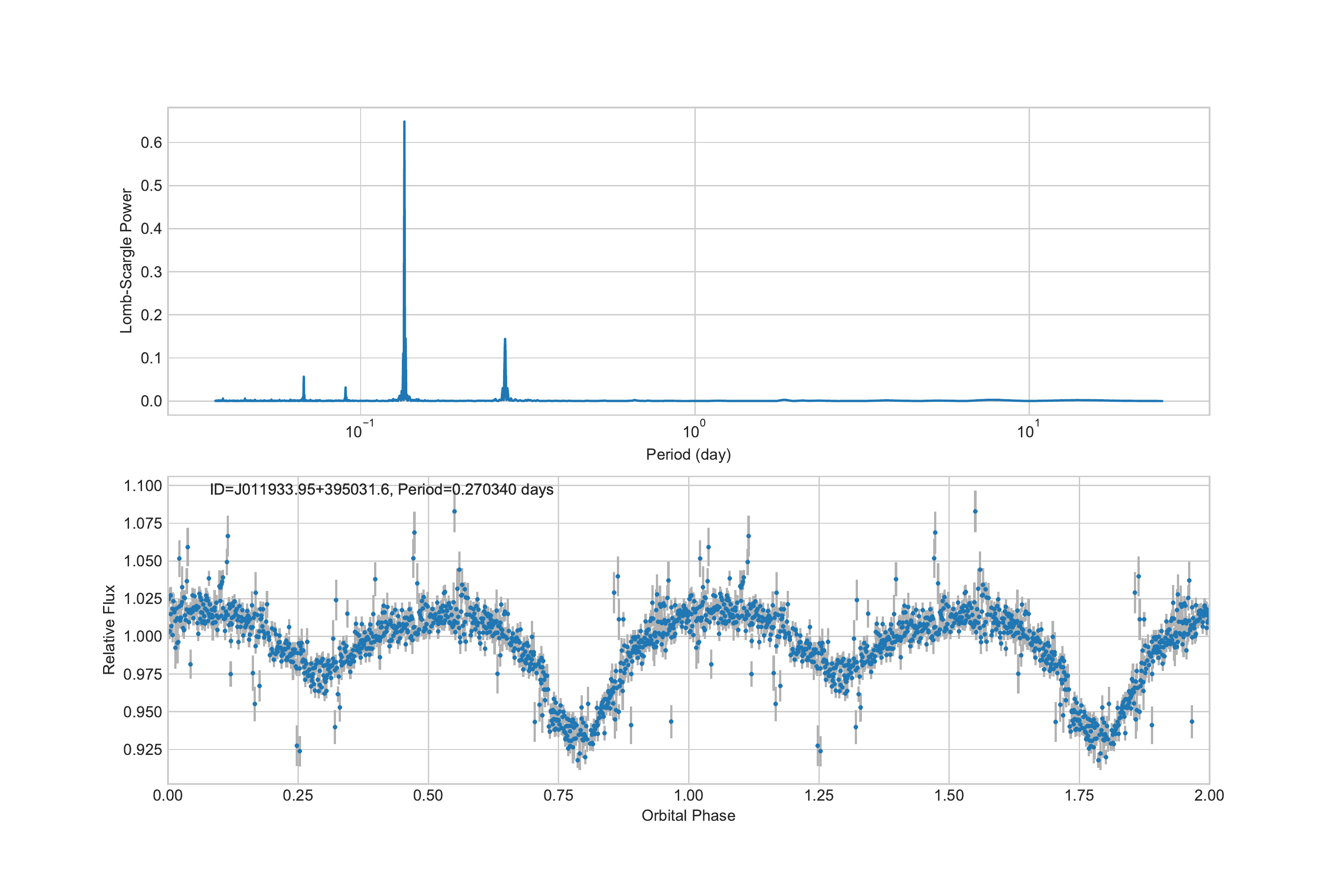}
 \includegraphics[angle=0,width=0.49\textwidth,height=0.2\textheight]{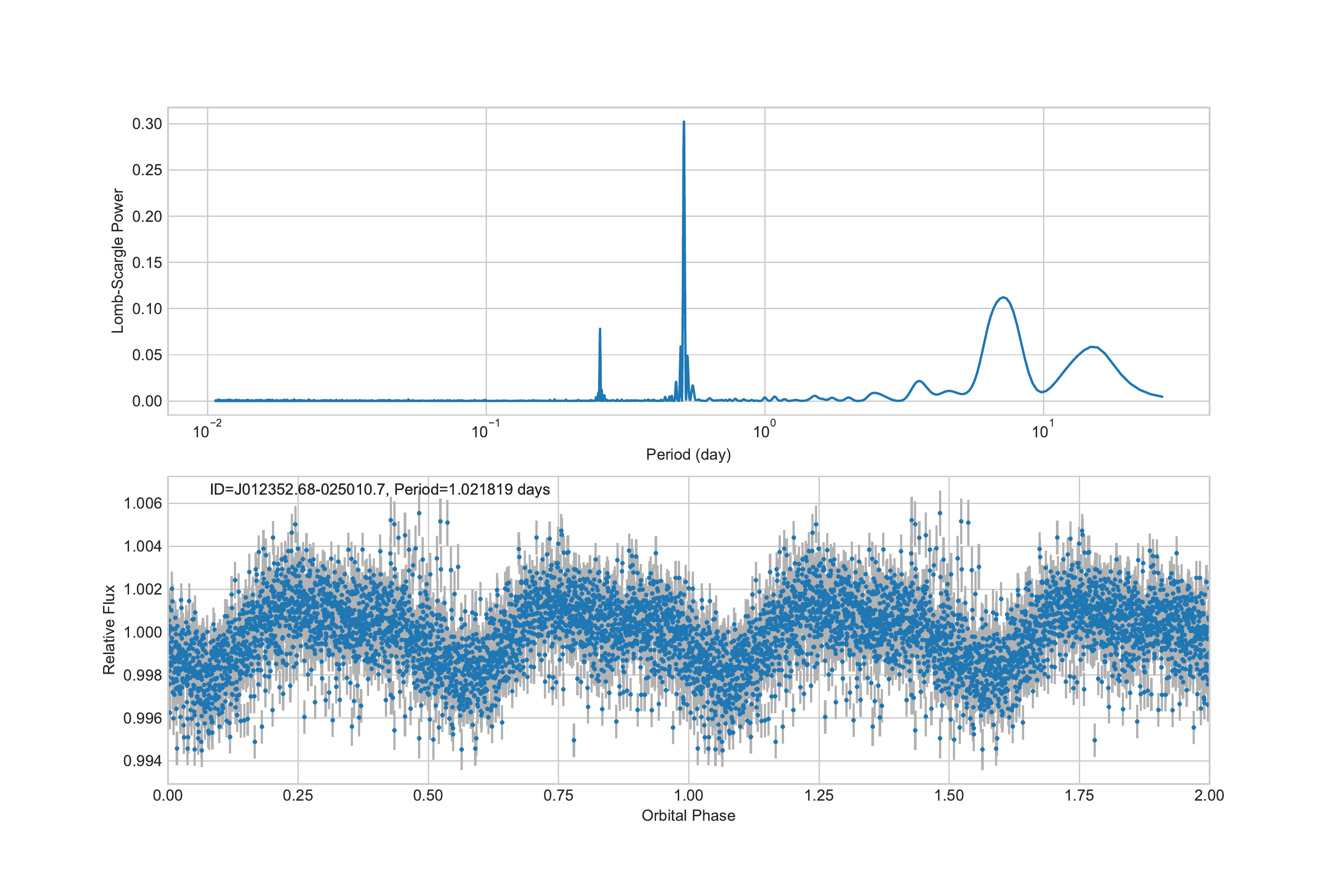}%
 \includegraphics[angle=0,width=0.49\textwidth,height=0.2\textheight]{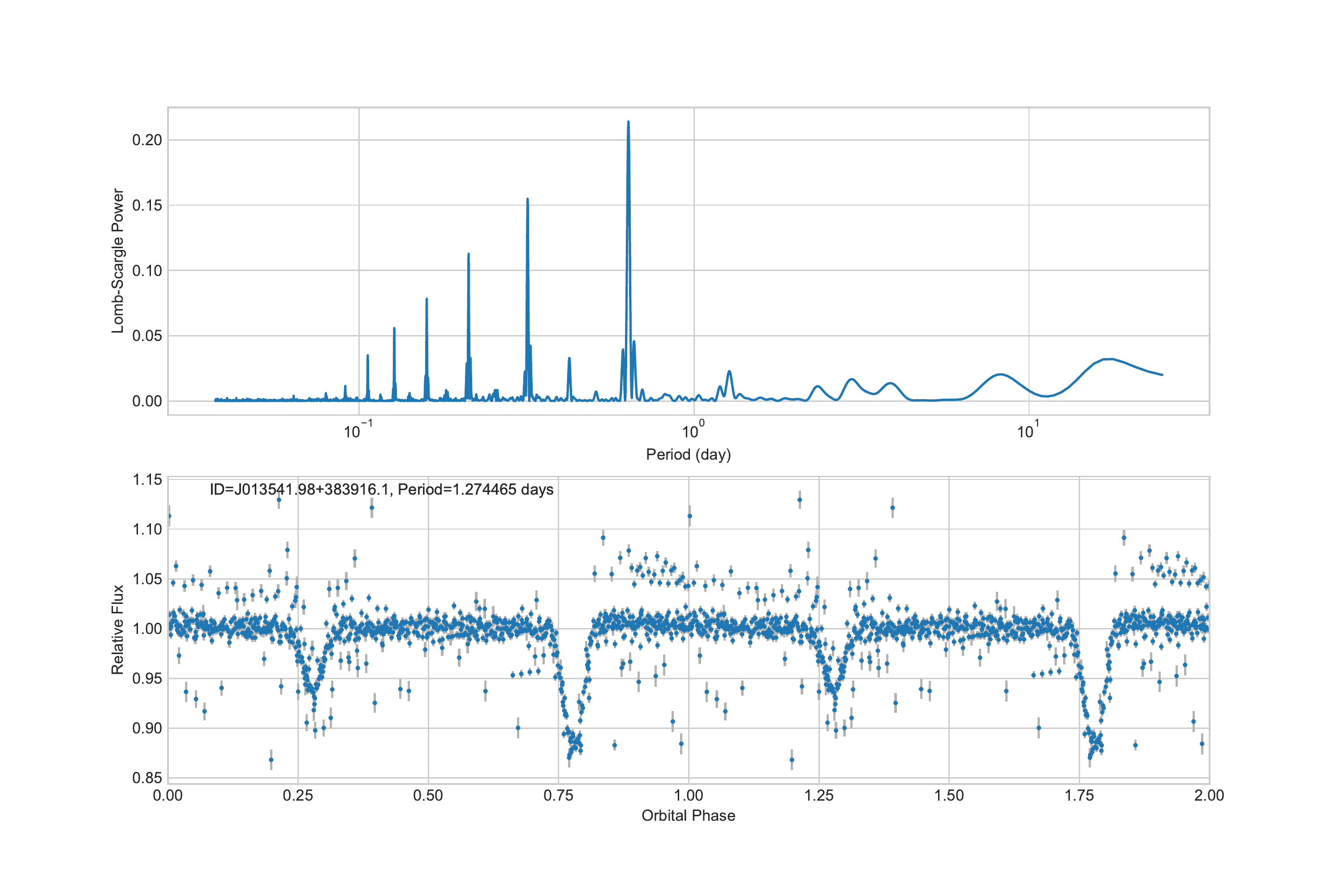}
 \includegraphics[angle=0,width=0.49\textwidth,height=0.2\textheight]{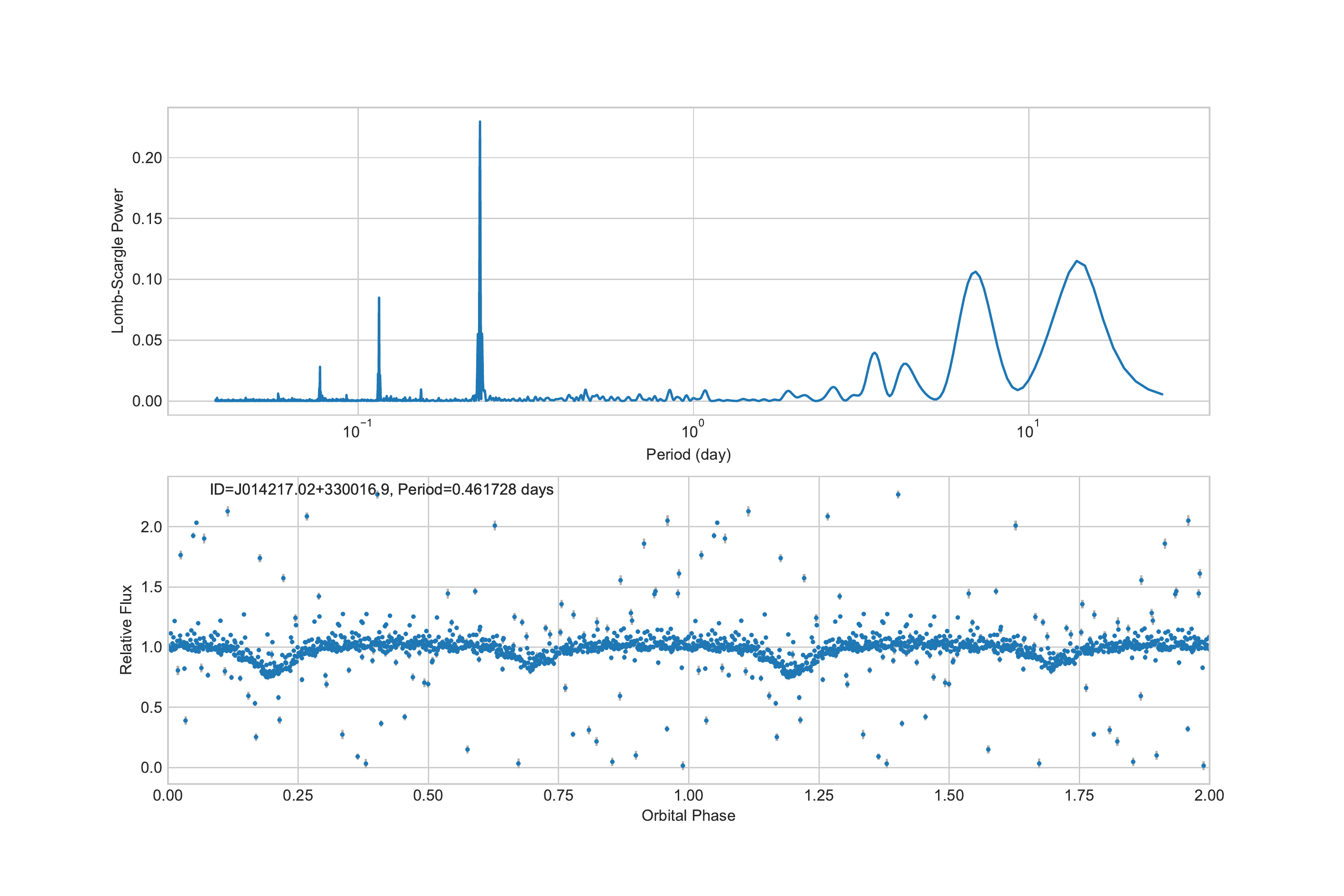}%
 \includegraphics[angle=0,width=0.49\textwidth,height=0.2\textheight]{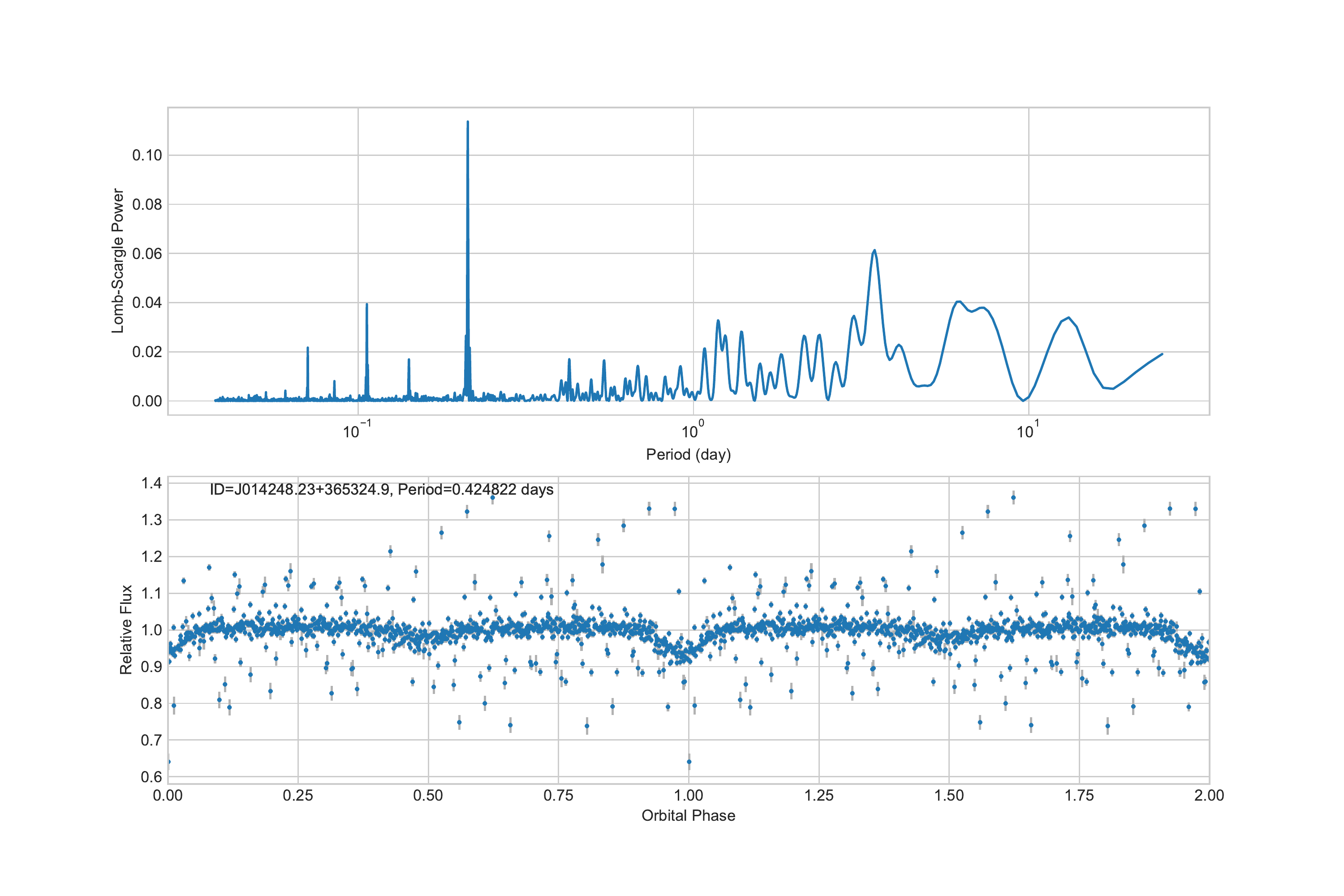}
 \includegraphics[angle=0,width=0.49\textwidth,height=0.2\textheight]{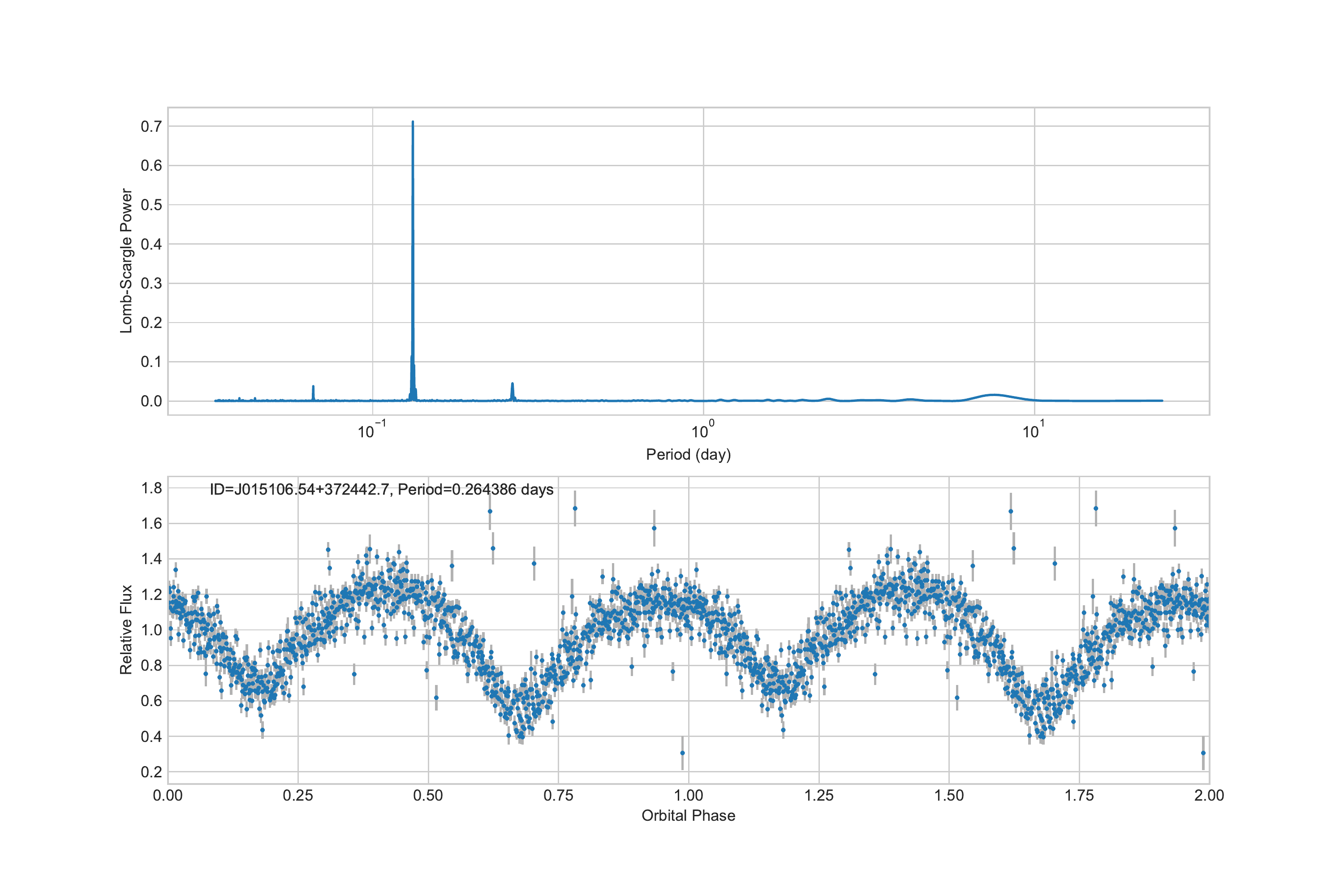}%
 \includegraphics[angle=0,width=0.49\textwidth,height=0.2\textheight]{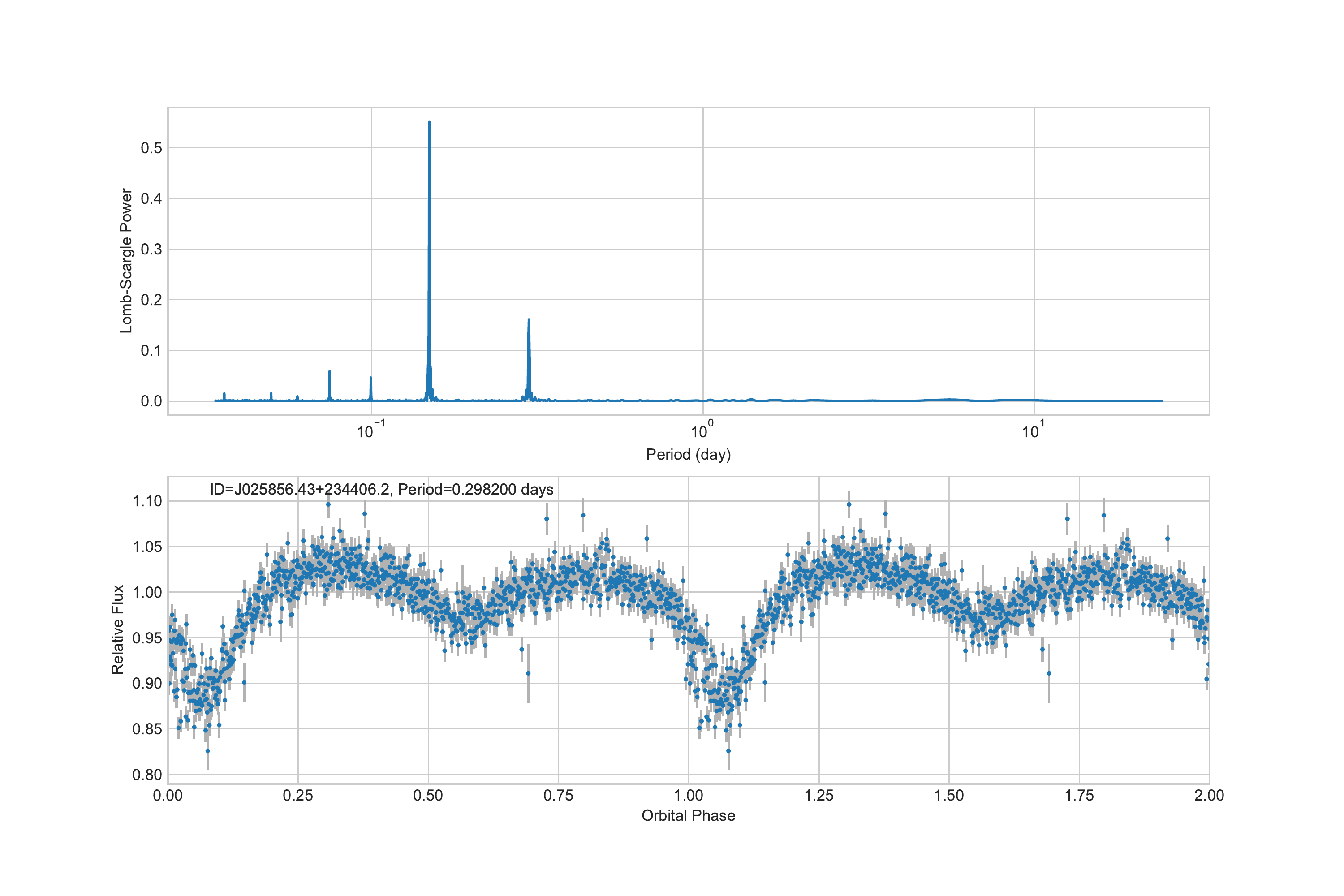}
 \caption{TESS light curves of the discarded eclipsing binaries.}
 \end{figure}
  \addtocounter{figure}{-1}
 \clearpage

 \begin{figure}
 \centering
 \includegraphics[angle=0,width=0.49\textwidth,height=0.2\textheight]{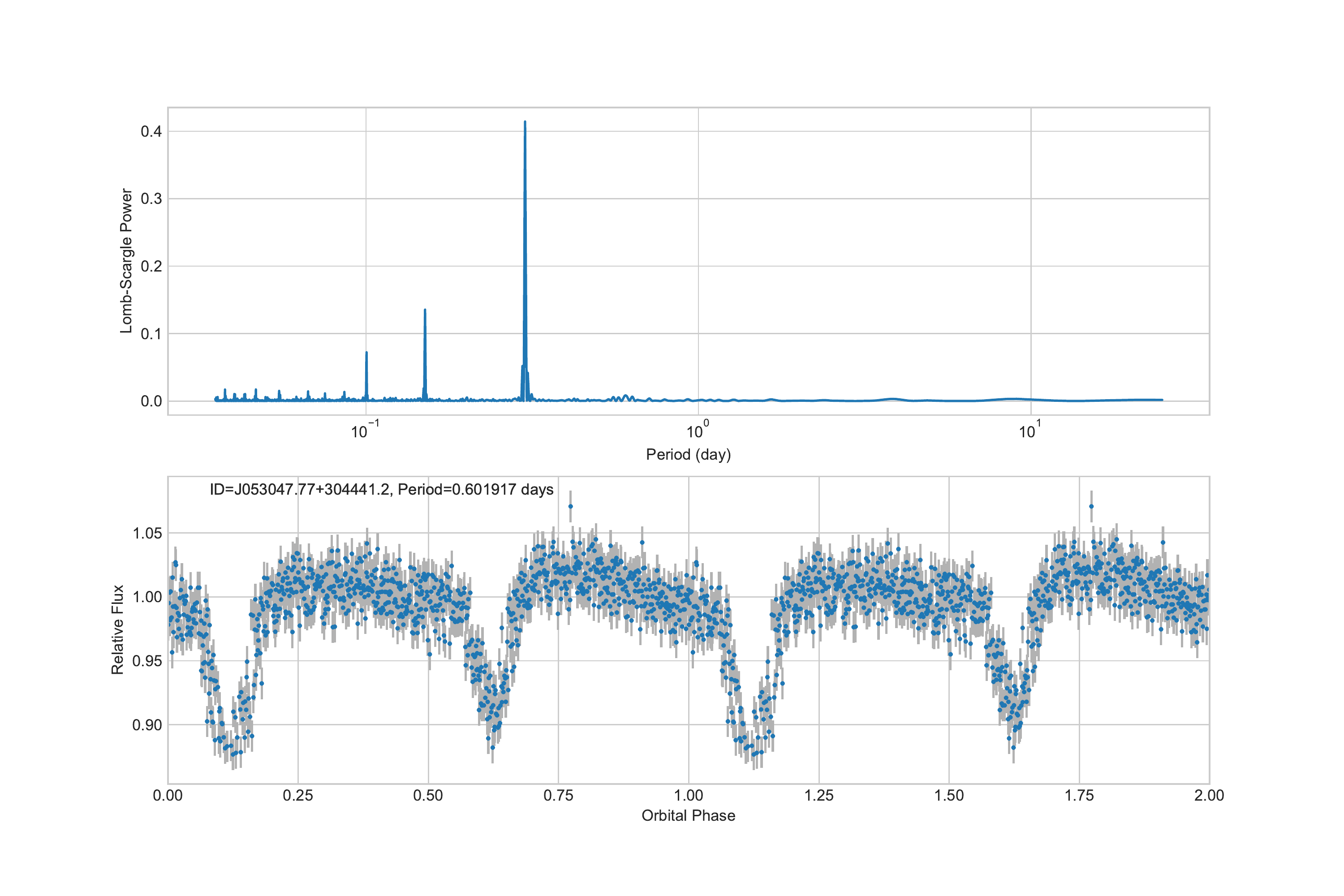}%
 \includegraphics[angle=0,width=0.49\textwidth,height=0.2\textheight]{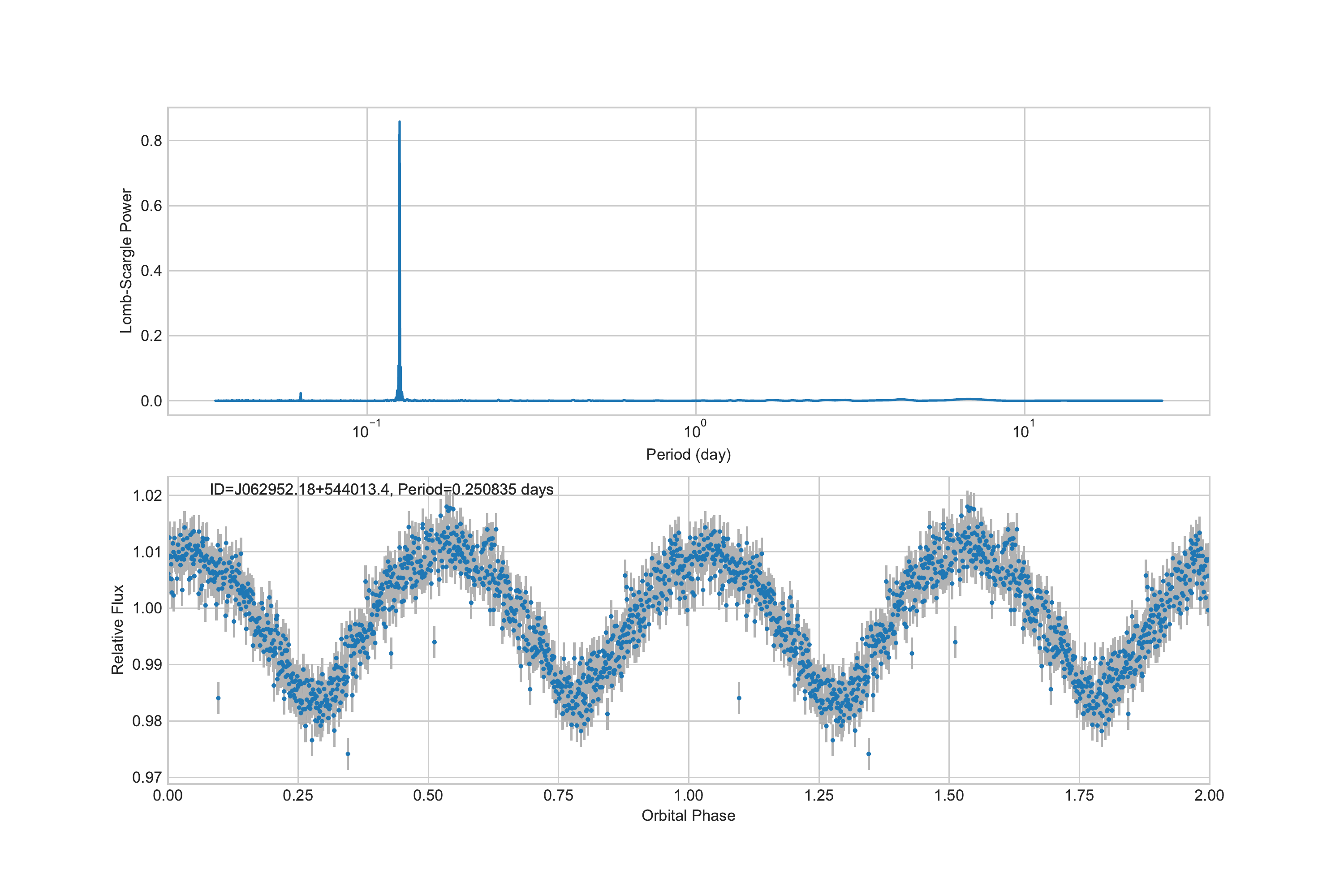}
 \includegraphics[angle=0,width=0.49\textwidth,height=0.2\textheight]{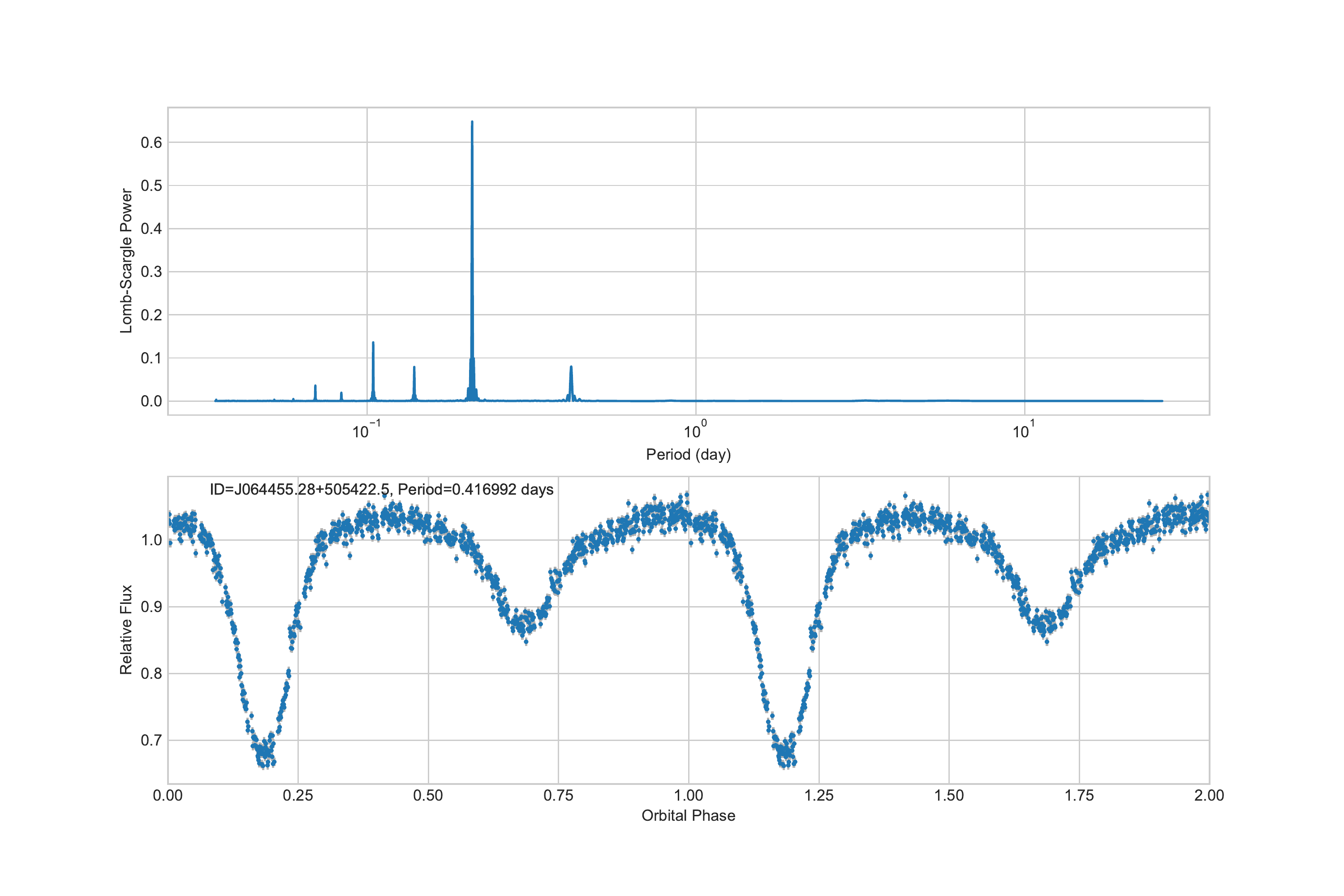}%
 \includegraphics[angle=0,width=0.49\textwidth,height=0.2\textheight]{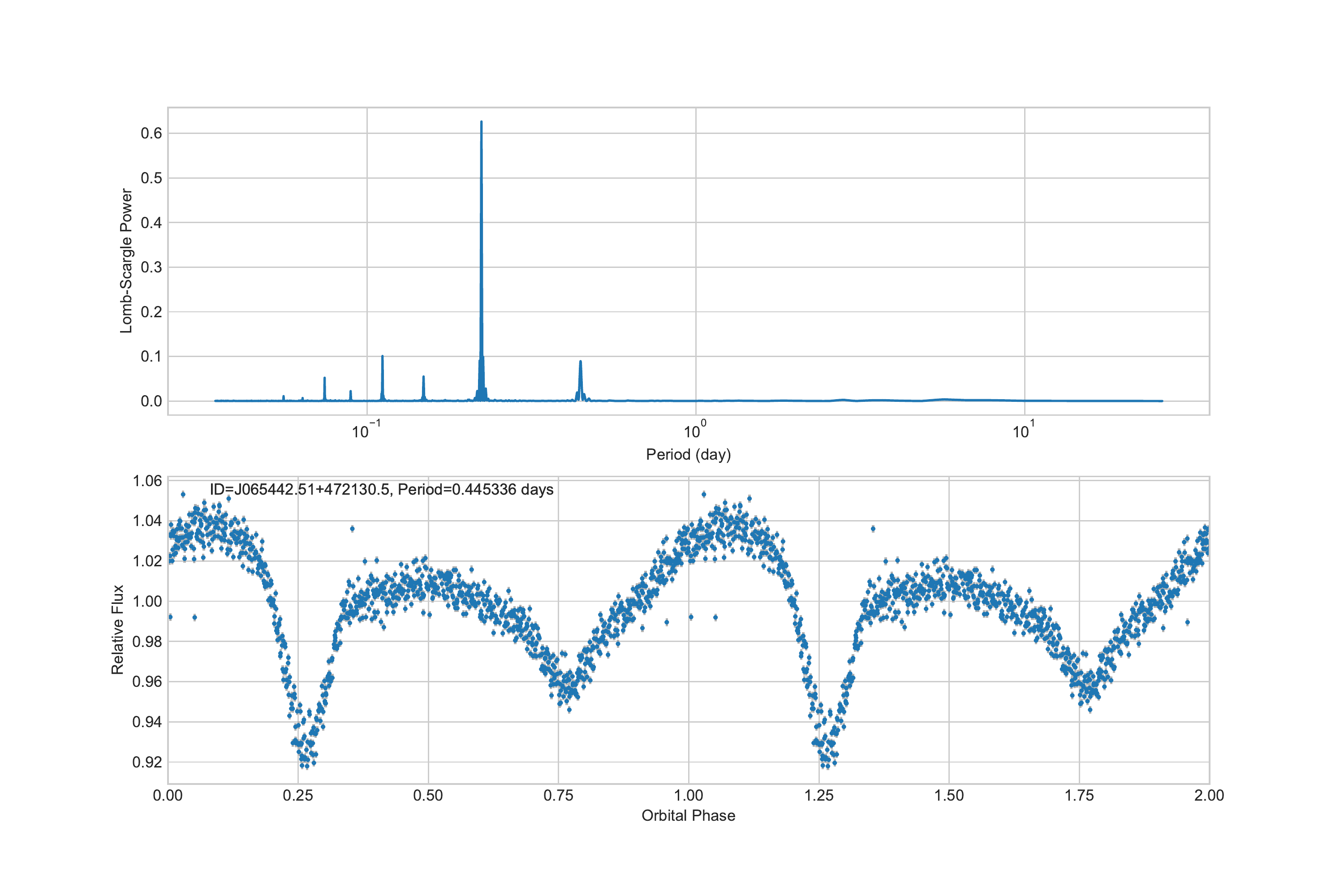}
 \includegraphics[angle=0,width=0.49\textwidth,height=0.2\textheight]{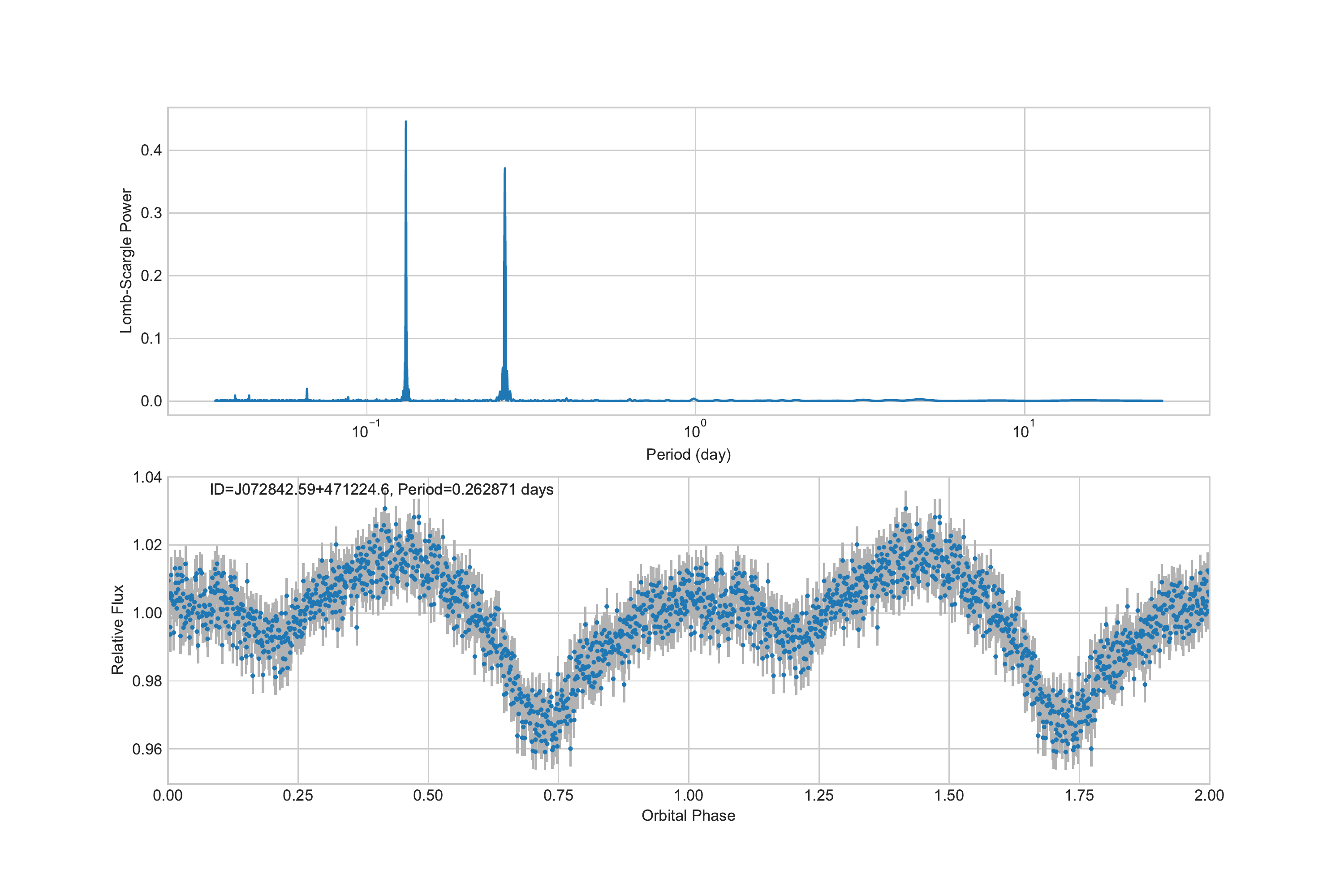}%
 \includegraphics[angle=0,width=0.49\textwidth,height=0.2\textheight]{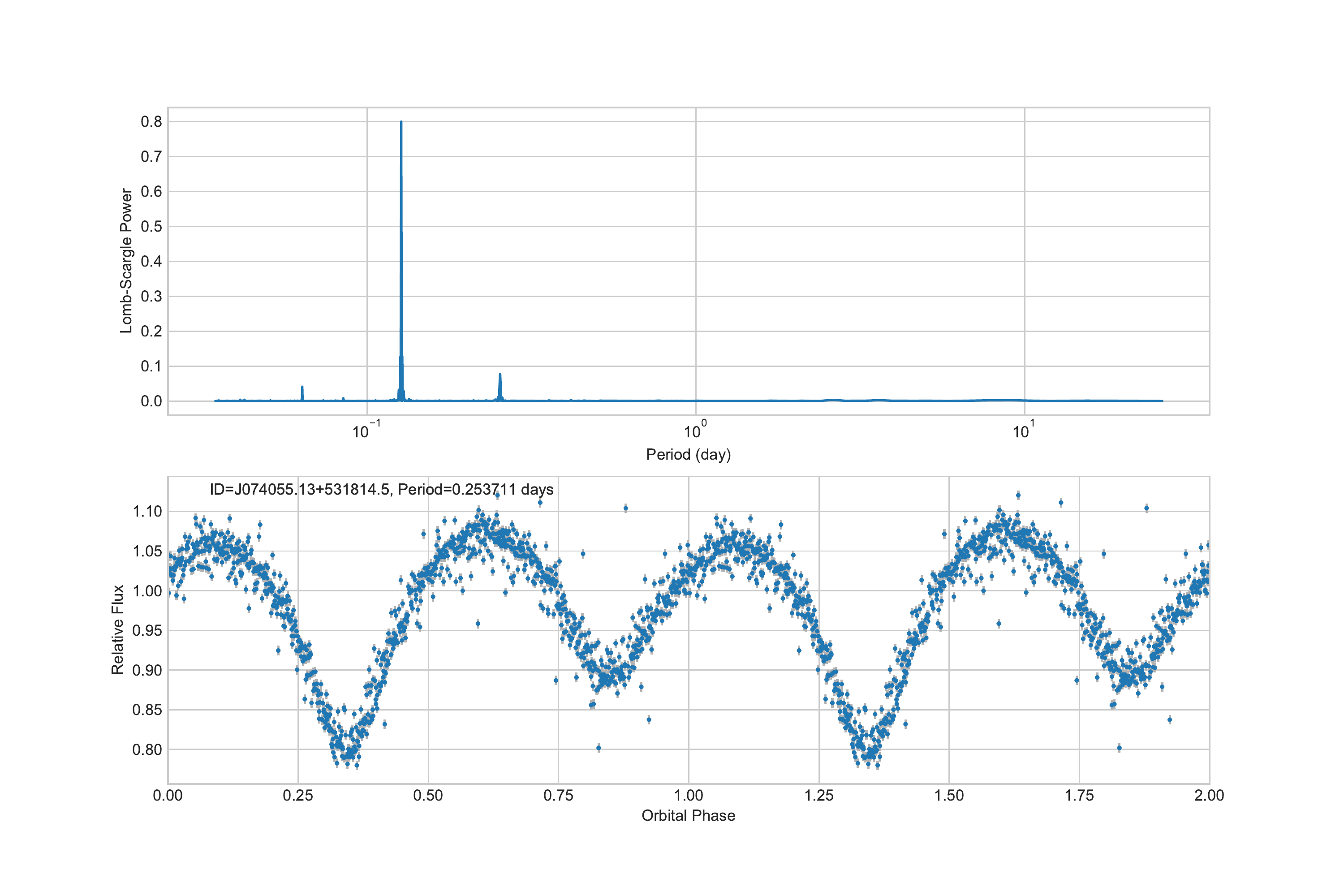}
 \includegraphics[angle=0,width=0.49\textwidth,height=0.2\textheight]{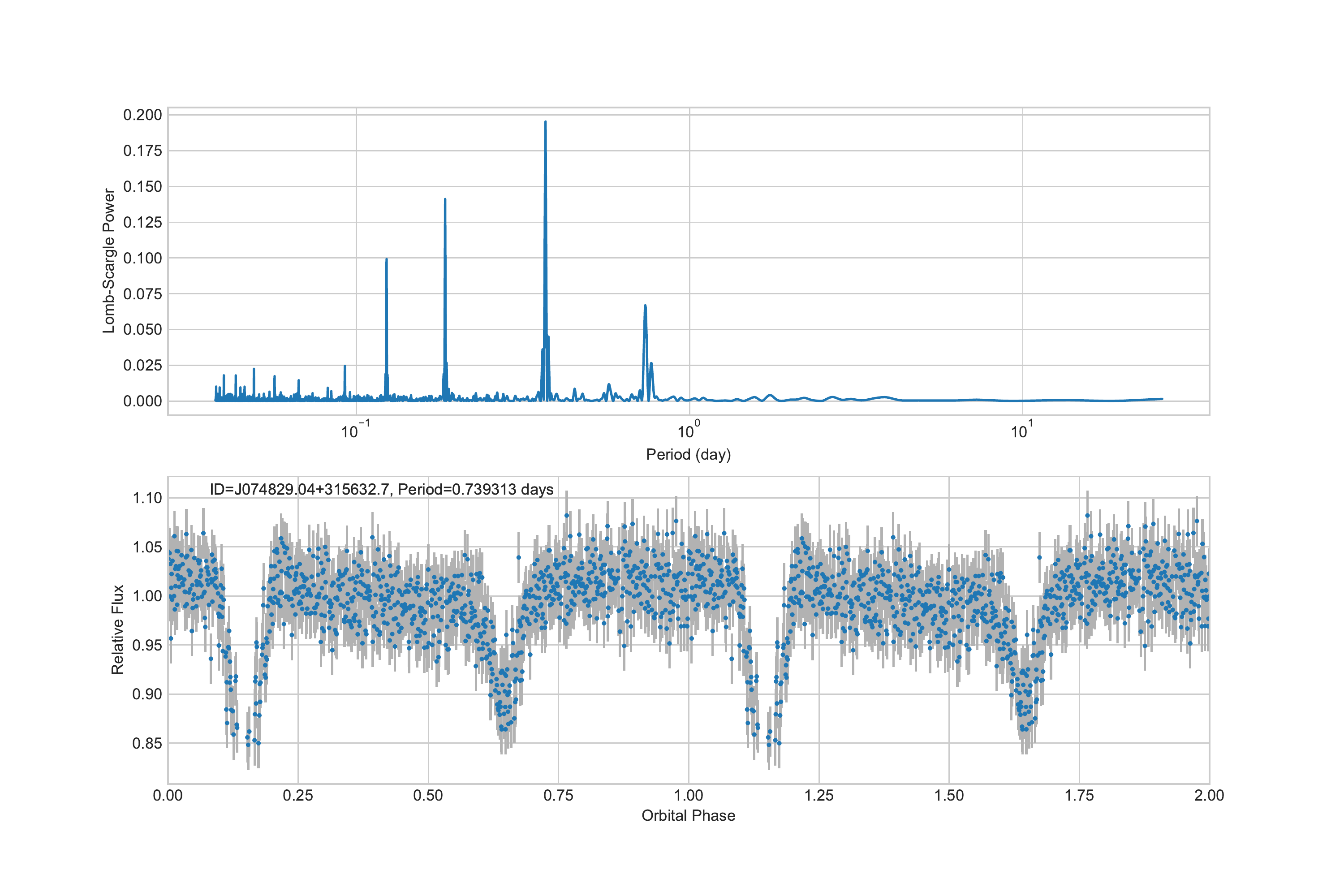}%
 \includegraphics[angle=0,width=0.49\textwidth,height=0.2\textheight]{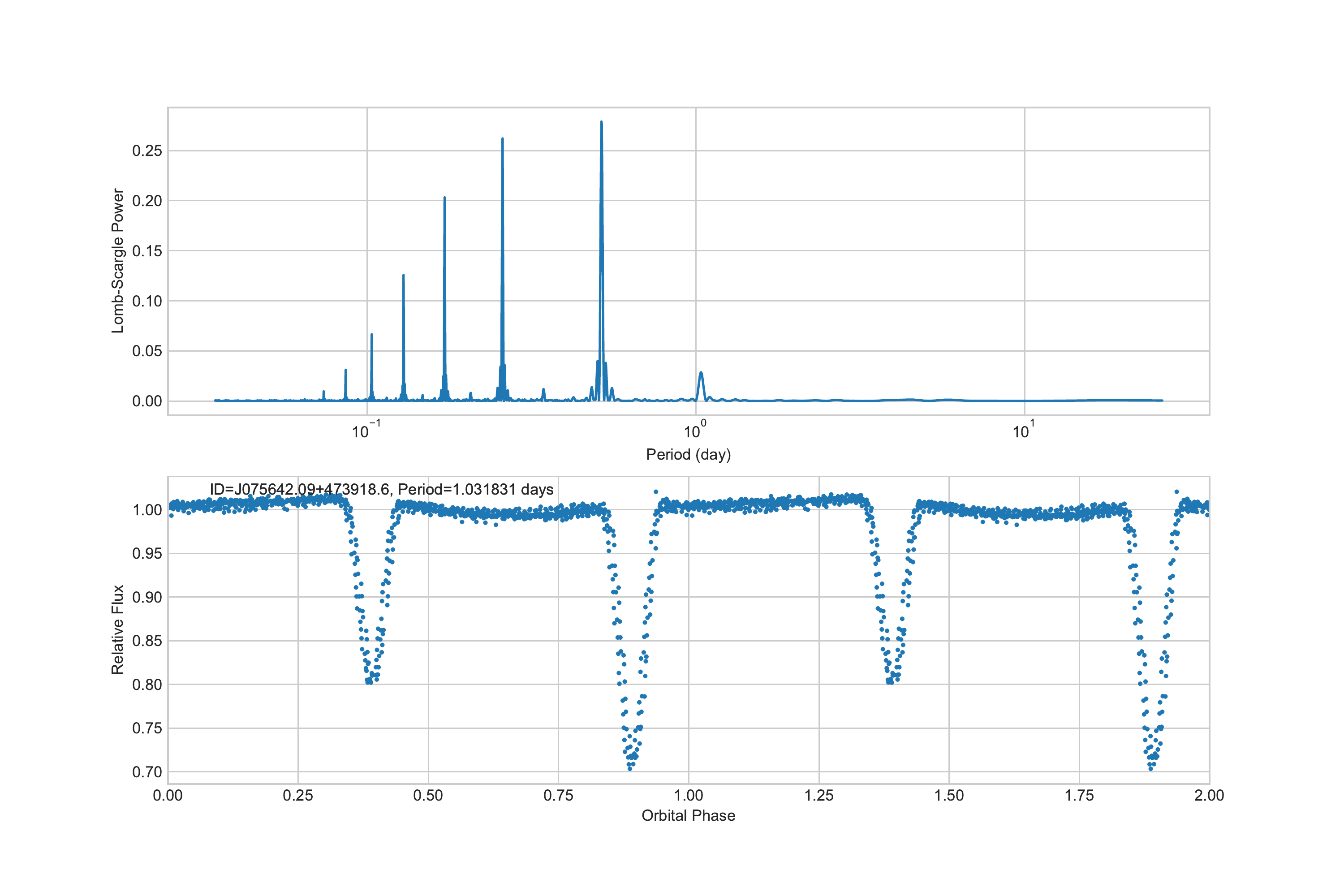}
 \caption{\it Continued.}
 \end{figure}
 \addtocounter{figure}{-1}

 \begin{figure}
 \centering
 \includegraphics[angle=0,width=0.49\textwidth,height=0.2\textheight]{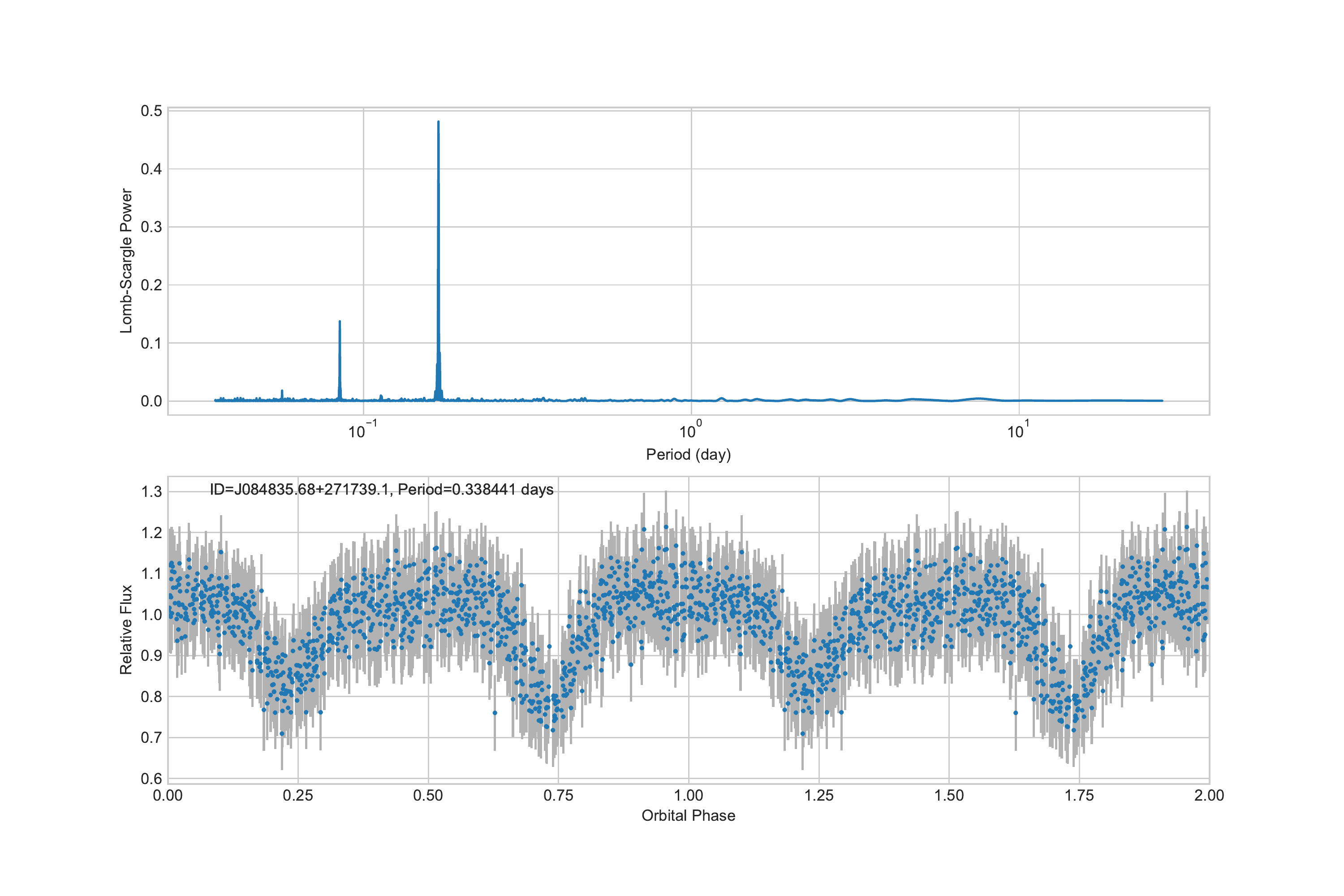}%
 \includegraphics[angle=0,width=0.49\textwidth,height=0.2\textheight]{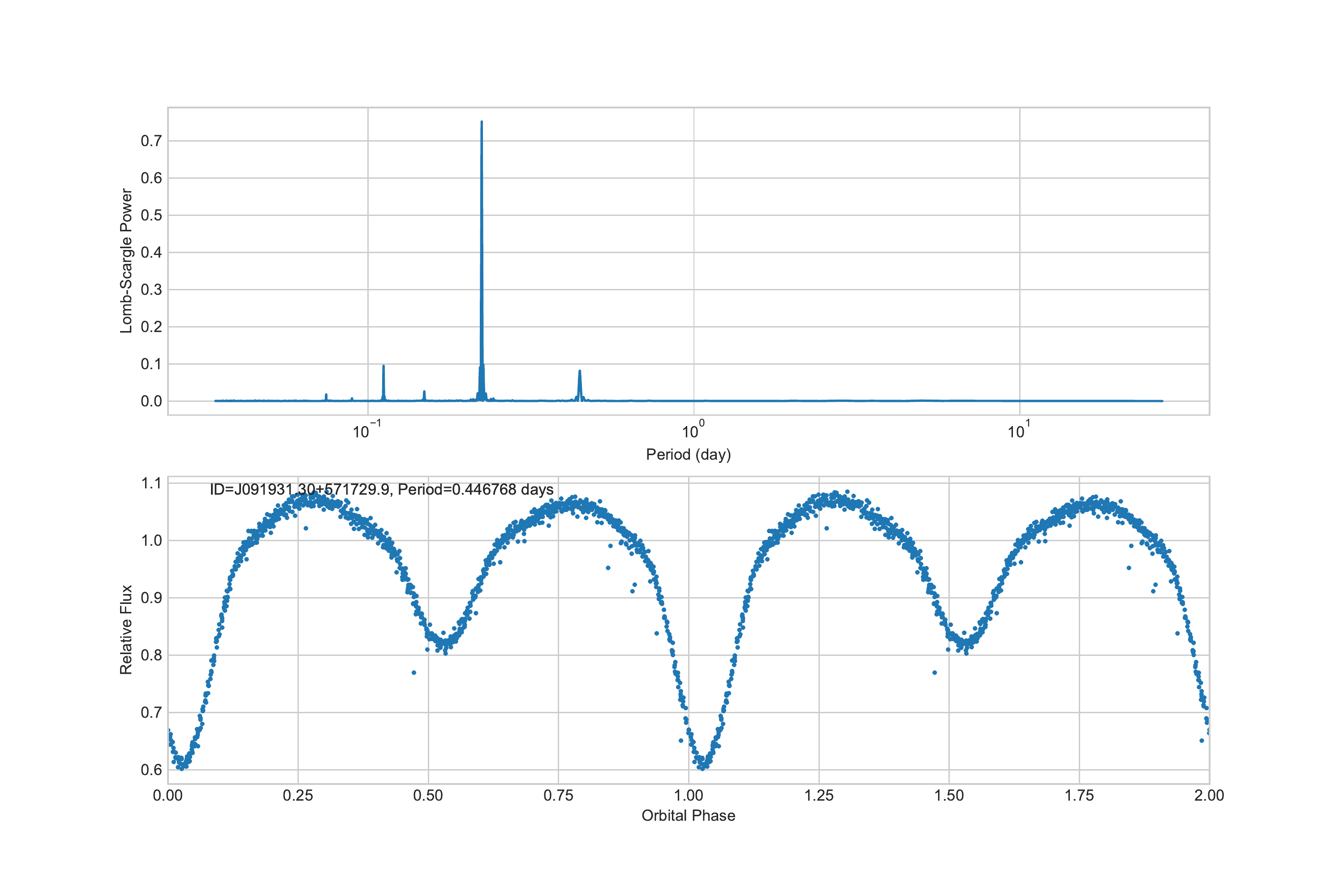}
 \includegraphics[angle=0,width=0.49\textwidth,height=0.2\textheight]{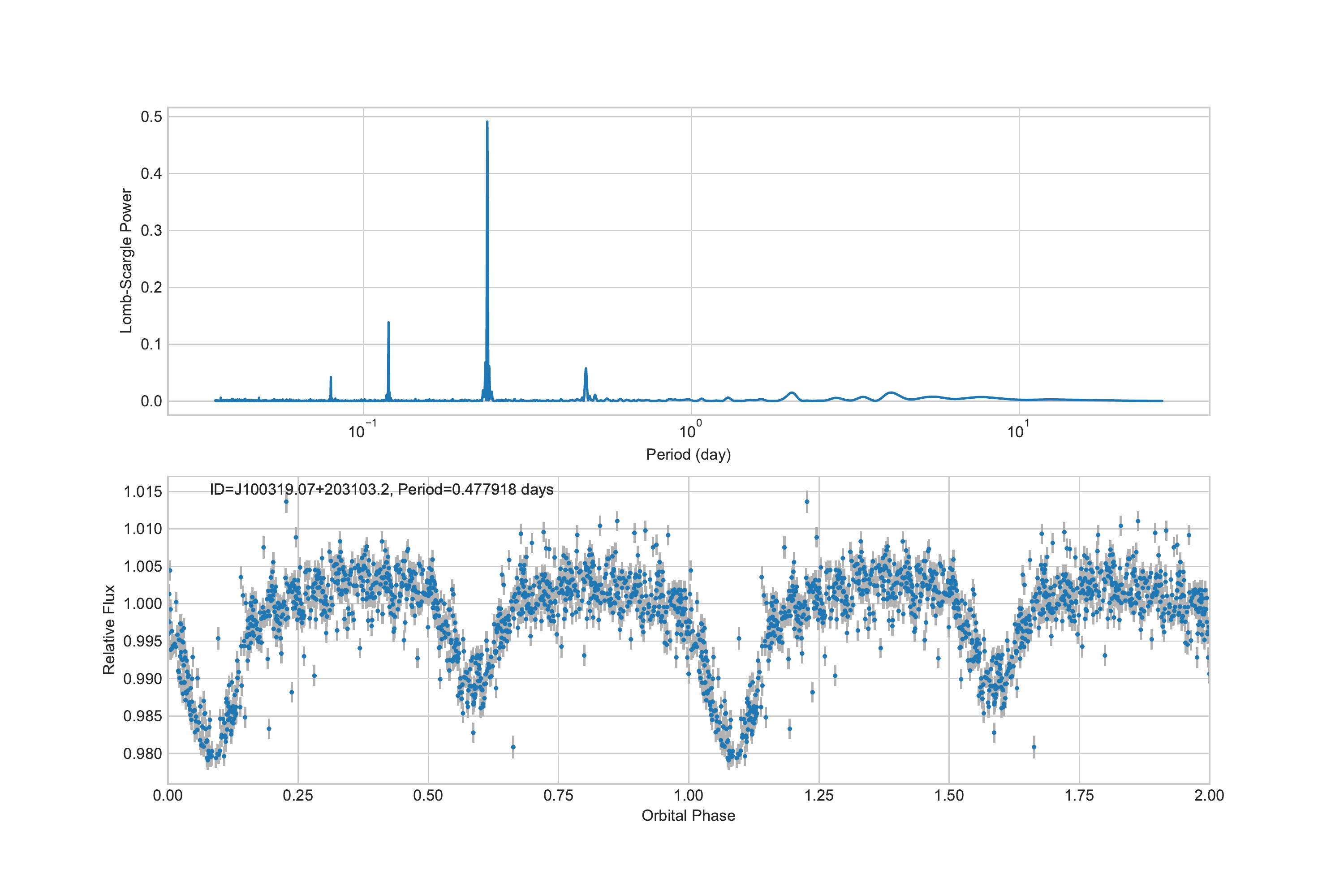}%
 \includegraphics[angle=0,width=0.49\textwidth,height=0.2\textheight]{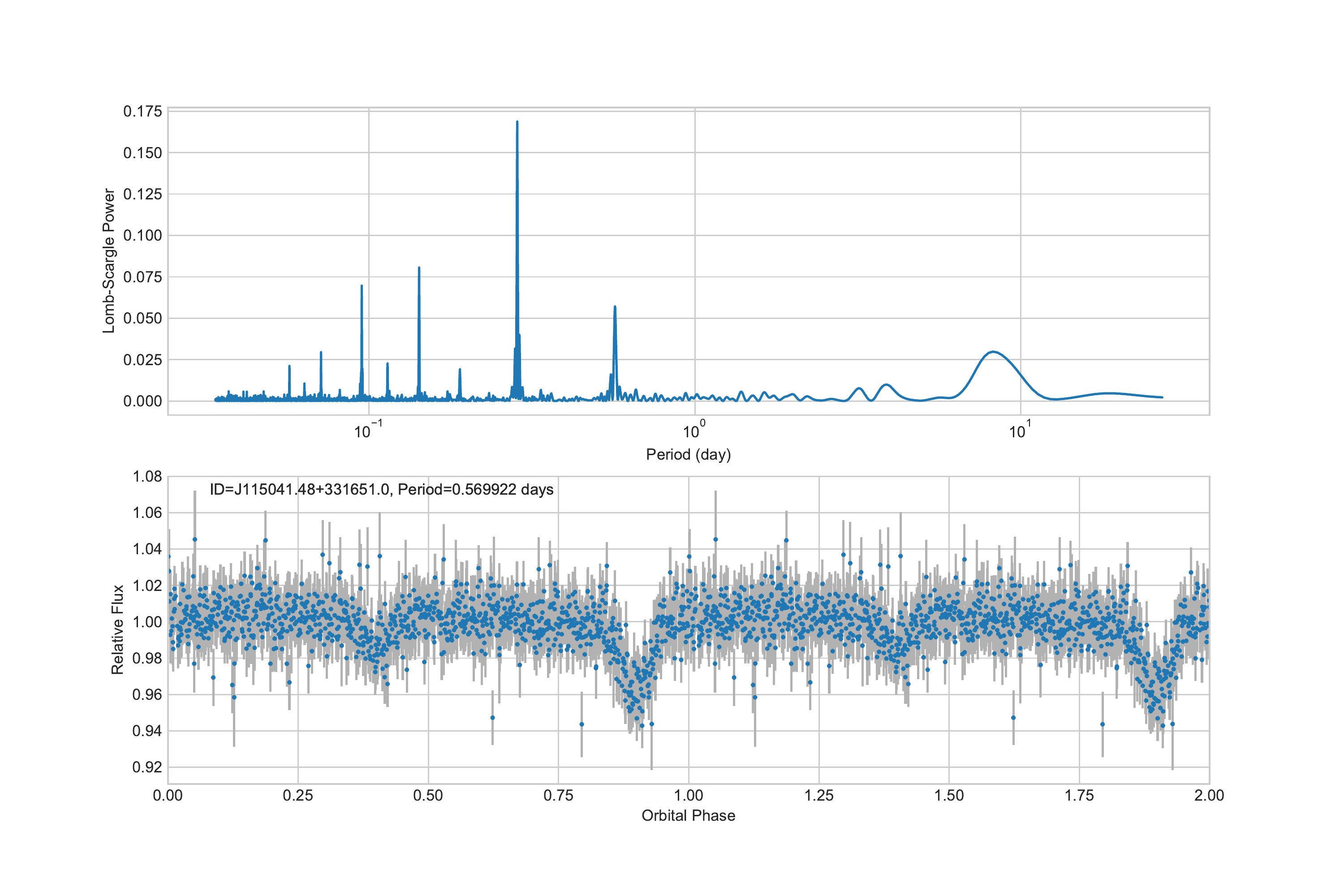}
 \includegraphics[angle=0,width=0.49\textwidth,height=0.2\textheight]{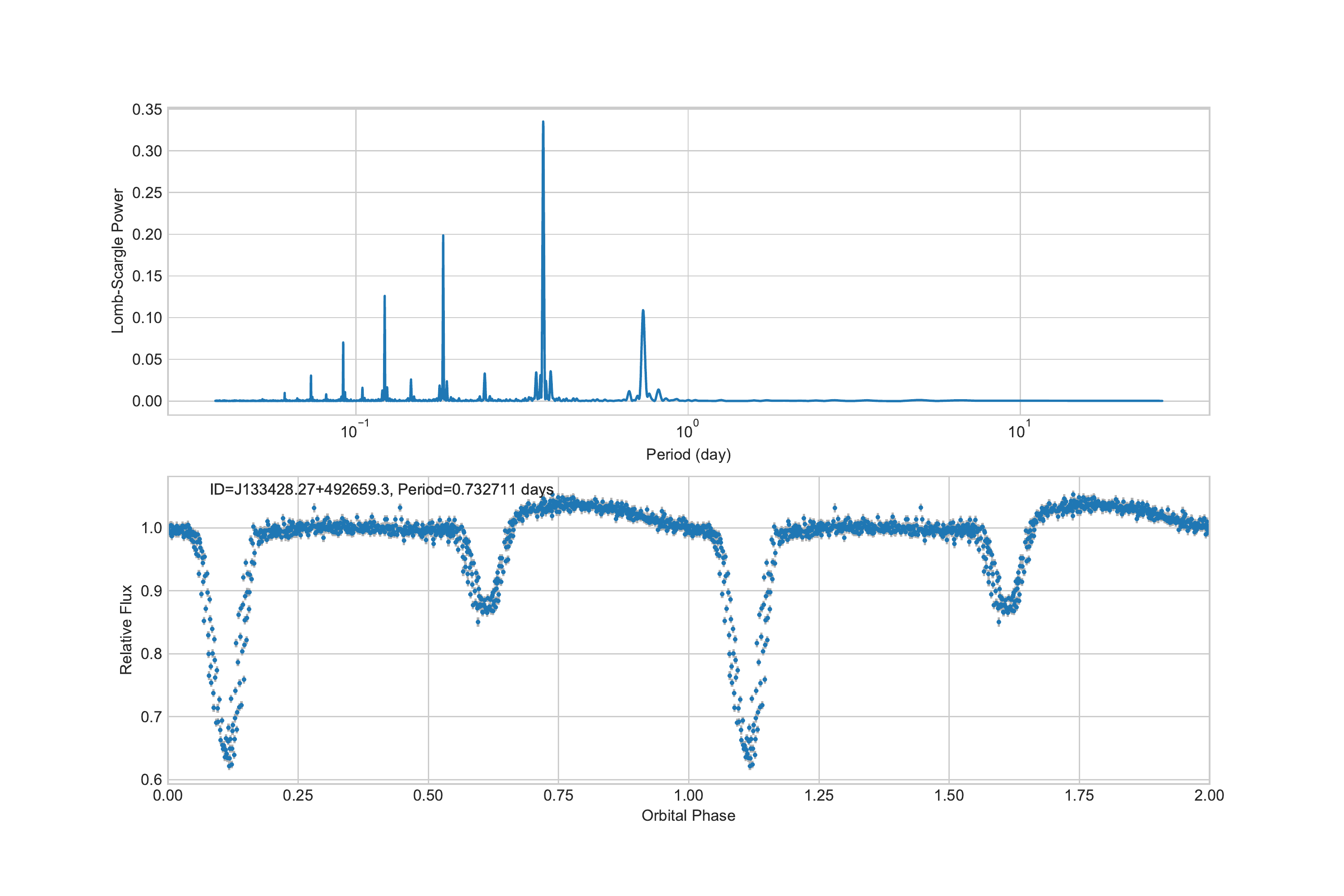}%
 \includegraphics[angle=0,width=0.49\textwidth,height=0.2\textheight]{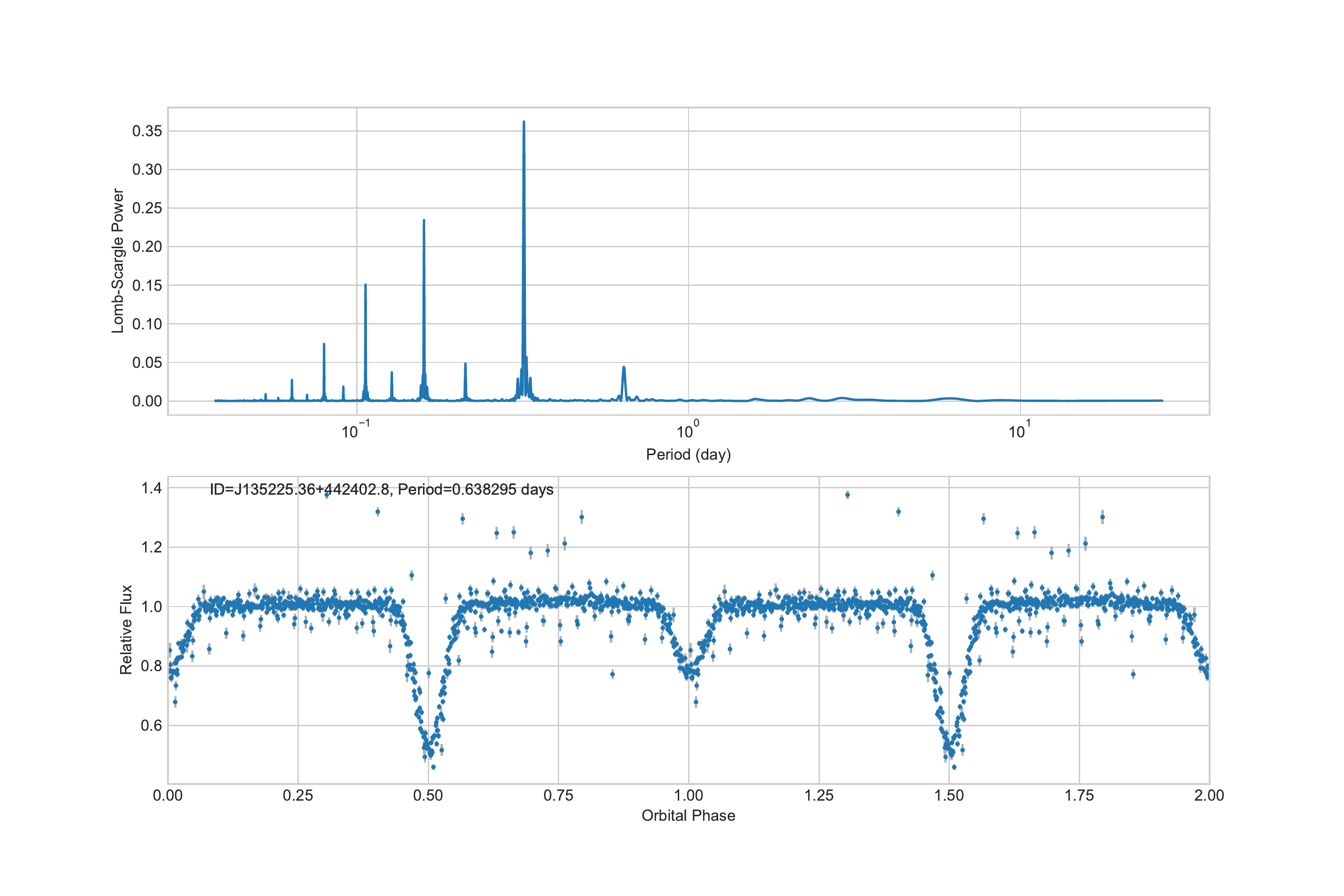}
 \includegraphics[angle=0,width=0.49\textwidth,height=0.2\textheight]{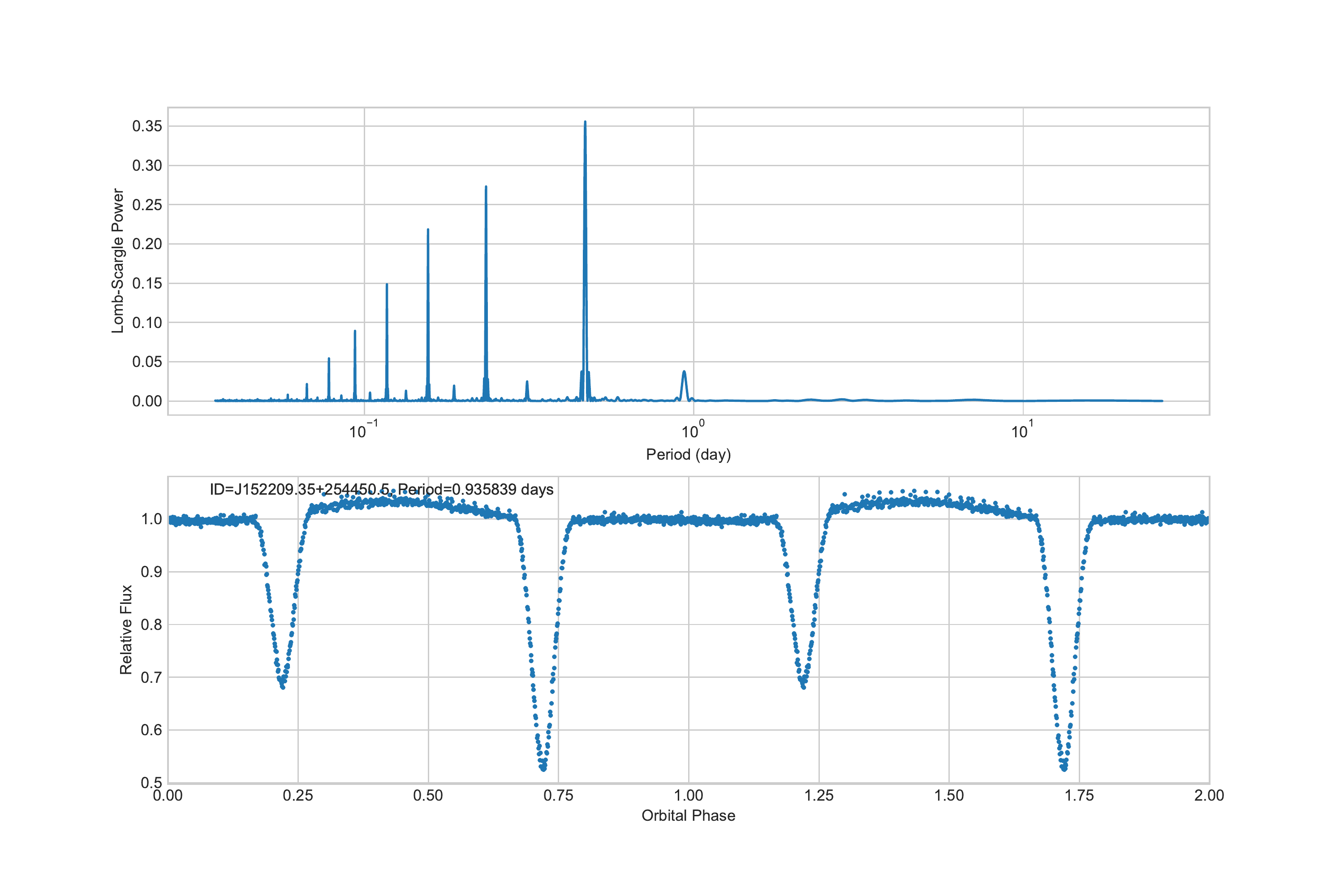}%
 \includegraphics[angle=0,width=0.49\textwidth,height=0.2\textheight]{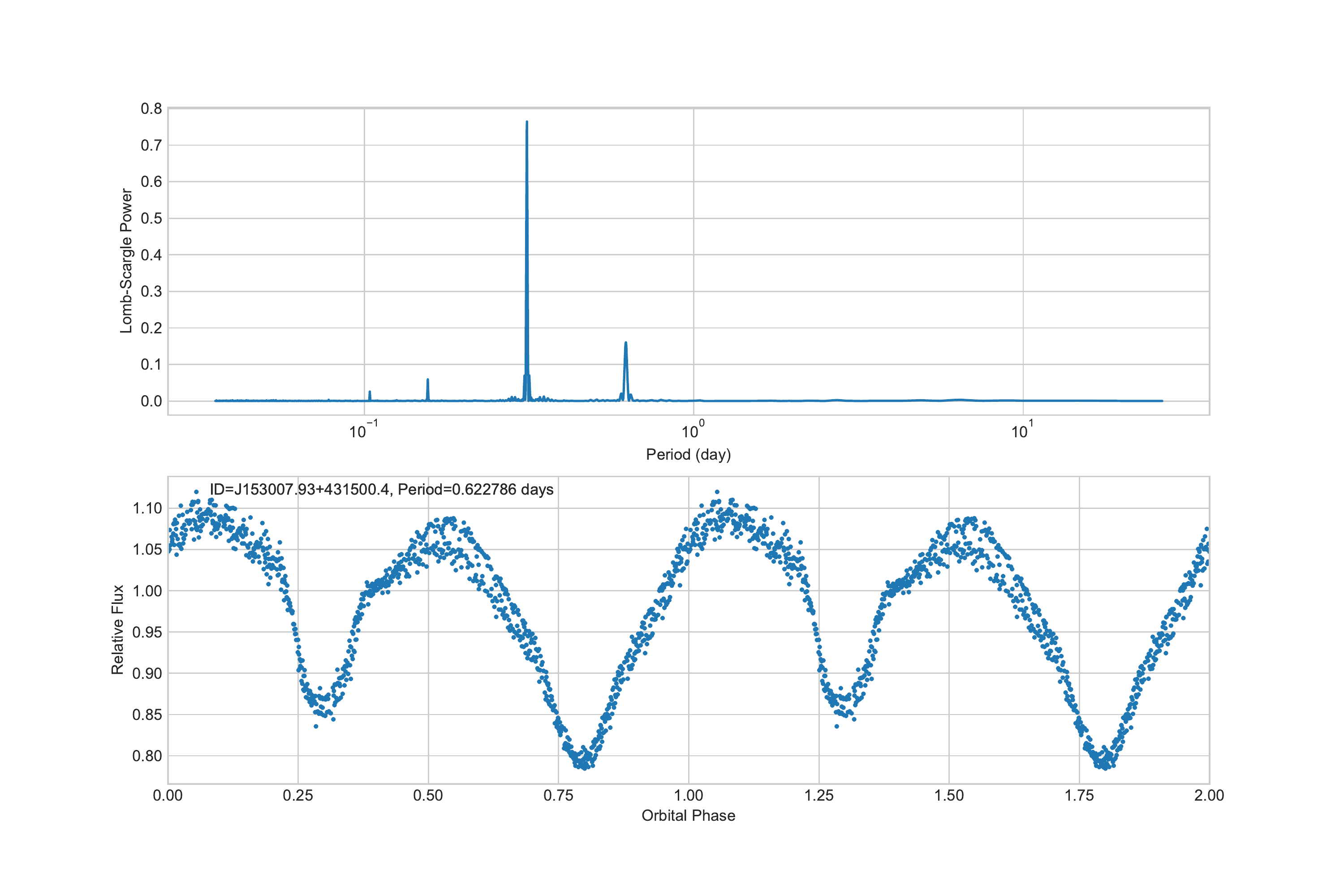}
 \caption{\it Continued.}
 \end{figure}
 \addtocounter{figure}{-1}

 \begin{figure}
 \centering
 \includegraphics[angle=0,width=0.49\textwidth,height=0.2\textheight]{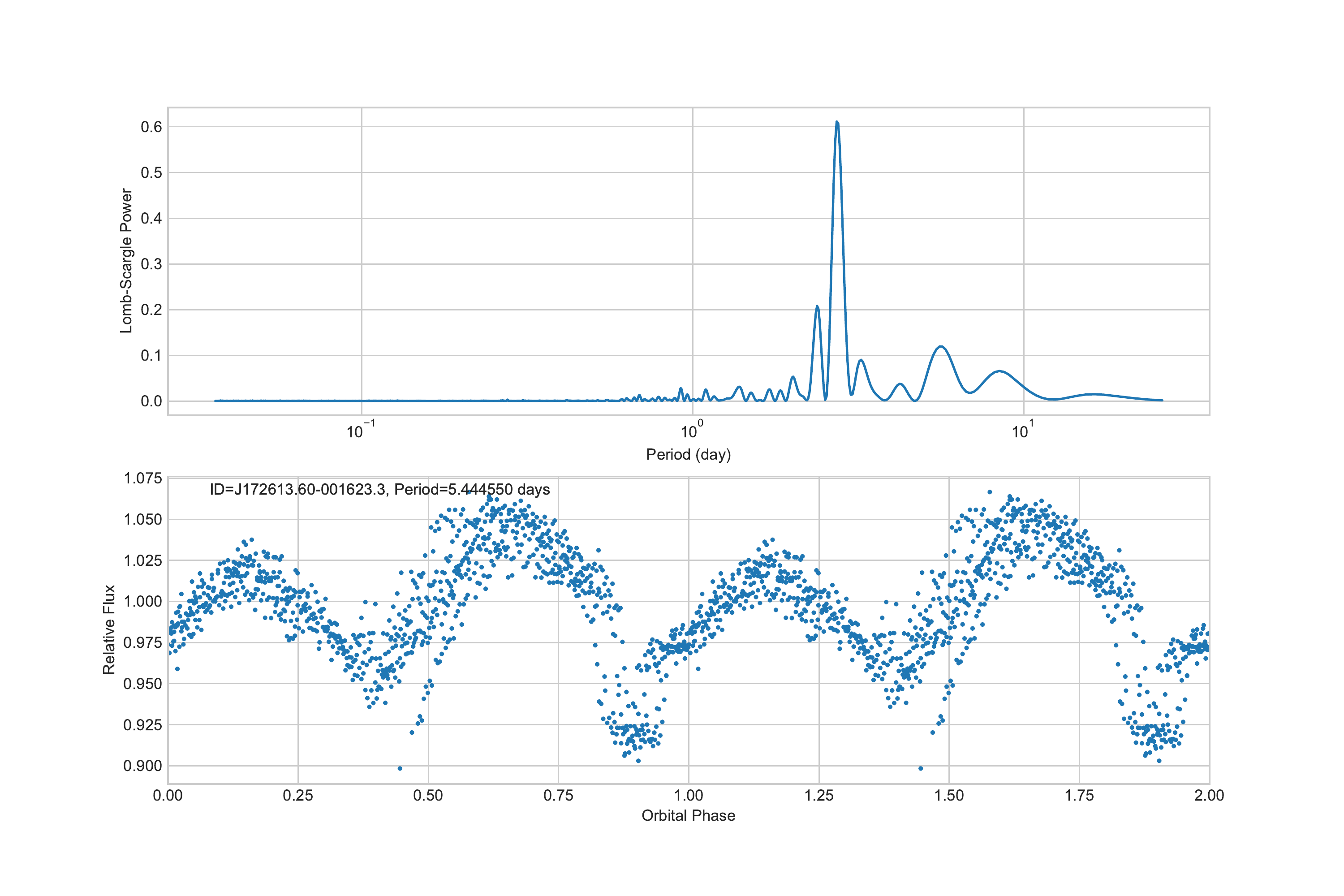}%
 \caption{\it Continued.}
 \end{figure}

 \clearpage

 \begin{figure}[H]
 \centering
 \includegraphics[angle=0,width=0.49\textwidth,height=0.2\textheight]{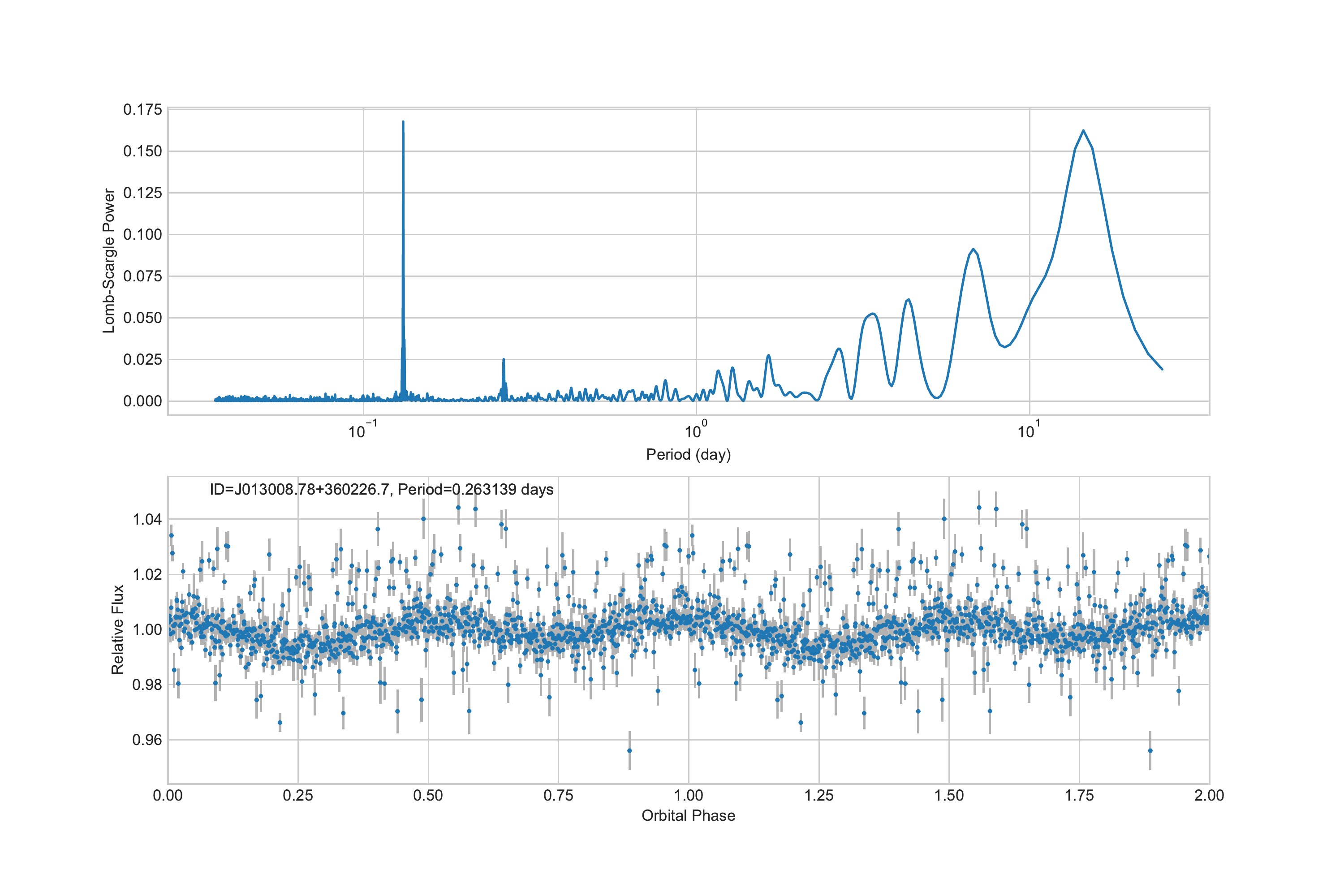}%
 \includegraphics[angle=0,width=0.49\textwidth,height=0.2\textheight]{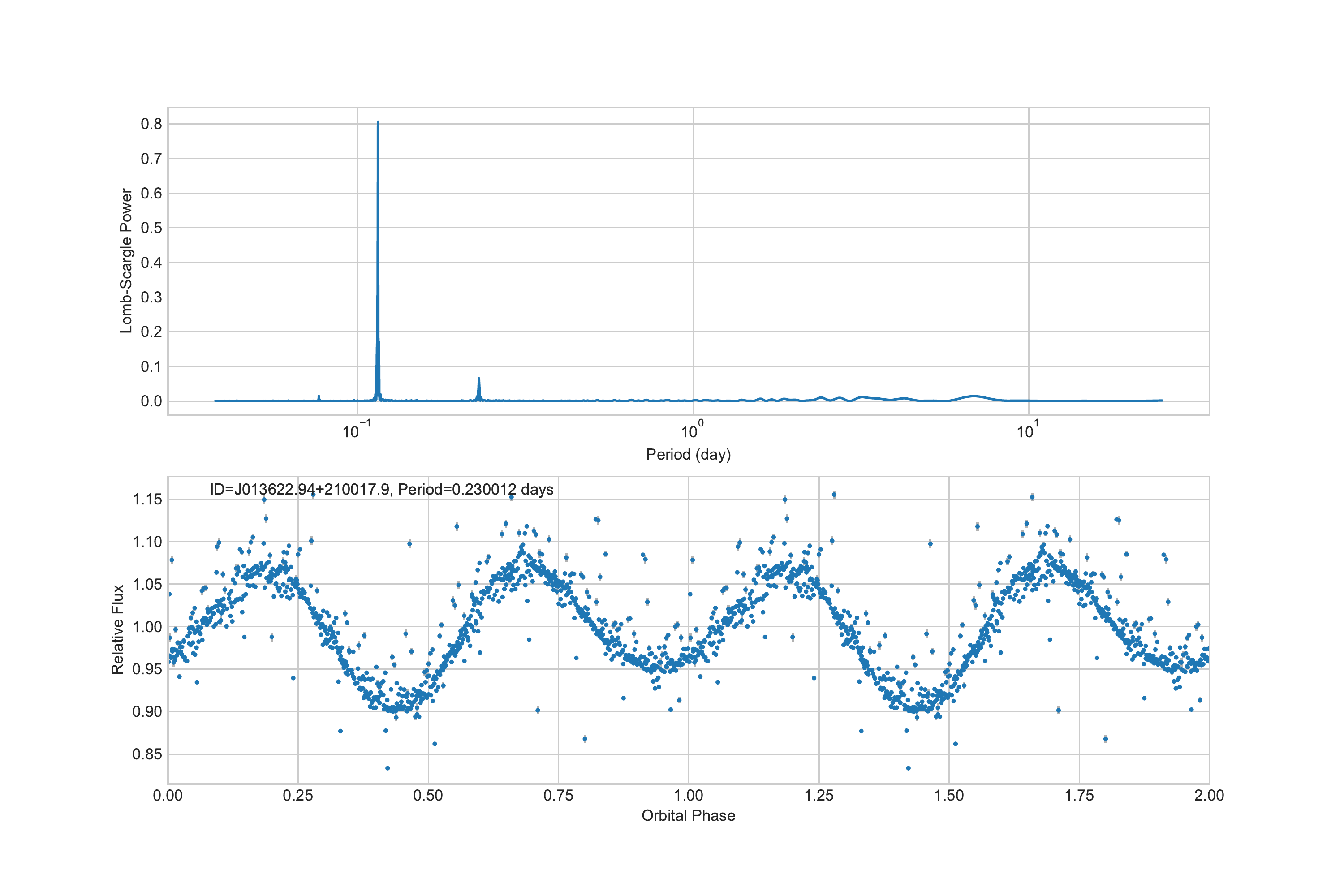}
 \includegraphics[angle=0,width=0.49\textwidth,height=0.2\textheight]{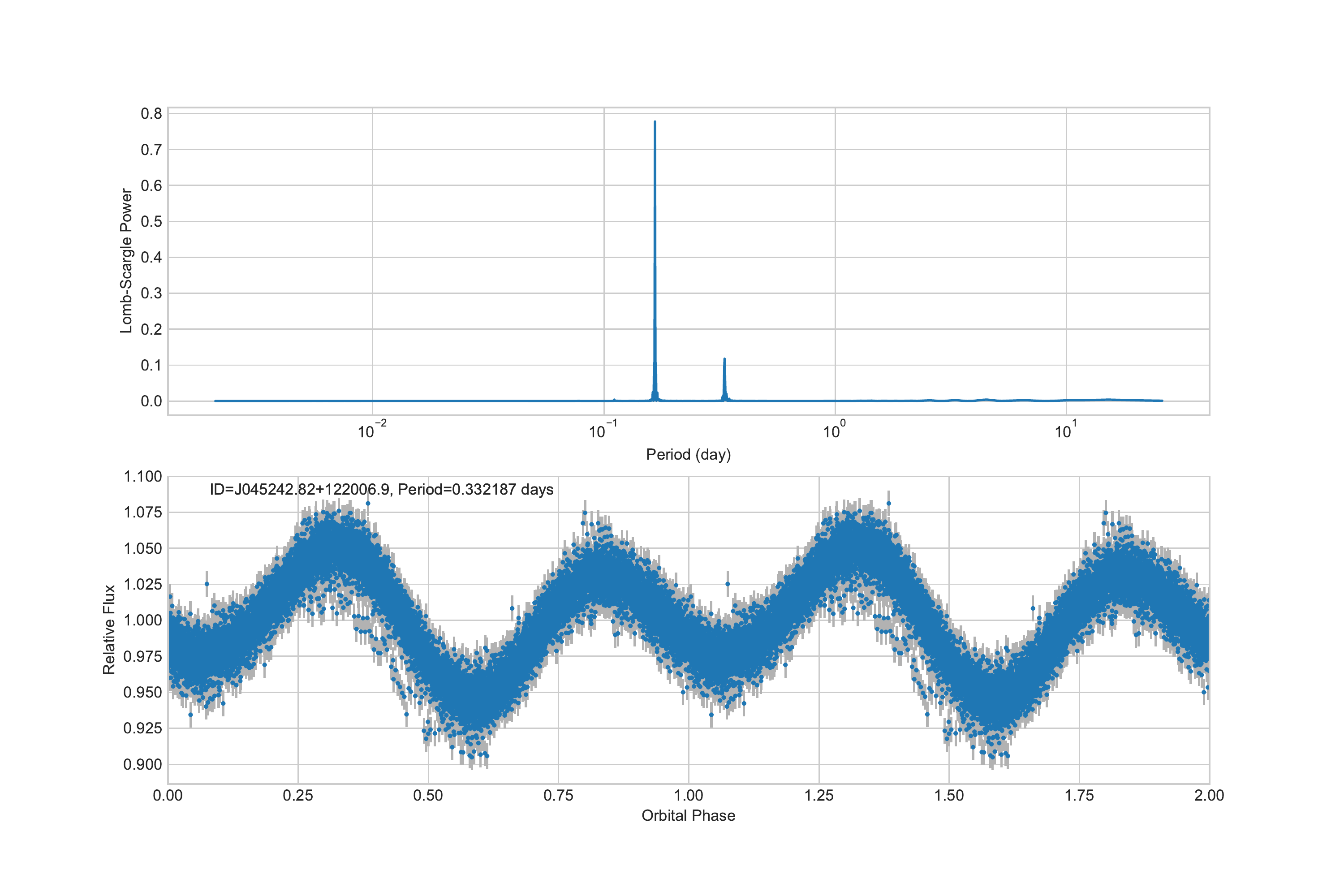}%
 \includegraphics[angle=0,width=0.49\textwidth,height=0.2\textheight]{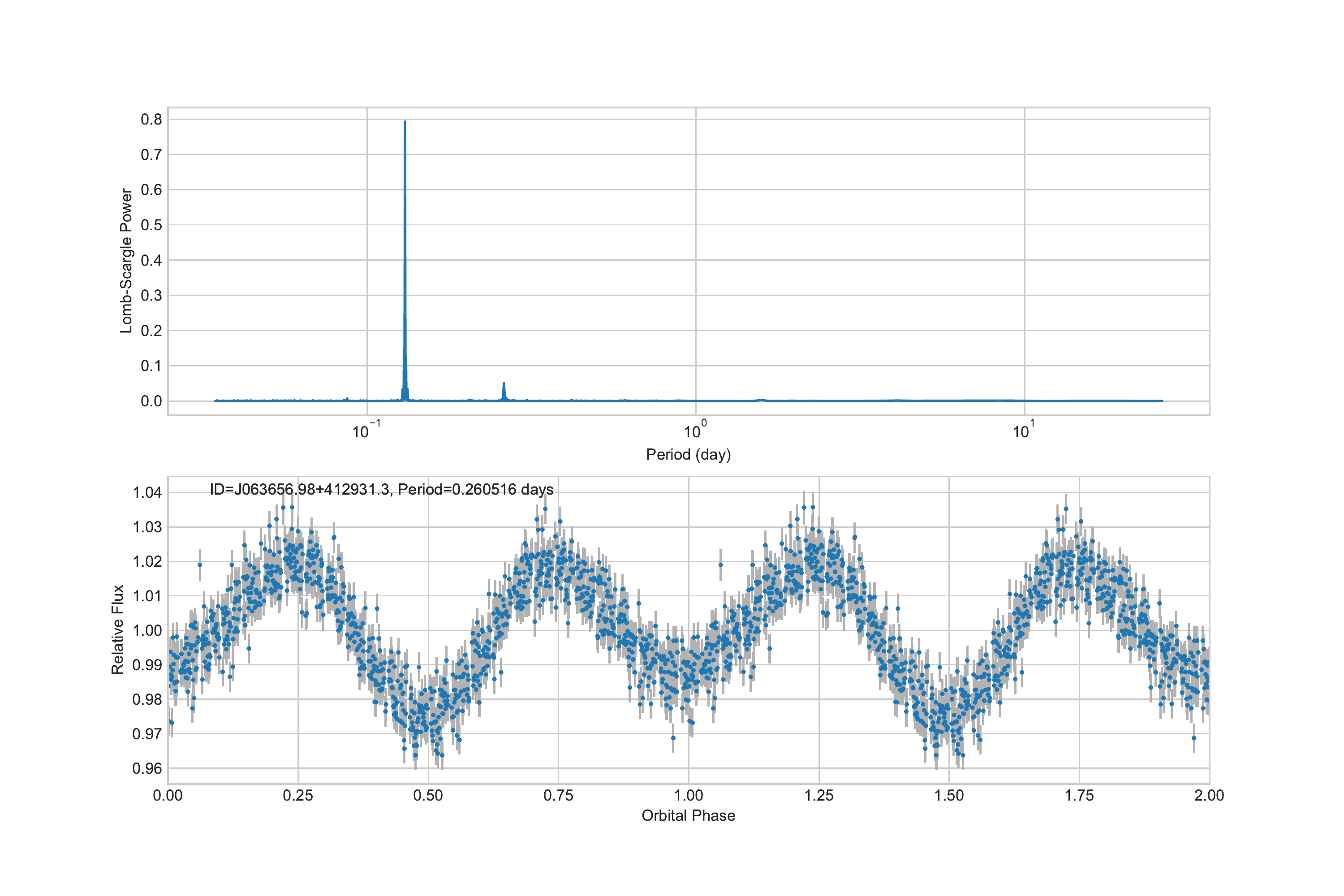}
 \includegraphics[angle=0,width=0.49\textwidth,height=0.2\textheight]{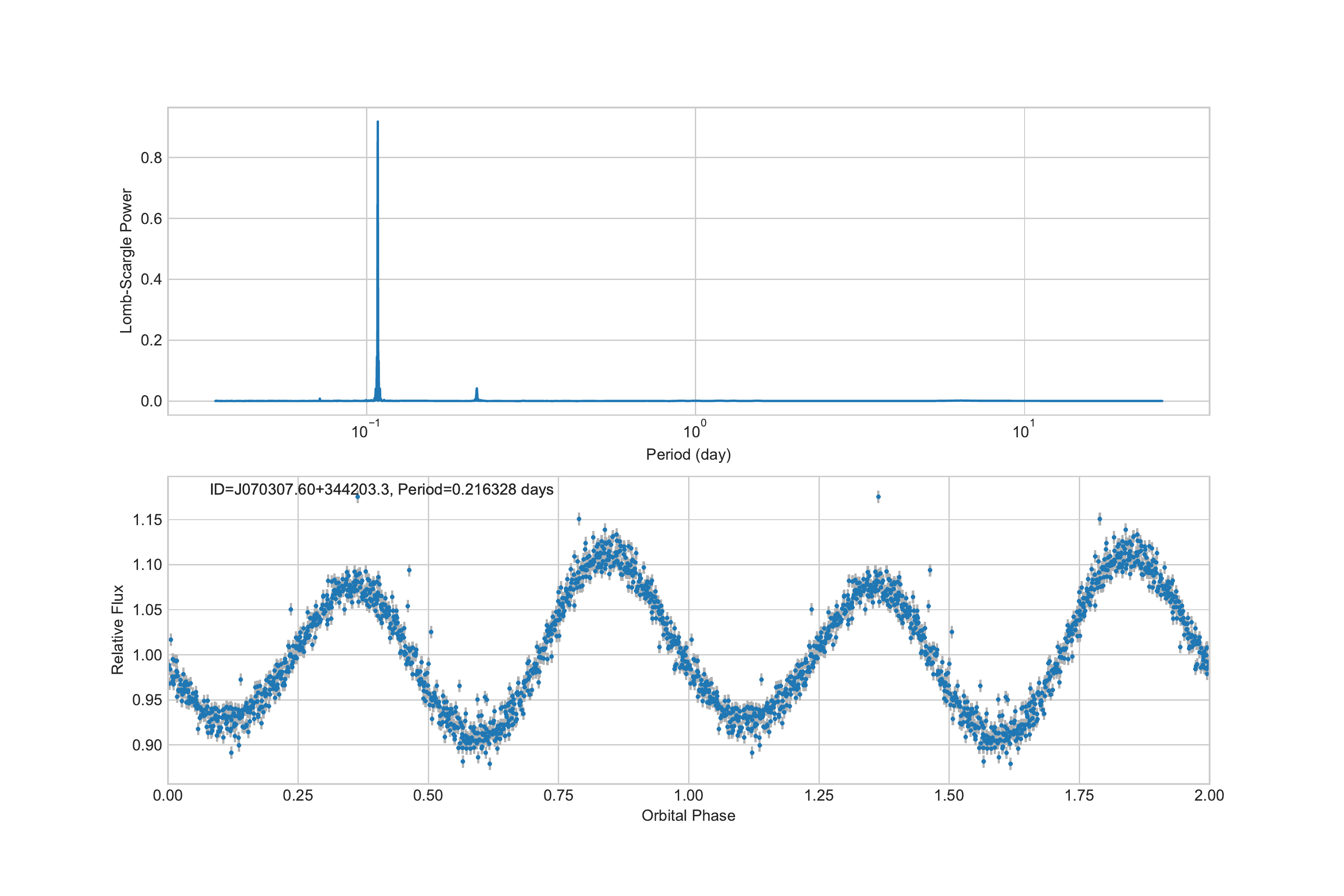}%
 \includegraphics[angle=0,width=0.49\textwidth,height=0.2\textheight]{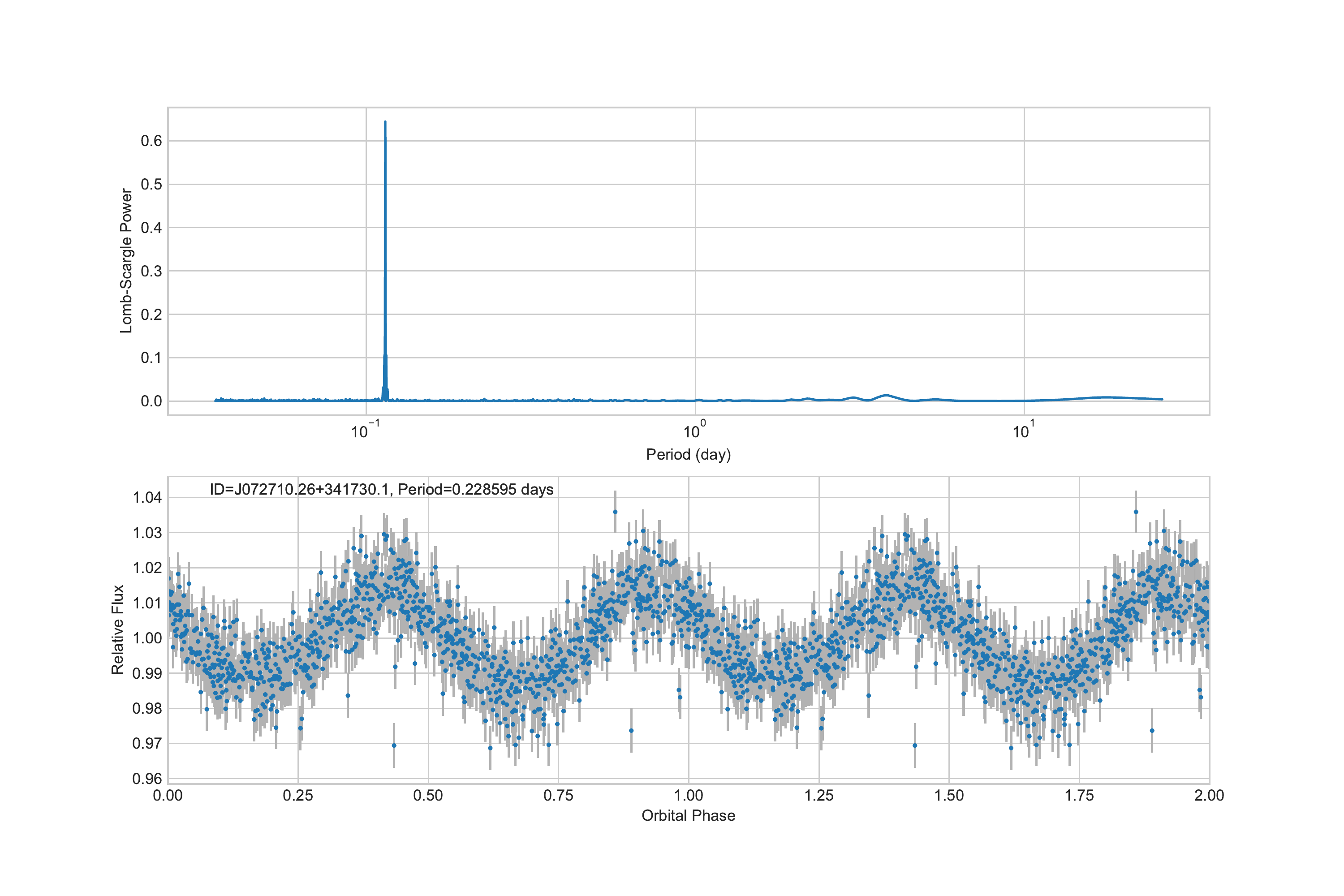}
 \includegraphics[angle=0,width=0.49\textwidth,height=0.2\textheight]{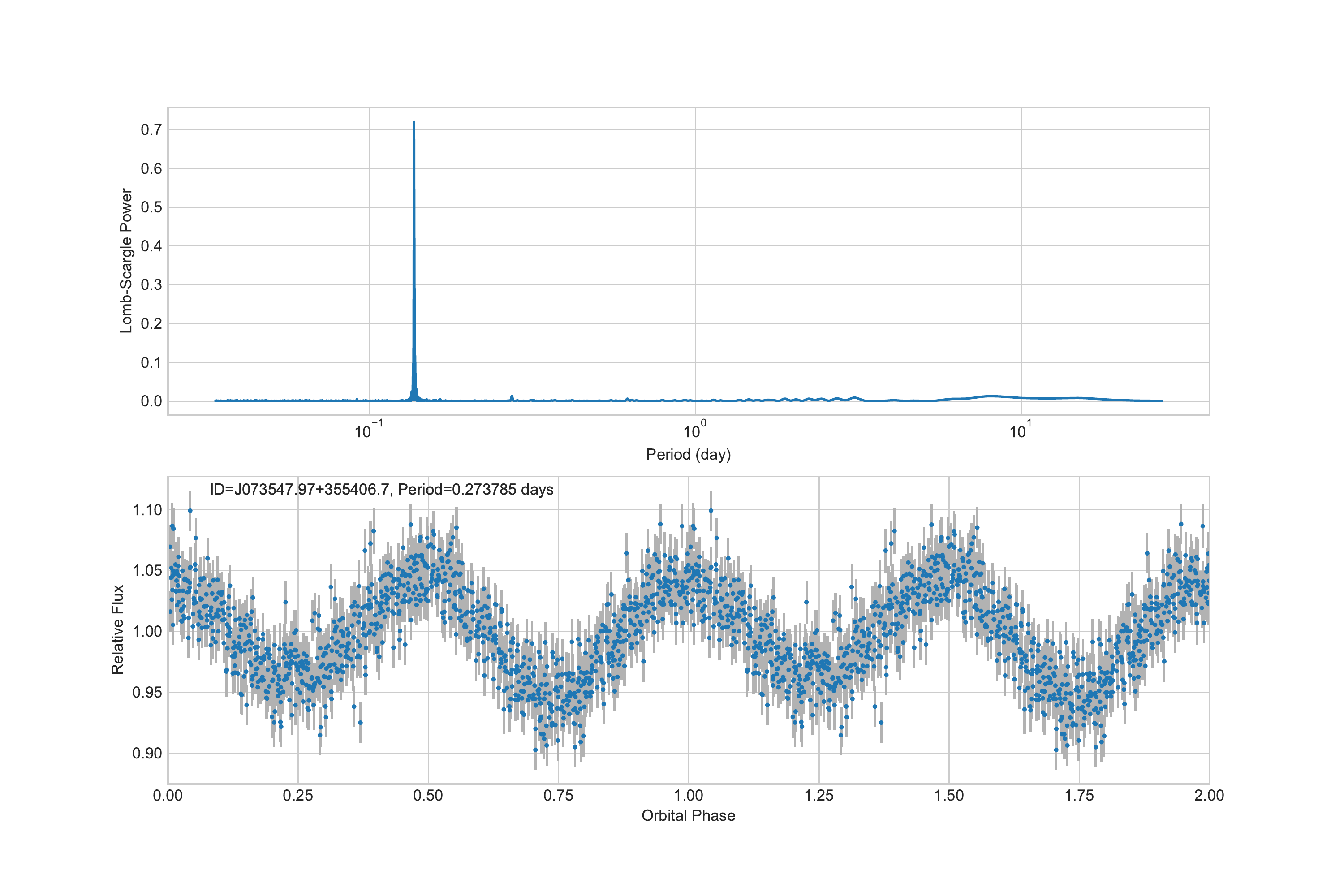}%
 \includegraphics[angle=0,width=0.49\textwidth,height=0.2\textheight]{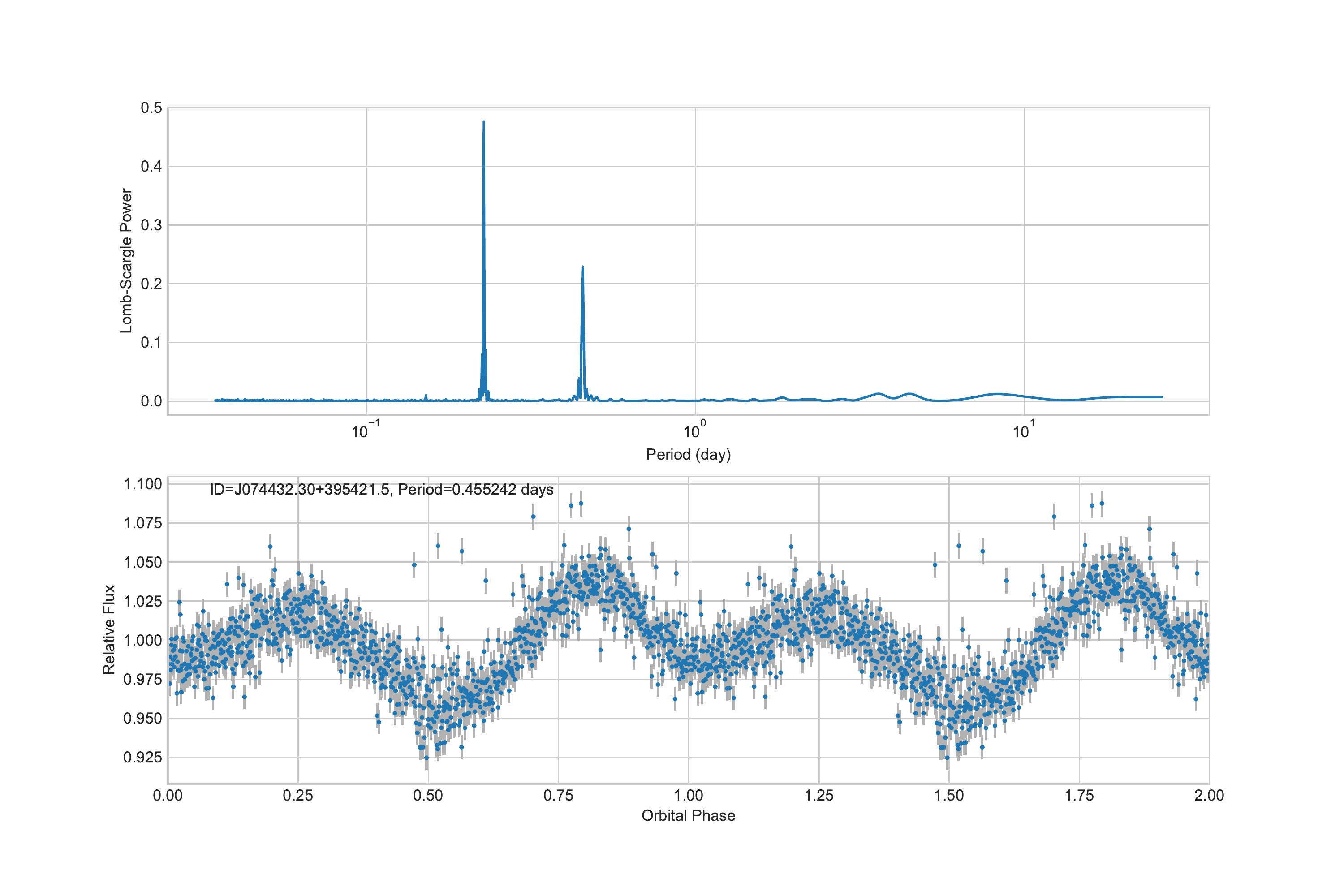}
 \caption{TESS light curves of ellipsoidal variables.}
 \end{figure}
  \addtocounter{figure}{-1}

 \begin{figure}
 \centering
 \includegraphics[angle=0,width=0.49\textwidth,height=0.2\textheight]{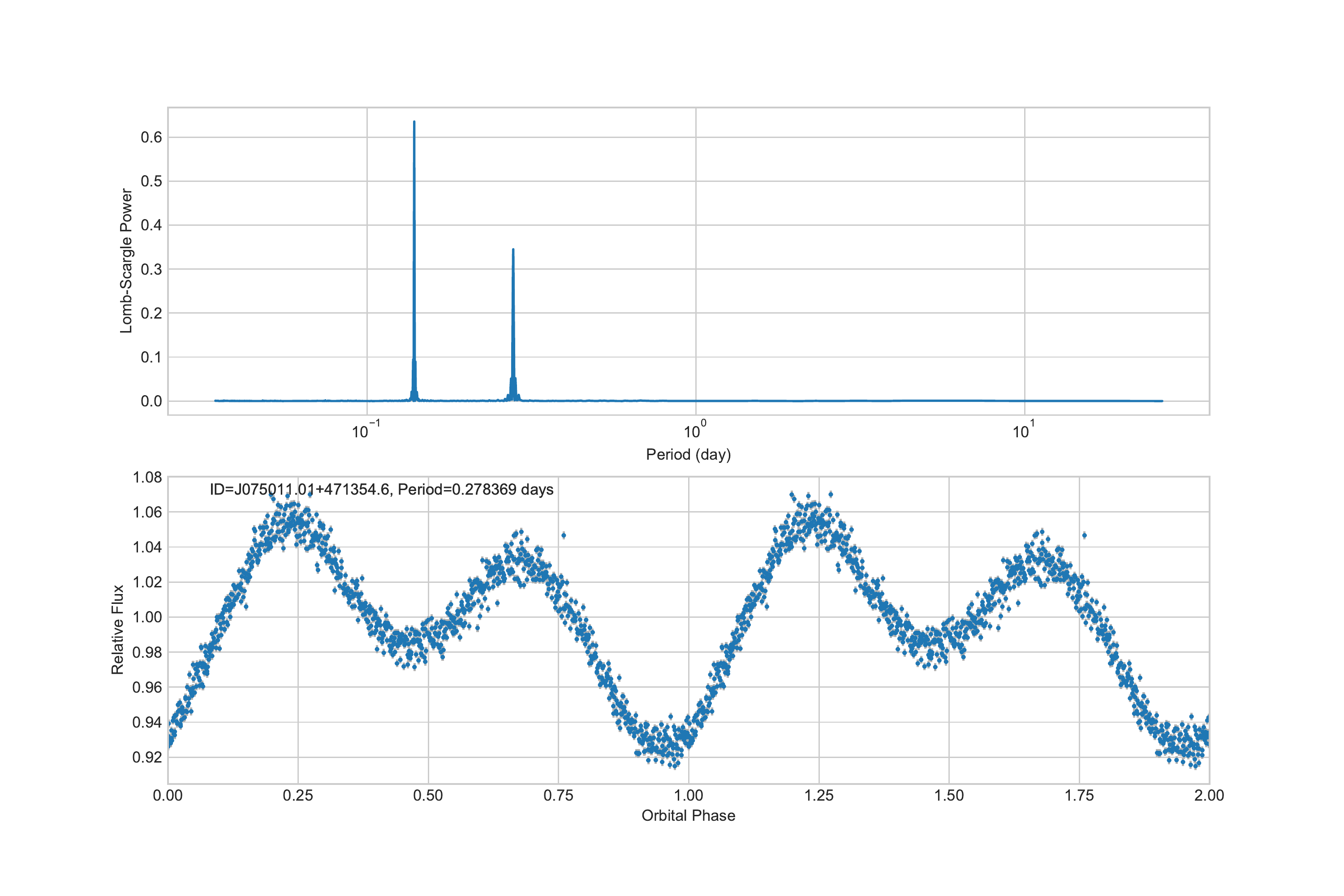}%
 \includegraphics[angle=0,width=0.49\textwidth,height=0.2\textheight]{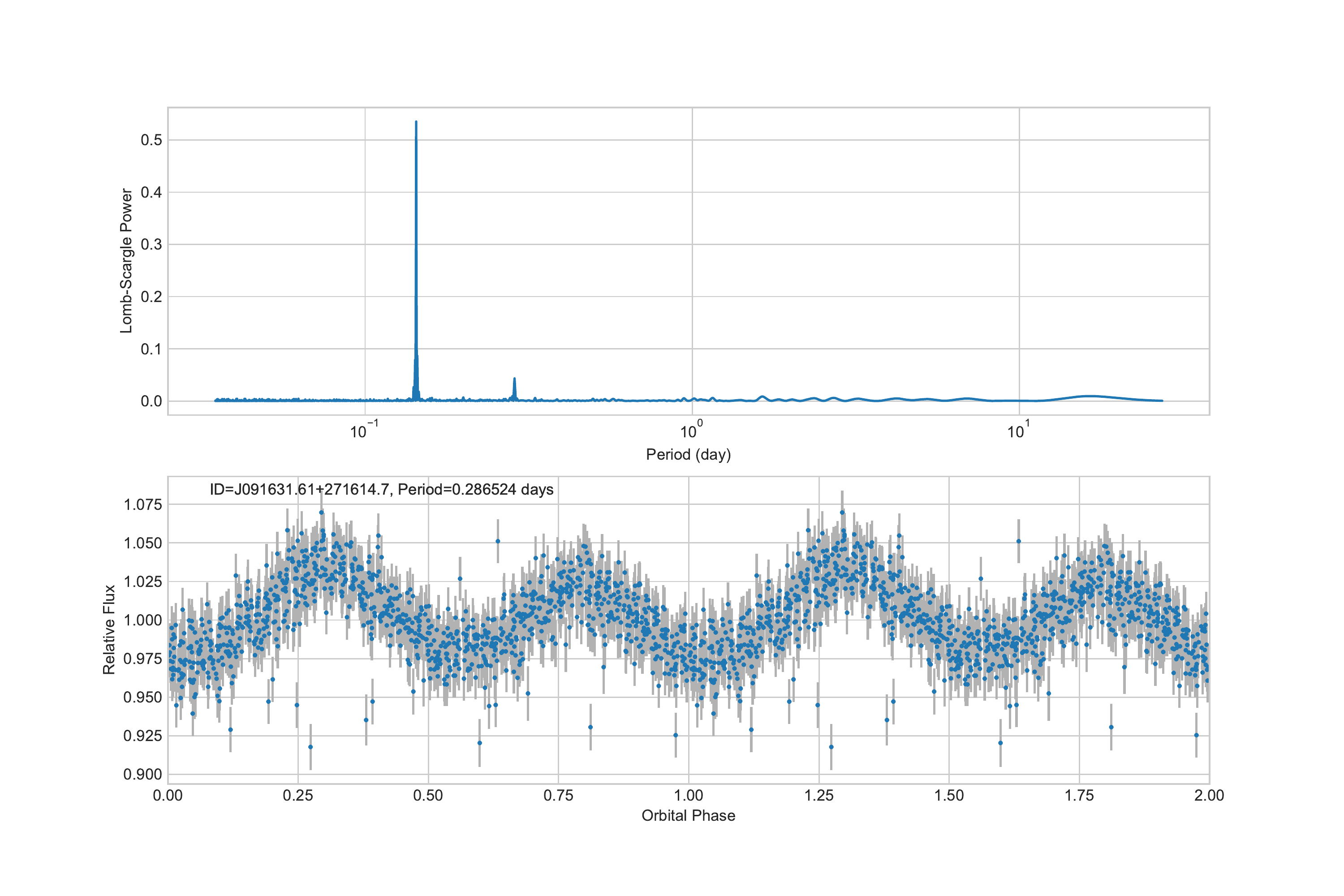}
 \includegraphics[angle=0,width=0.49\textwidth,height=0.2\textheight]{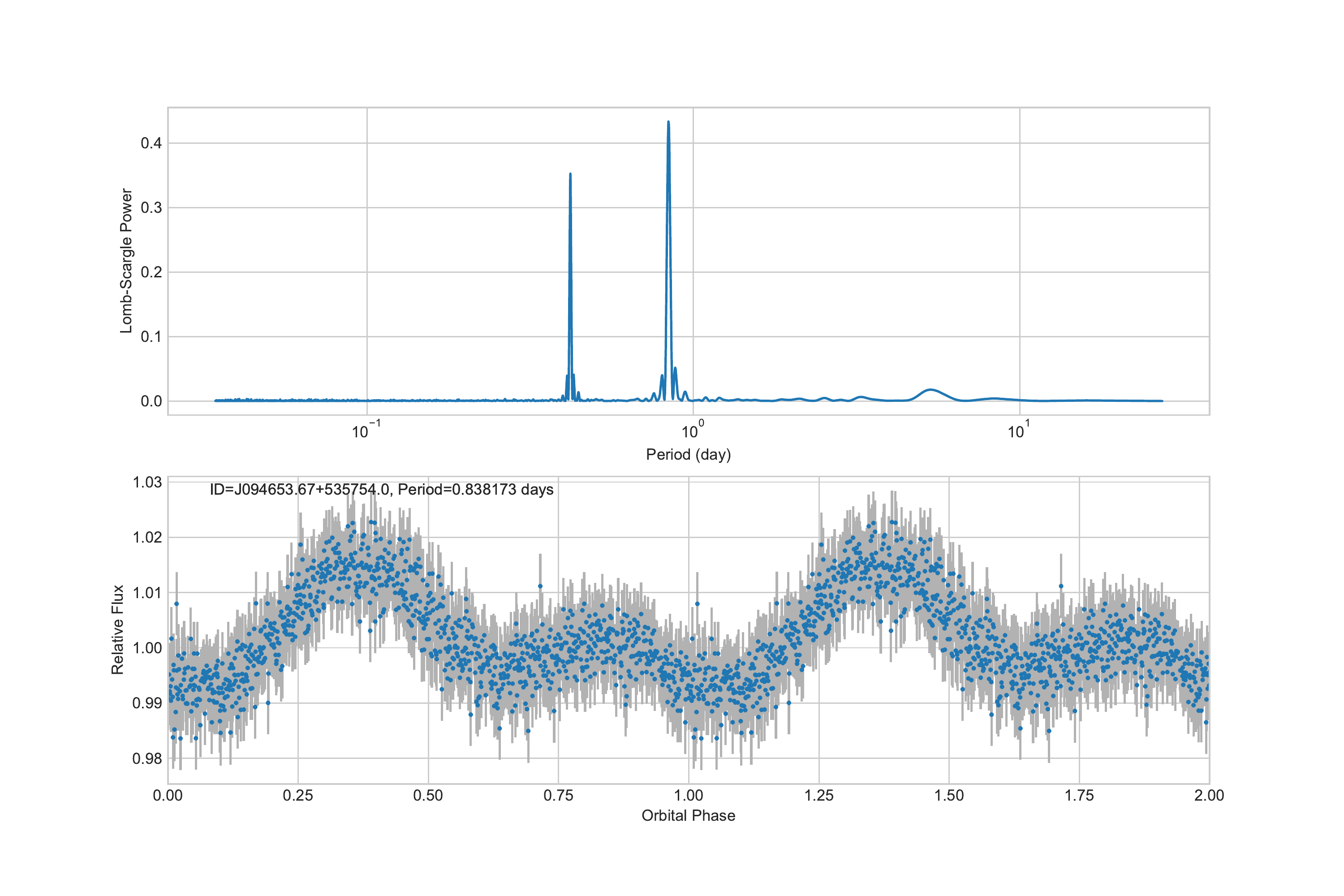}%
 \includegraphics[angle=0,width=0.49\textwidth,height=0.2\textheight]{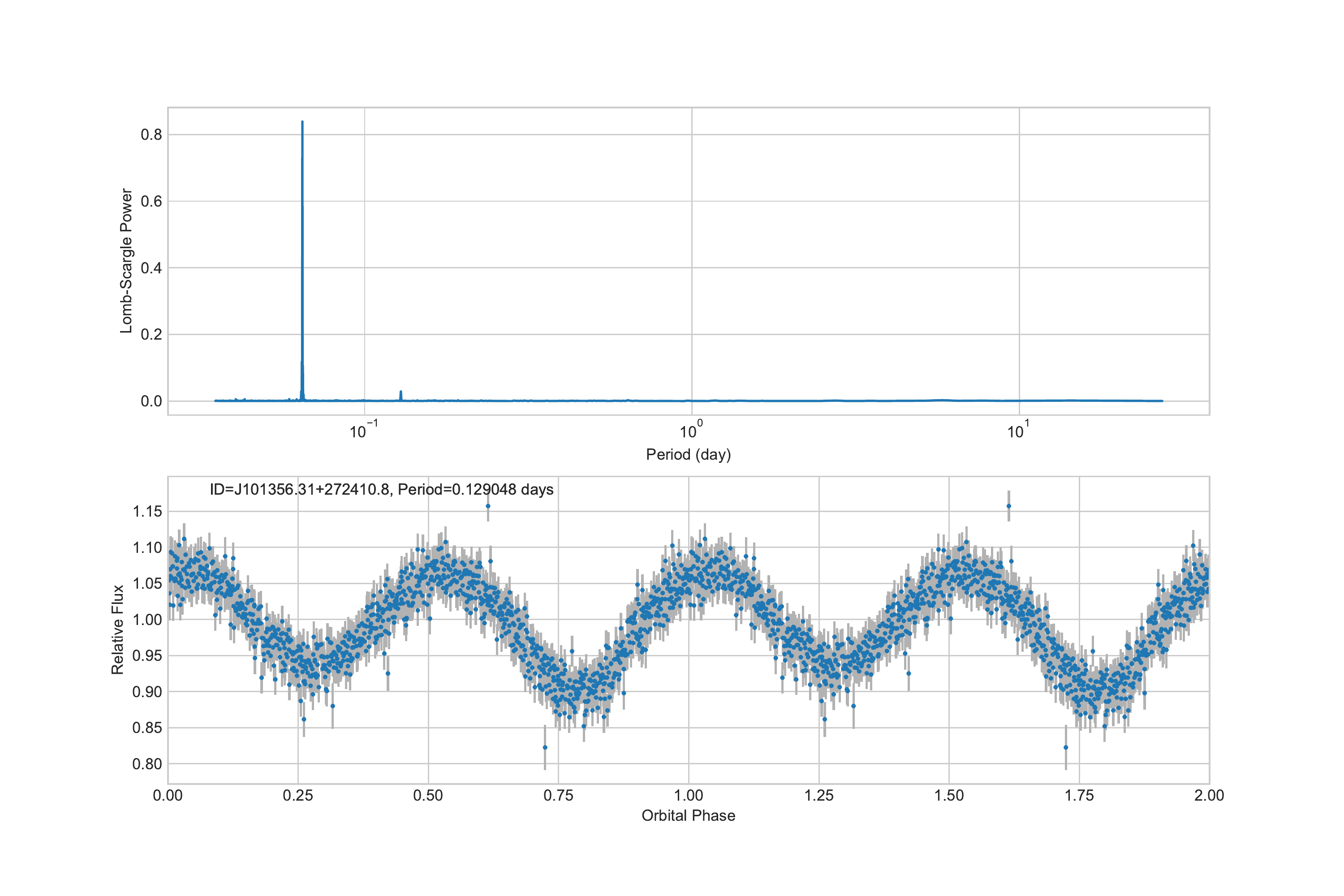}
 \includegraphics[angle=0,width=0.49\textwidth,height=0.2\textheight]{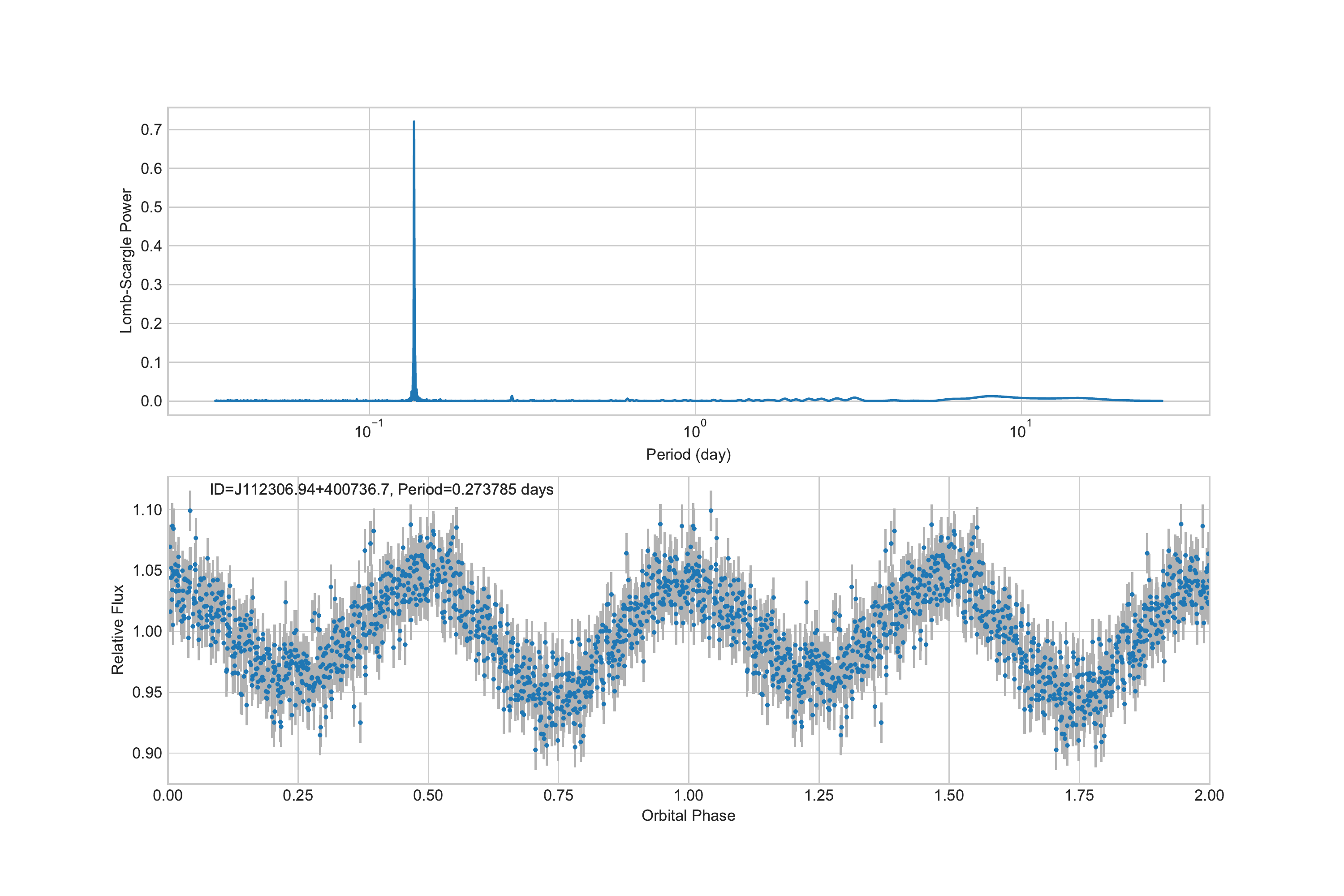}%
 \includegraphics[angle=0,width=0.49\textwidth,height=0.2\textheight]{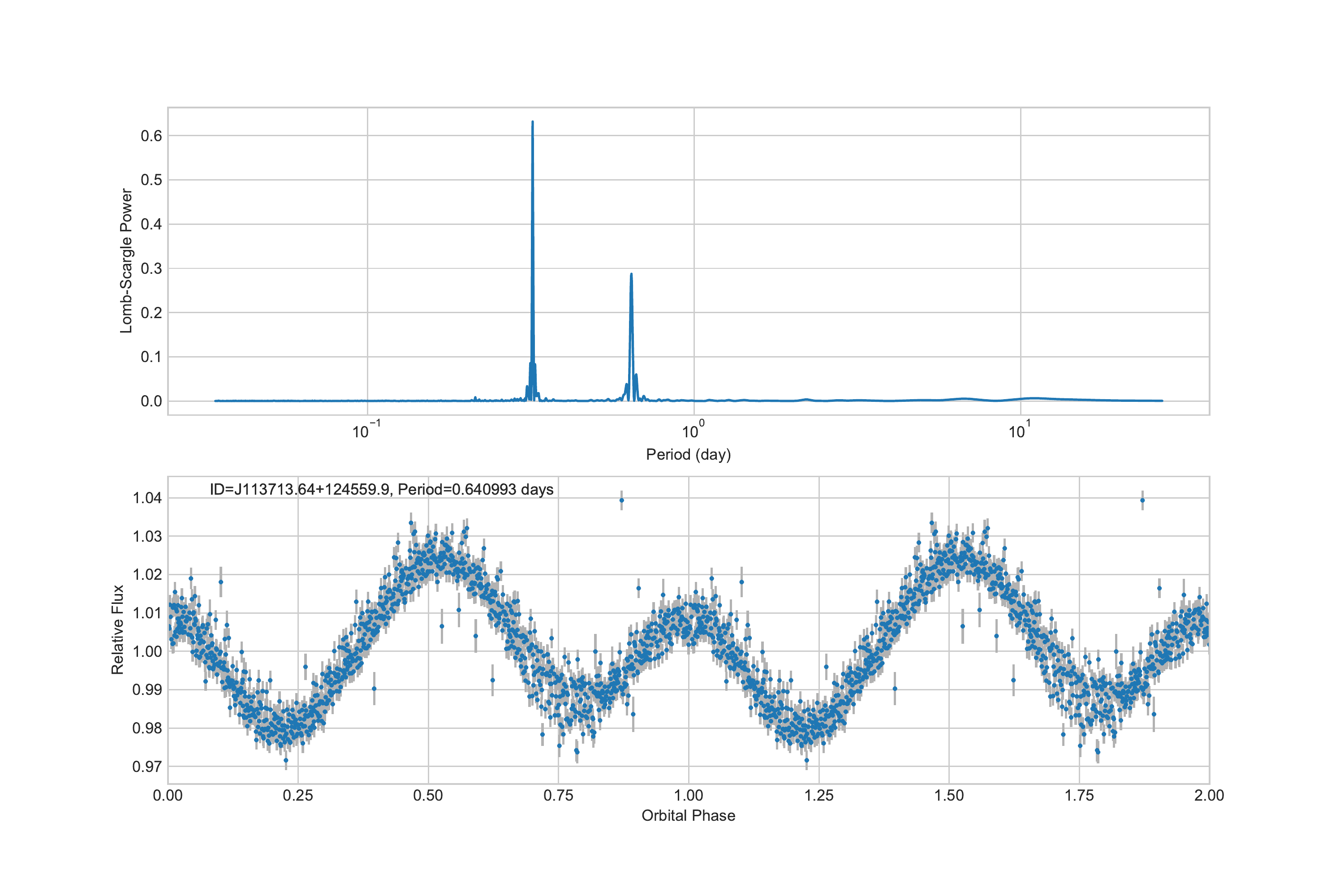}
 \includegraphics[angle=0,width=0.49\textwidth,height=0.2\textheight]{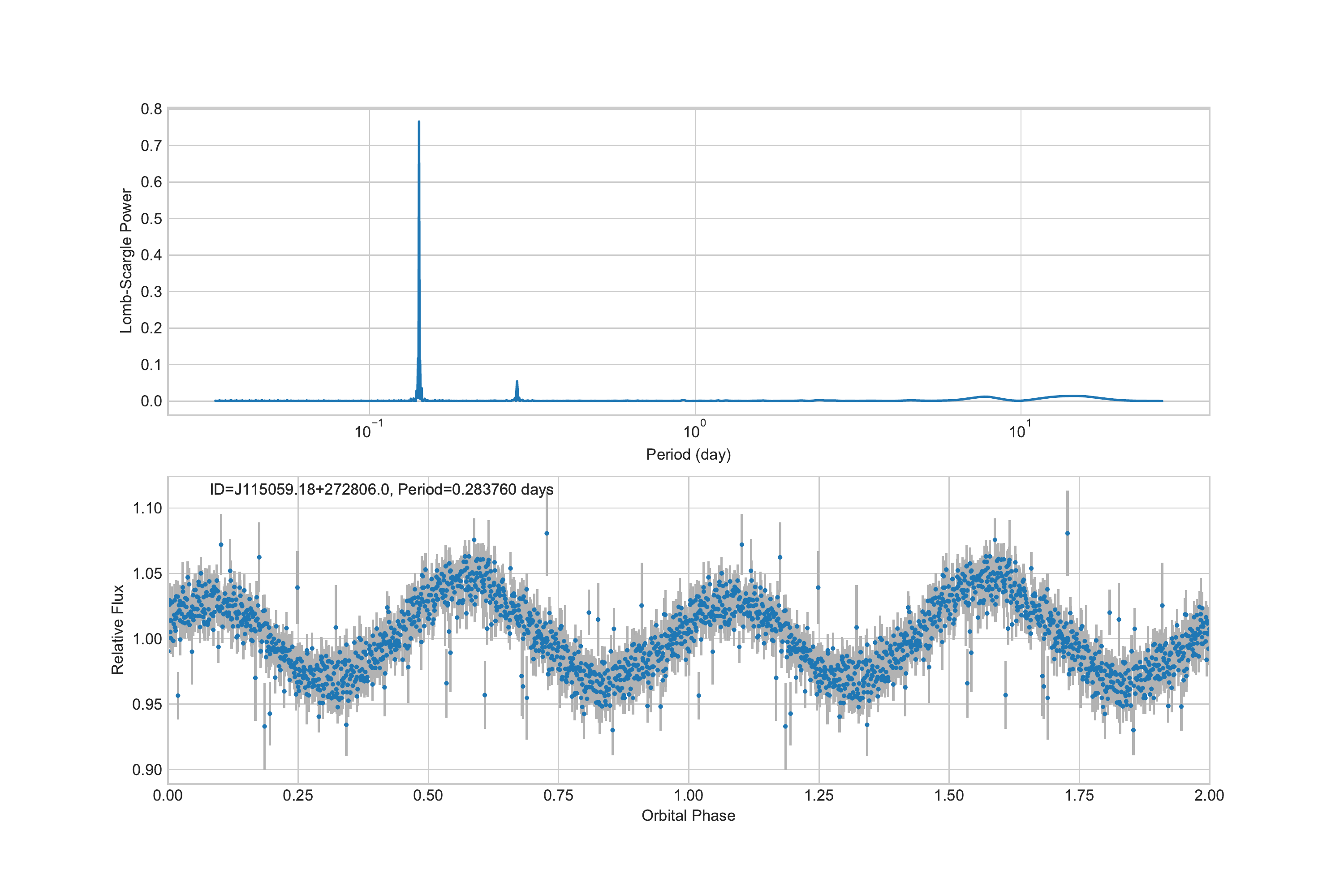}%
 \includegraphics[angle=0,width=0.49\textwidth,height=0.2\textheight]{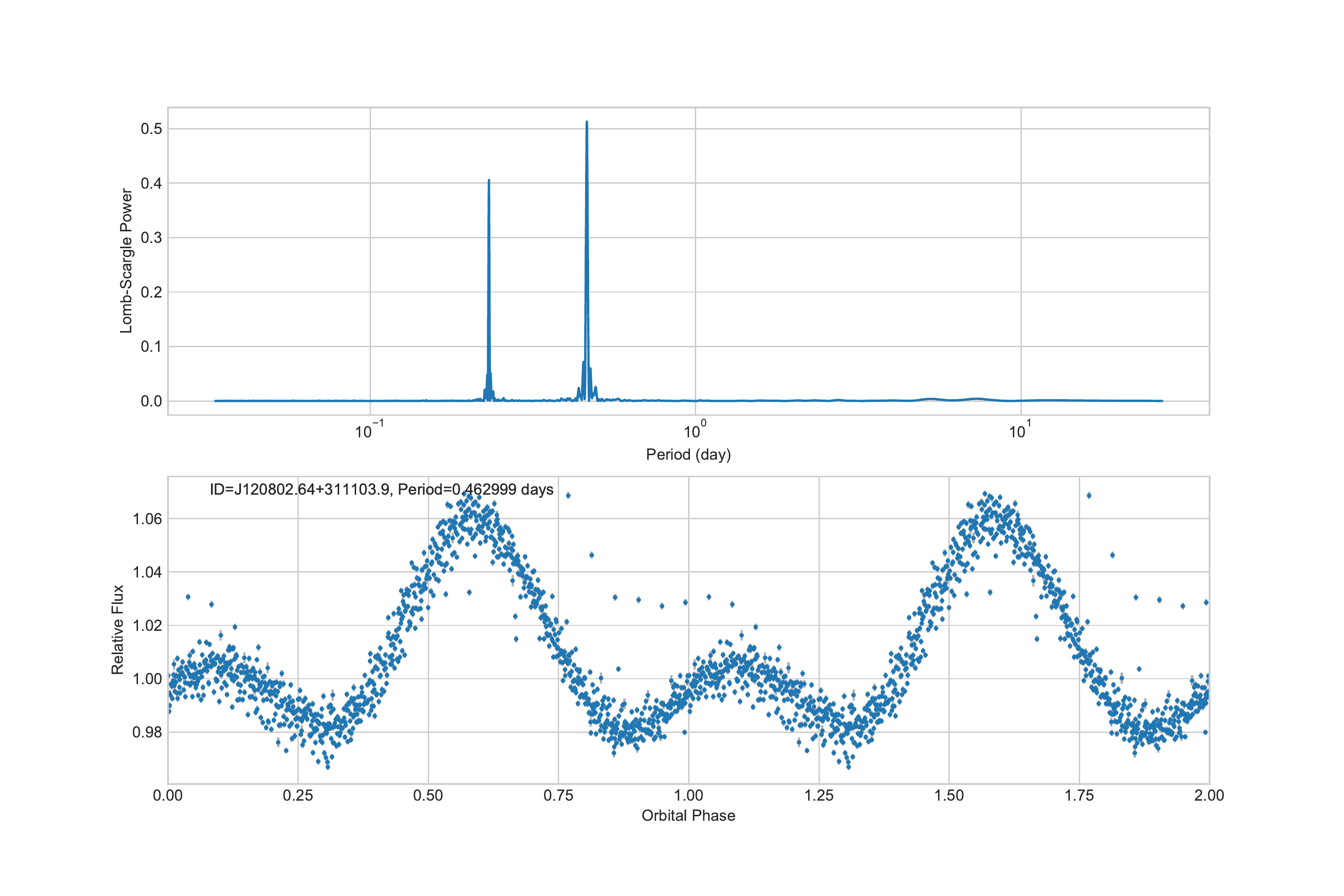}
 \caption{\it Continued.}
 \end{figure}
  \addtocounter{figure}{-1}

 \begin{figure}
 \centering
 \includegraphics[angle=0,width=0.49\textwidth,height=0.2\textheight]{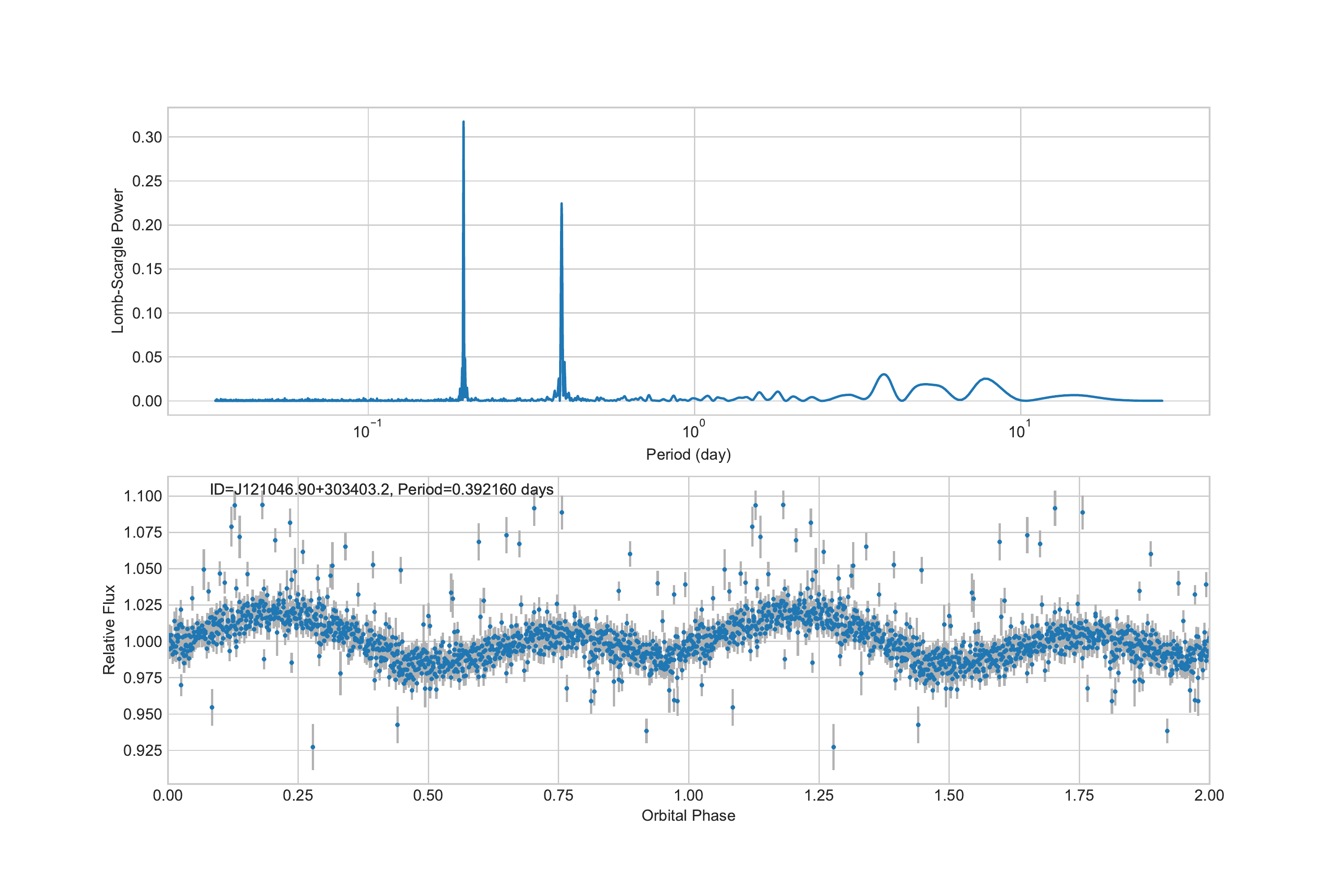}%
 \includegraphics[angle=0,width=0.49\textwidth,height=0.2\textheight]{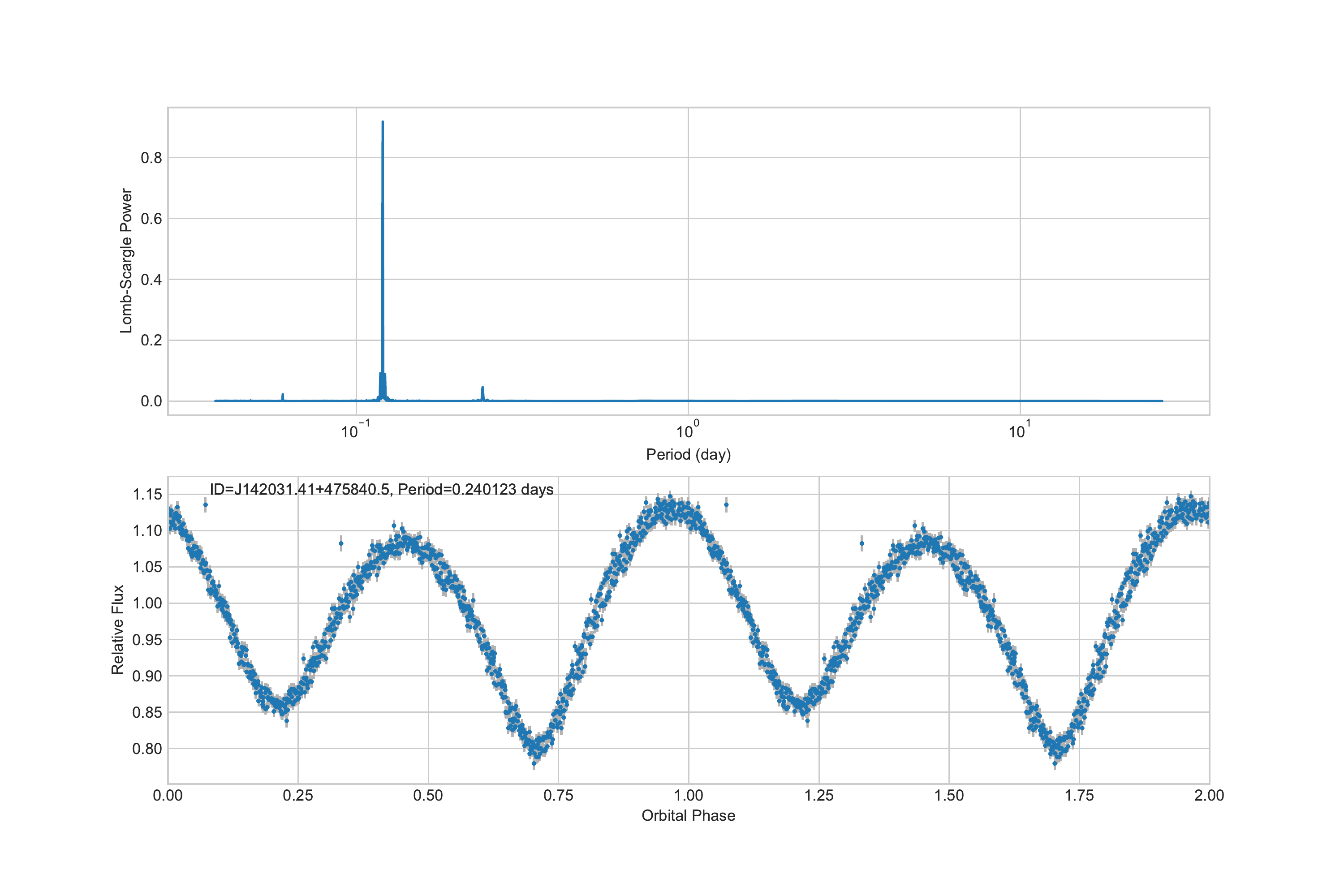}
 \includegraphics[angle=0,width=0.49\textwidth,height=0.2\textheight]{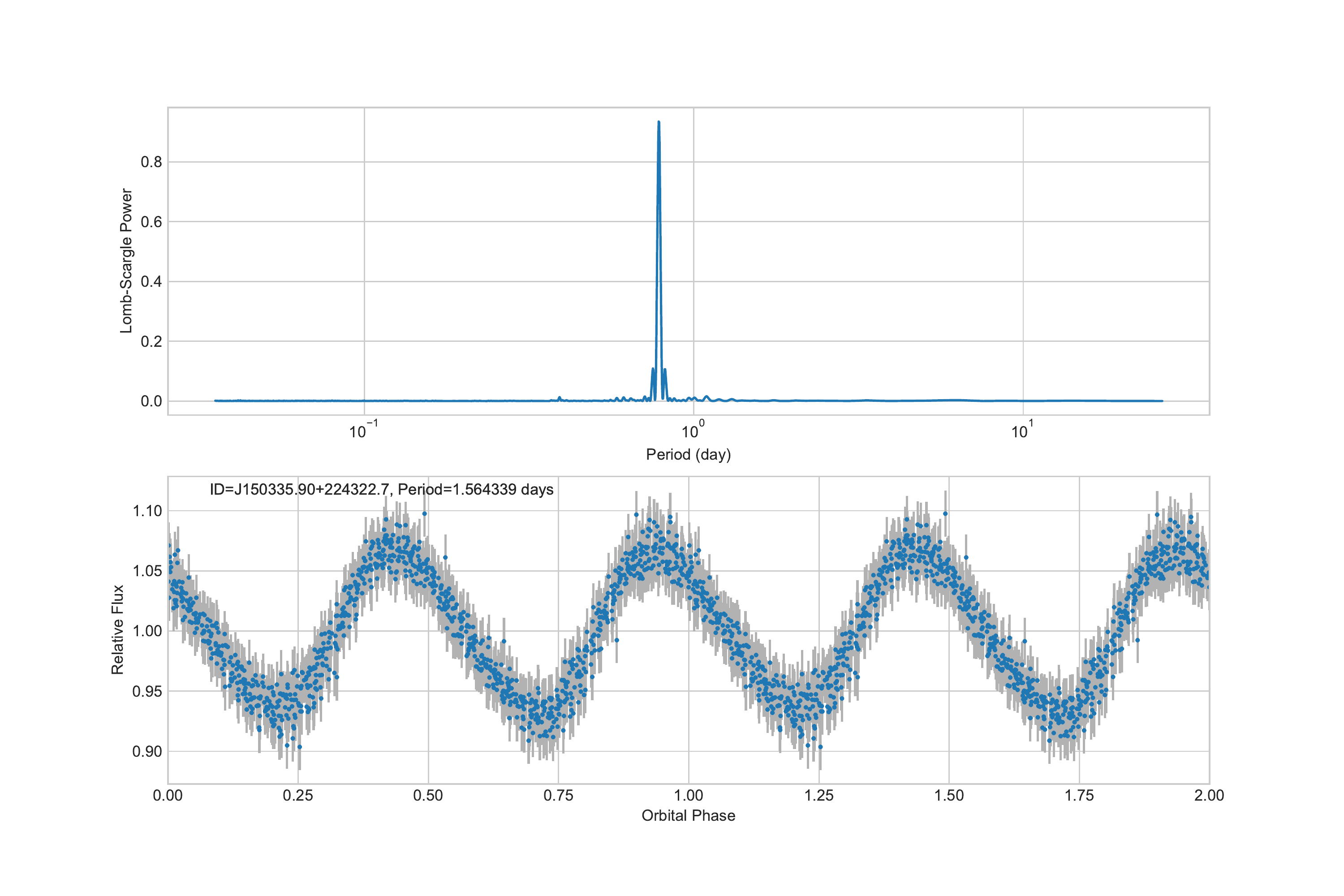}%
 \caption{\it Continued.}
 \end{figure}

 \clearpage

 \begin{figure}[H]
 \centering
 \includegraphics[angle=0,width=1.0\textwidth,height=0.9\textheight]{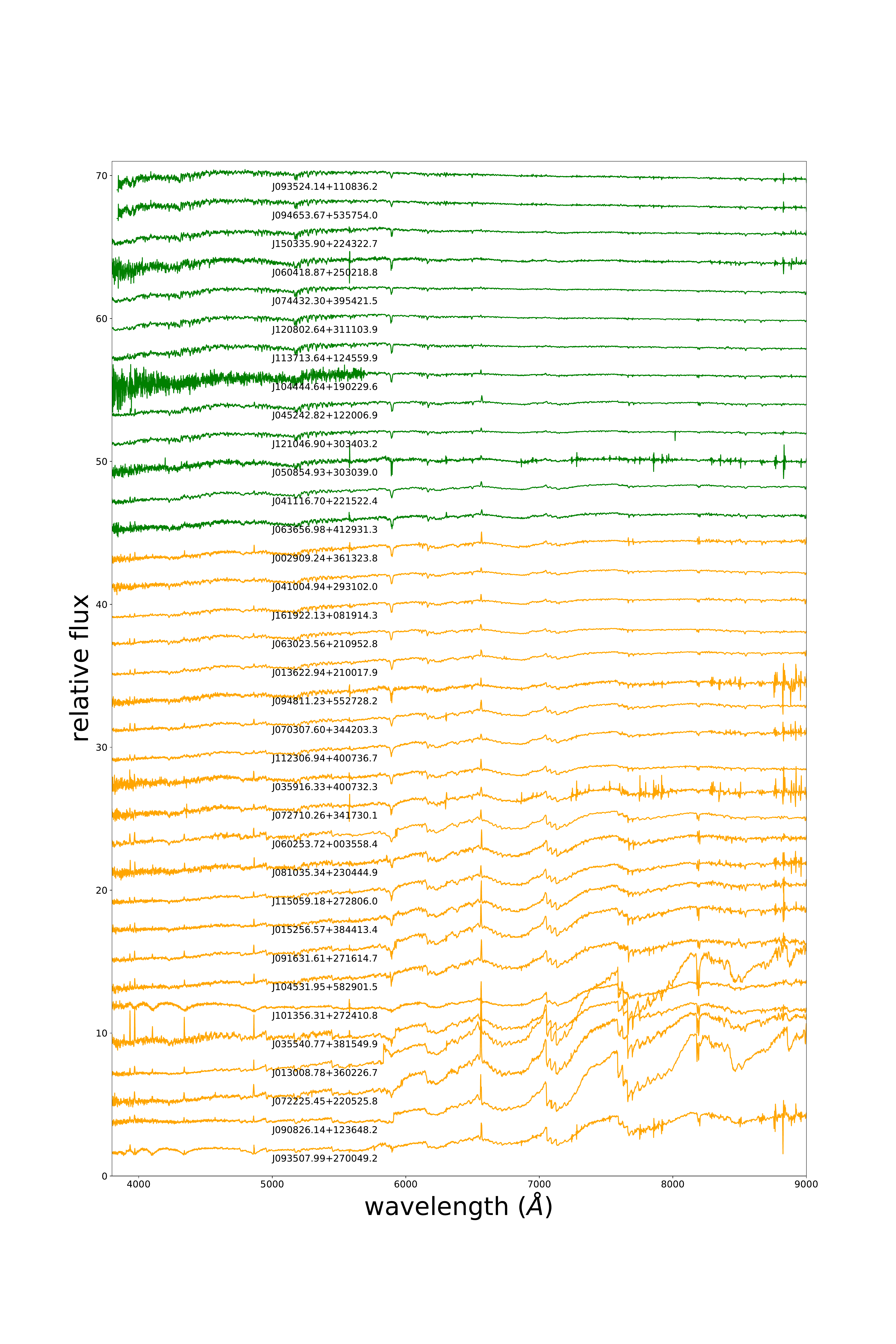}%
 \caption{The combined spectra of the $35$ sources provided by LAMOST.
 The olive spectra above and the orange spectra below represent the K-type and M-type of the PyHammer classification, respectively.}
 \end{figure}

 \begin{table}[t]
 \footnotesize
 \caption{Gaia astrometric quality flags for K/M-dwarfs in our sample.}
 \label{T-new}
 \tabcolsep 12pt
 \centering
 \begin{tabular*}{\textwidth}{cccc|cccc}
  \hline
 Designation&\texttt{ruwe}&$\epsilon_{i}$&$BP/RP$& Designation&\texttt{ruwe}&$\epsilon_{i}$& $BP/RP$ \\
 &&& flux excess& &&& flux excess\\\hline
 $[1]$&$[2]$ &$[3]$ &$[4]$ &$[5]$&$[6]$ &$[7]$ &$[8]$\\\hline
 J001057.66+470307.2&0.954&0&1.3&J074432.27+395421.8&1.045&0.031&1.25\\
 J002909.24+361323.8&1.046&0.112&1.338&J074829.03+315632.7&1.014&0&1.286\\
 J003750.29+452754.1&0.947&0&1.303&J075011.01+471354.6&1.007&0&1.275\\
 J011933.95+395031.7&1.066&0.089&1.246&J075642.26+473917.9&0.926&0&1.252\\
 J012352.68-025010.7&1.407&0.223&1.266&J081035.34+230444.9&1.033&0.052&1.394\\
 J013008.78+360226.7&1.133&0.13&1.505&J084835.68+271739.1&0.978&0&1.205\\
 J013541.98+383916.1&1.484&0.276&1.297&J090826.14+123648.2&1.144&0.121&1.529\\
 J013622.94+210017.9&1&0&1.339&J091631.62+271614.8&1.013&0&1.416\\
 J014217.02+330016.9&8.18&2.469&1.532&J091931.31+571730.2&1.419&0.119&1.227\\
 J014248.23+365324.9&1.018&0.072&1.277&J093507.99+270049.2&1.08&0.021&1.433\\
 J015106.54+372442.7&1.032&0.059&1.37&J093524.14+110836.2&1.151&0.171&1.251\\
 J015256.57+384413.4&1.006&0&1.399&J094653.65+535754.5&1.006&0&1.294\\
 J025856.43+234406.2&1.062&0.089&1.281&J094811.23+552728.2&1.04&0&1.402\\
 J033655.85+192321.5&1.023&0&1.268&J100319.08+203103.2&0.989&0&1.253\\
 J034210.62+384219.9&0.994&0&1.277&J101356.31+272410.8&1.072&0.159&1.416\\
 J035540.77+381549.9&1.058&0.101&1.463&J104444.64+190229.6&1.053&0.076&1.28\\
 J035829.70+232454.3&0.954&0&1.258&J104531.95+582901.5&1.007&0&1.389\\
 J035916.33+400732.3&0.959&0&1.374&J112306.94+400736.7&1.008&0&1.365\\
 J040441.99+022425.1&0.989&0&1.402&J113326.70+313108.1&1.03&0.028&1.29\\
 J041004.94+293102.0&1.498&0.383&1.385&J113713.64+124559.9&1.023&0&1.246\\
 J041116.70+221522.4&1.012&0&1.391&J114114.74-050825.5&1.067&0.099&1.31\\
 J045242.82+122006.9&1.092&0.026&1.324&J115041.48+331651.0&1.039&0.769&1.423\\
 J050854.93+303039.0&1.018&0.037&1.344&J115059.18+272805.8&1.086&0.086&1.385\\
 J053047.77+304441.2&1.029&0&1.31&J120803.04+311103.9&1.025&0.13&1.253\\
 J053757.36+220843.0&1.01&0.057&1.276&J121046.90+303403.4&1.29&0.175&1.292\\
 J060253.72+003558.4&1.206&0.092&1.424&J123151.59+252230.8&0.908&0&1.257\\
 J060418.87+250218.8&0.979&0&1.306&J123529.01+255425.4&0.946&0&1.273\\
 J062952.18+544013.4&1.17&0.147&1.244&J133428.27+492659.3&1.095&0.038&1.28\\
 J063023.56+210952.8&0.979&0&1.346&J133751.25+360653.0&4.242&0.479&1.288\\
 J063656.98+412931.3&1.006&0&1.347&J135225.36+442403.1&1.02&0&1.278\\
 J064455.26+505423.6&1.342&0.195&1.317&J140408.29+035337.0&1.044&0.067&1.253\\
 J065442.51+472130.5&3.927&0.515&1.239&J142031.41+475840.5&1.991&0.327&1.263\\
 J070307.60+344203.3&1.057&0.063&1.344&J150335.90+224322.7&1.021&0.038&1.26\\
 J071349.17+154644.4&0.982&0&1.238&J150916.03+360200.1&0.996&0.059&1.224\\
 J071934.24+321719.4&0.972&0&1.273&J152209.35+254450.5&1.34&0.103&1.283\\
 J072225.45+220525.8&0.918&0&1.449&J152843.82+205140.2&1.04&0.047&1.272\\
 J072652.92+272232.1&8.265&1.512&1.48&J153007.93+431500.4&1.084&0.076&1.201\\
 J072710.26+341730.1&0.978&0&1.362&J153854.15+361453.3&0.998&0.028&1.286\\
 J072842.59+471224.6&0.934&0&1.325&J161922.13+081914.3&1.049&0&1.32\\
 J073547.97+355406.7&1.034&0.054&1.248&J172613.60-001623.3&0.898&0&1.281\\
 J074055.15+531814.4&1.054&0.048&1.302&&&&\\
 \hline
 \end{tabular*}
 \begin{tablenotes}
 \item[*]
 $[1]$ and $[5]$: target designation.\\
 $[2]$ and $[6]$: the renormalised unit weight error.\\
 $[3]$ and $[7]$: excess noise of the source: astrometric\_excess\_noise.\\
 $[4]$ and $[8]$: BP/RP flux excess factor: phot\_bp\_rp\_excess\_factor.\\
 \end{tablenotes}
 \end{table}


\end{document}